\documentclass[aps,prl,superscriptaddress, onecolumn, amsmath, amssymb]{revtex4}
\usepackage{graphicx}
\usepackage{amsmath,amssymb}
\usepackage{dcolumn}
\usepackage{mathrsfs}
\usepackage{physics}
\usepackage{float}
\usepackage{booktabs}
\usepackage{svg}
\usepackage{comment}

\usepackage[american]{babel}
\addto\captionsenglish{}

\usepackage{color}

\begin{document}
\title{Spectra of laser diodes}
\date{\today}
\author{Bjarne Tromborg}
\affiliation{Department of Electrical and Photonics Engineering, Technical University of Denmark, Building 343, 2800 Kongens Lyngby, Denmark}	
\author{Palle Jeppesen}
\email[]{pjep@fotonik.dtu.dk}
\affiliation{Department of Electrical and Photonics Engineering, Technical University of Denmark, Building 343, 2800 Kongens Lyngby, Denmark}

\begin{abstract}
This paper provides an introduction to the theory of semiconductor laser diodes, with special focus on their noise properties. It may be considered an additional chapter to the textbook \cite{Jeppesen-2023}. As such, it will also refer to equations in that book.
\end{abstract}
\maketitle

\section{Introduction}

The performance of an optical communication system may be strongly
influenced by the noise properties of the applied light source(s).
There are many types of rival light sources used in optical communication
systems (laser diodes, fiber laser, solid state lasers, ...) but semiconductor
lasers or laser diodes are still by far the most popular. In this
chapter we will consider a simple mathematical model for a generic
semiconductor laser that incorporates many essential features of the
static and dynamic behavior of real devices. In particular, the model
allows calculation of important spectra (field power spectra, relative
intensity noise spectra and frequency noise spectra) that characterize
noise properties of laser diodes. The topic of semiconductor lasers
or laser diodes is treated in many textbooks \cite{Agrawal-1993,Coldren-1995,Yariv-1997,Zory-1993,Ohtsubo-2010}.
We refer especially to \cite{Agrawal-1993,Coldren-1995} for detailed
introductions to the theory of laser diodes and the basics about electronic
structure and optical properties of semiconductor materials. However,
the detailed theory is not a condition for understanding our model.

This paper serves as an Appendix to the textbook \cite{Jeppesen-2023}. References to figures and equations in that book are indicated as Fig.\ X-\cite{Jeppesen-2023}, and Eq.\ X-\cite{Jeppesen-2023}, respectively.

\section{Rate equations for carrier and photon numbers\label{sec: rate equations}}

\begin{figure}
\begin{centering}
\includegraphics[scale=0.4]{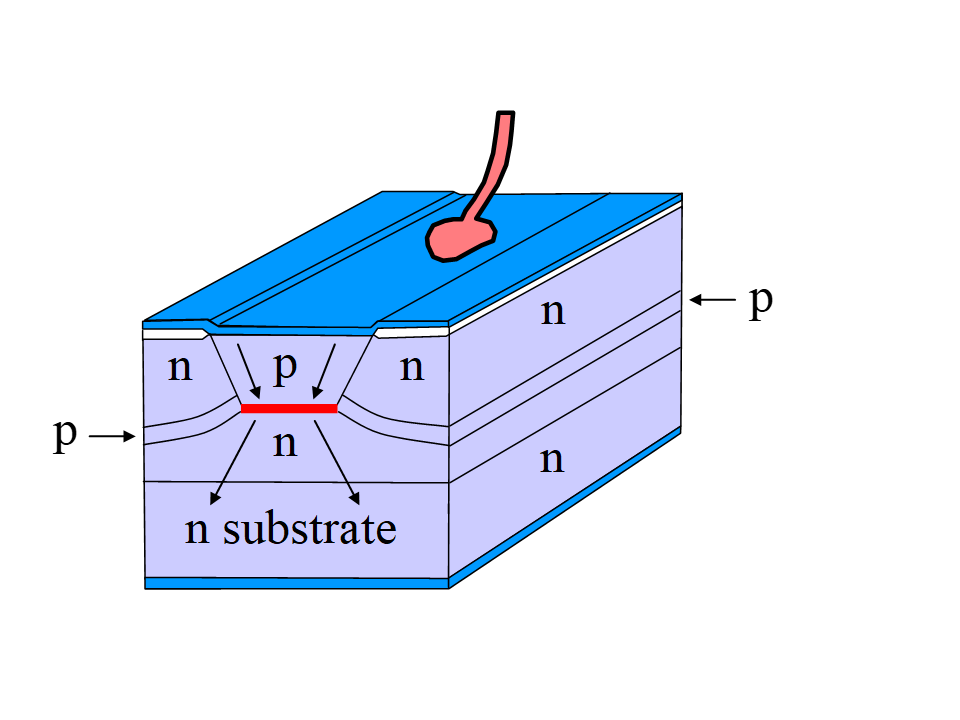}
\par\end{centering}

\caption{Schematic of an edge emitting laser diode. The red layer is the active
layer sandwiched between p-doped and n-doped host material of InP.
The current is guided through the active layer by current blocking
layers of n- and p-type InP. The figure is not showing the relative
scales correctly. The active layer is typically around 2 \foreignlanguage{american}{\textmu{}m}
wide and 0.2 \foreignlanguage{american}{\textmu{}m} thick while the
laser chip is around 300 \foreignlanguage{american}{\textmu{}m} long,
150 \foreignlanguage{american}{\textmu{}m} wide and 10 \foreignlanguage{american}{\textmu{}m}
high.}

\label{Edge emitting laser diode}
\end{figure}

The basic principles of operation of a laser diode can be illustrated
by the device in Figure \ref{Edge emitting laser diode}. It shows
a schematic of an edge emitting laser based on the semiconductor material
InP - a binary alloy of Indium (In) and Phosphor (P). It is a diode
and is made by epitaxial growth of layers of n-type and p-type doped
InP on an n-type doped substrate of InP. At the pn-junction there
is a thin layer of quaternary alloy InGaAsP that also contains Gallium
(Ga) and Arsenide (As). It has a lower band gap than InP and is the
active layer where photons are generated. There are metal contacts
at top and bottom. When the diode is forward biased, a current flows
from top to bottom; it is carried by valence band holes in the p-type
layer and in the n-type layers it is carried by conduction band electrons
flowing opposite to the current.

\begin{figure}
\begin{centering}
\includegraphics[scale=0.6]{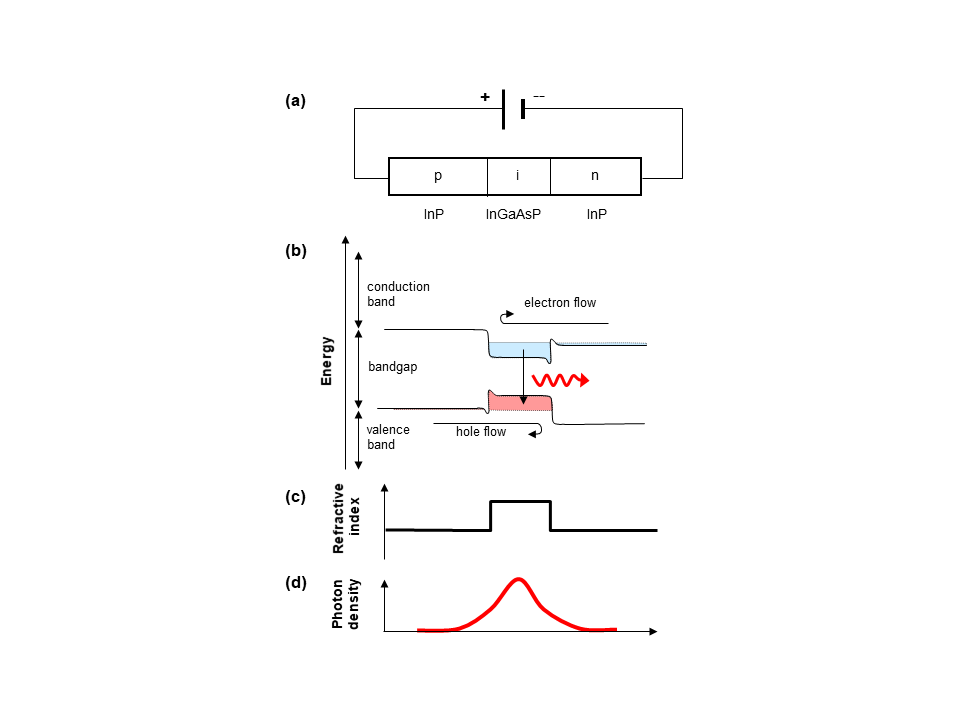}
\par\end{centering}

\caption{(a) Cross section through a forward biased active layer of undoped
InGaAsP sandwiched between p-type and n-type doped InP. (b) Schematic
of conduction and valence band edges across the active layer. (c)
Variation of refractive index and (d) of photon density.}

\label{pn junction in laser diode}
\end{figure}

Figure \ref{pn junction in laser diode} shows a schematic of the
variation of the valence and conduction band edges across the forward
biased diode. The holes and electrons meet in the active layer where
the lower band gap makes it act as a potential well for both holes
and electrons. The presence of both carriers confined by the well
in the same region allows electron-hole pairs to recombine and release
their energy difference by radiative recombination, i.e. by emission
of a photon with that energy. The relation between photon energy and
angular frequency $\omega$ is $E=\hbar\omega$ where $\hbar$ is
Planck's constant divided by $2\pi$. In order to improve the efficiency
of the laser diode it is usually constructed in such a way that the
current is restricted to pass the junction and the active layer in
a narrow stripe that is typically a couple of microns wide. An example
of a current confining structure is shown in Figure \ref{Edge emitting laser diode}
which has been made by etching away the p-type InP and active layer
outside a narrow central mesa and subsequent regrowth of current blocking
layers.

At the frequency of the emitted light the active layer has a higher
refractive index than the surrounding higher band gap material (see
Figure \ref{pn junction in laser diode}(c)). The layer therefore
forms a waveguide for the photons that are emitted within a certain
angle close to the direction of the waveguide. Radiative recombination
can be either spontaneous or stimulated. In stimulated emission the
recombination rate is proportional to the density of photons at the
energy of the electron-hole pair and the emitted photon belongs to
the same waveguide mode as the photons that stimulate the emission.
The opposite process where a photon is absorbed by exciting an electron
from the valence band to the conduction band is called stimulated
absorption. The rate of this type of absorption is also proportional
to the density of photons at the transition energy. If the rate of
stimulated emission is larger than the rate of stimulated absorption
a photon that is spontaneously emitted into the waveguide can start
a cascade of stimulated emissions of photons, that propagate along
the waveguide as an electromagnetic wave with increasing amplitude.
The wave is partly transmitted and partly reflected at the end facet
of the laser diode. The transmitted wave appears as output light from
the laser facet and the reflected light moves back along the waveguide
and is again amplified by stimulated emissions. The state of stable
laser operation is achieved when the light reflected at the facets
propagate back and forth along the waveguide while stimulated emissions
compensate for loss at the facets or by scattering or absorption along
the waveguide. At the same time the current has to replenish the carriers
in the active layer. The main processes that control the balance can
be represented by terms in a simple system of rate equations for the
carriers and photons.

Let $N$ be the number of electrons in the conduction band of the
active layer. For an un-doped active layer the condition of charge
neutrality implies that $N$ is also the number of holes in the valence
band of the active layer. If $I$ is the current into the laser diode
and $q$ is the electron charge, then $I/q$ is the number of carriers
injected into the laser per second. Some of these carriers may avoid
being captured in the active layer either by bypassing it through
the blocking layers or by leakage over the potential well barrier
(see Figure \ref{pn junction in laser diode}(b)). The rate of carrier
injection into the active laser is then $J=\eta I/q$ where $\eta$
is the fraction of carriers that reach the active layer. It is named
the internal quantum efficiency and can be close to $100\%$. The
number of carriers in the active layer decay due to various recombination
processes so the rate of change of $N$ is the difference between
the injection rate and the total rate of recombination $R_{rec}$,
i.e.
\begin{equation}
\frac{d}{dt}N=J-R_{rec}\,.\label{N rate 1}
\end{equation}
It is convenient to write $R_{rec}$ as the sum $R_{rec}=R(N)+R_{st}$
where $R_{st}$ is the net rate of stimulated transitions, i.e. the
rate of stimulated emission minus the rate of stimulated absorption.
$R(N)$ then describes all other recombination processes such as spontaneous
emission, recombination via defects or recombination that involves
phonons. It is often approximated by a 3rd order polynomial with $R(0)=0$,
but we will use the simple approximation $R(N)=N/\tau_{e}$ where
$\tau_{e}$ is of the order of 1 nsec and is called the effective
carrier lifetime. The net rate of stimulated transitions $R_{st}$
is taken to be of the form
\begin{equation}
R_{st}=a(N-N_{tr}){\cal {P}}\label{Linear Rst}
\end{equation}
where ${\cal {P}}$ is the number of photons guided by the active
layer waveguide and $a$ is a constant. For $N<N_{tr}$ the rate is
negative, which means that stimulated absorption dominates over stimulated
emission. Since $R_{st}(N_{tr})=0$, the carrier number $N_{tr}$
is called the carrier number at transparency. The rate equation for
the carrier number is then
\begin{equation}
\frac{d}{dt}N=J-\frac{N}{\tau_{e}}-a(N-N_{tr}){\cal {P}}\,.\label{N rate 2}
\end{equation}
The rate equation for the photon number ${\cal {P}}$ is correspondingly
given by
\begin{equation}
\frac{d}{dt}{\cal {P}}=R_{st}+R_{sp}-\frac{{\cal {P}}}{\tau_{p}}\label{P rate 1}
\end{equation}
where $R_{sp}$ is the rate of spontaneous emission into the waveguide
mode and ${\cal {P}}/\tau_{p}$ is the loss rate due to output at
the facets and to absorption in and scattering out of the waveguide.
$\tau_{p}$ is a measure of the photon lifetime in the waveguide;
it is of the order of a few picoseconds. Stimulated transitions do
not change the sum $N+{\cal {P}}$ since each stimulated annihilation
of an electron-hole pair creates a photon and each stimulated absorption
of a photon creates an electron-hole pair. $R_{st}$ must therefore
appear in the rate equations for $N$ and ${\cal {P}}$ with opposite
sign. Inserting (\ref{Linear Rst}) in (\ref{P rate 1}) the rate
equation becomes
\begin{equation}
\frac{d}{dt}{\cal {P}}=\left(a(N-N_{tr})-\frac{1}{\tau_{p}}\right){\cal {P}}+R_{sp}\,.\label{P rate 2}
\end{equation}
The nonlinear equations (\ref{N rate 2}) and (\ref{P rate 2}) can
be used to calculate ${\cal {P}}(t)$ and $N(t)$ as a function of
the carrier injection rate $J(t)=\frac{\eta}{q}I(t)$ for given parameter
values. In the following numerical calculations we will use the example
parameters of Table 1 adopted from \cite{Agrawal-1993}.
\begin{center}
\begin{tabular}{llll}
\hline
Parameter  & Symbol  & Value  & Unit\tabularnewline
\hline
Cavity length  & $L$  & 300  & \selectlanguage{american}%
\textmu{}m\selectlanguage{english}%
\tabularnewline
Stripe width  & $w$  & 2  & \selectlanguage{american}%
\textmu{}m\selectlanguage{english}%
\tabularnewline
Active layer thickness  & $d$  & 0.2  & \selectlanguage{american}%
\textmu{}m\selectlanguage{english}%
\tabularnewline
Facet reflectivity  & $r_{1}$, $r_{2}$  & 0.55  & \selectlanguage{american}%
\selectlanguage{english}%
\tabularnewline
Internal quantum efficiency  & $\eta$  & 1  & \selectlanguage{american}%
\selectlanguage{english}%
\tabularnewline
Transparency carrier number  & $N_{tr}$  & $10^{8}$  & \selectlanguage{american}%
\selectlanguage{english}%
\tabularnewline
Gain rate  & $a$  & $5.6\cdot10^{3}$  & $\textrm{s}^{-1}$\tabularnewline
Internal absorption  & $\alpha_{i}$  & 40  & $\textrm{c\ensuremath{m^{-1}}}$ \tabularnewline
Carrier lifetime  & $\tau_{e}$  & 2.2  & ns\tabularnewline
Photon lifetime  & $\tau_{p}$  & 1.6  & ps\tabularnewline
Rate of spontaneous emission  & $R_{sp}$  & $1.25\cdot10^{12}$  & $\textrm{s}^{-1}$\tabularnewline
Effective refractive index  & $n$  & 3.4  & \selectlanguage{american}%
\selectlanguage{english}%
\tabularnewline
Effective group index  & $n_{g}$  & 4  & \selectlanguage{american}%
\selectlanguage{english}%
\tabularnewline
Linewidth enhancement factor  & $\alpha$  & 5  & \selectlanguage{american}%
\selectlanguage{english}%
\tabularnewline
Confinement factor  & $\Gamma$  & 0.3  & \selectlanguage{american}%
\selectlanguage{english}%
\tabularnewline
\hline
\end{tabular}
\end{center}
\begin{center}
Table 1: Example parameters of an edge emitting laser diode operating
at a wavelength of 1.3 \foreignlanguage{american}{\textmu{}m} \cite{Agrawal-1993}.\label{Table laser diode}
\par
\end{center}

\section{Stationary solutions}

\label{Stationary solutions} 
\begin{figure}
\begin{centering}
\includegraphics[scale=0.5]{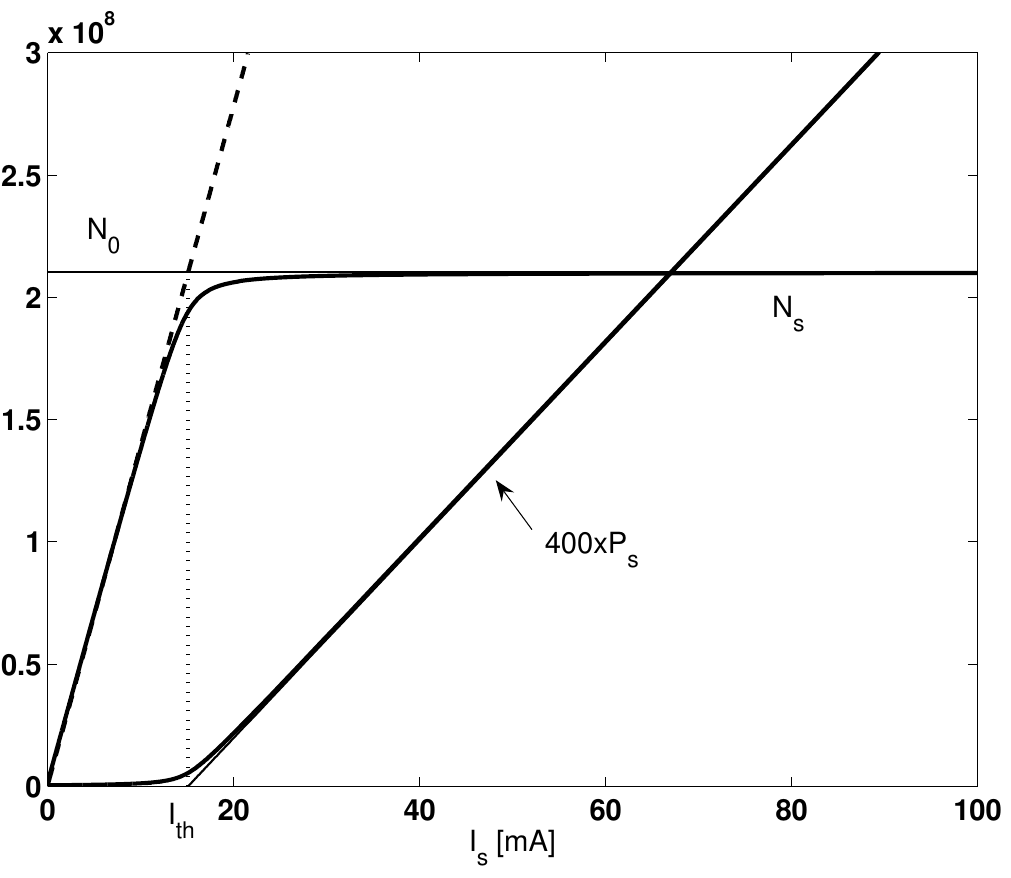}
\par\end{centering}
\caption{\selectlanguage{american}%
Thick solid curves show stationary carrier number $N_{s}$ and photon
number $P_{s}$ as a function of current $I_{s}$. Thin horizontal
line at the clamped carrier number $N_{0}$ and vertical dotted line
at $I_{th}$. Thick dashed line is $N_{s}=(\tau_{e}\eta/q)I_{s}$.\selectlanguage{english}%
}
\selectlanguage{american}%
\label{Light-current-curves} \selectlanguage{english}%
\end{figure}

We will first determine the stationary
solutions ${\cal {P}}(t)=P_{s}$ and $N(t)=N_{s}$ for constant injection
current $I(t)=I_{s}$ and hence constant carrier injection rate $J(t)=J_{s}=\frac{\eta}{q}I_{s}$.
The rate $R_{sp}$ of spontaneous emission into the waveguide mode
is a small parameter. If we ignore $R_{sp}$ in (\ref{P rate 2})
the steady state solutions to (\ref{N rate 2}) and (\ref{P rate 2})
will have to satisfy the equations
\begin{equation}
J_{s}=\frac{N_{s}}{\tau_{e}}+a(N_{s}-N_{tr})P_{s}\label{Steady N}
\end{equation}
and
\begin{equation}
\left(a(N_{s}-N_{tr})-\frac{1}{\tau_{p}}\right)P_{s}=0\,.\label{Steady P}
\end{equation}
From the equations we can determine $N_{s}$ and $P_{s}$ as functions
of $J_{s}$ or $I_{s}$. Eq. (\ref{Steady P}) has the solutions $P_{s}=0$
or $a(N_{s}-N_{tr})=1/\tau_{p}$. For $P_{s}=0$ the corresponding
solution to (\ref{Steady N}) is $N_{s}=\tau_{e}J_{s}$ shown as the
thick dashed line in Figure \ref{Light-current-curves}.

For $a(N_{s}-N_{tr})=1/\tau_{p}$ the carrier number is clamped at
\begin{equation}
N_{0}=N_{tr}+\frac{1}{a\tau_{p}}\label{Clamped N0}
\end{equation}
indicated by the thin horizontal line in Figure \ref{Light-current-curves}.
The corresponding solution to (\ref{Steady N}) is
\begin{equation}
P_{0}=\tau_{p}\left(J_{s}-J_{th}\right)=\frac{\tau_{p}\eta}{q}\left(I_{s}-I_{th}\right)\label{P0 vs I0}
\end{equation}
where $J_{th}=N_{0}/\tau_{e}=(\eta/q)I_{th}$ is the threshold carrier
injection rate and $I_{th}$ is the threshold current. The photon
number is positive so the solution is unphysical for $I_{s}<I_{th}$.
$P_{0}$ versus $I_{s}$ is shown as the thin solid line starting
at $I_{s}=I_{th}$ in Figure \ref{Light-current-curves}.

For $R_{sp}\neq0$, the photon rate equation (\ref{P rate 2}) gives
the relation
\begin{equation}
P_{s}=\frac{R_{sp}}{\frac{1}{\tau_{p}}-a(N_{s}-N_{tr})}=\frac{R_{sp}}{a(N_{0}-N_{s})}\label{P vs N}
\end{equation}
between the steady state photon and carrier numbers $P_{s}$ and $N_{s}$.
Inserting $N_{s}$ and $P_{s}$ in (\ref{Steady N}) and using (\ref{P vs N})
we then get
\begin{equation}
J_{s}=\frac{N_{s}}{\tau_{e}}+\frac{(N_{s}-N_{tr})R_{sp}}{N_{0}-N_{s}}\label{I vs N}
\end{equation}
from which we can calculate $N_{s}$ as a function of $J_{s}$. The
relation between $P_{s}$ and $J_{s}$ is subsequently obtained from
(\ref{P vs N}). The thick solid curves in Figure \ref{Light-current-curves}
are calculated for a value of $R_{sp}$ which is taken to be a factor
around $10^{3}$ too high to exaggerate its effect. The thick solid
curve for $N_{s}$ versus $I_{s}$ follows the dashed line below threshold
and the clamped carrier number $N_{0}$ above threshold. The output
power from the laser is proportional to $P_{s}$, so the model predicts
the output power to be close to zero until the current reaches a threshold
current $I_{th}$ above which it increases almost linearly with current.
This linear light-current relation is a key characteristics of laser
diodes.

\section{Small-signal analysis}

The rate equations (\ref{N rate 2}) and (\ref{P rate 2}) can be
solved explicitly in the case where the injection current $I(t)$
is a constant bias current $I_{s}$ plus a small modulating current
$I_{1}(t)$, i.e.~$I(t)=I_{s}+I_{1}(t)$. The corresponding carrier
injection rate is $J(t)=J_{s}+J_{1}(t)=\tfrac{\eta}{q}(I_{s}+I_{1}(t))$.
For $|I_{1}(t)|$ sufficiently small compared to $I_{s}$ we assume
that the solution to the rate equations is of the form ${\cal {P}}(t)=P_{s}+P_{1}(t)$
and $N(t)=N_{s}+N_{1}(t)$, where $P_{s}$ and $N_{s}$ are the steady
state solutions to (\ref{P vs N}) and (\ref{I vs N}) for $J=J_{s}$
and where $|P_{1}(t)|\ll P_{s}$ and $|N_{1}(t)|\ll N_{s}$. We use
the index $"1"$ to indicate that it is the 1st order term in a perturbation
expansion. Higher order terms are introduced below expansions (193) and (194) in subection "Langevin Noise Functions"\ref{sec:Langevin-noise-functions}.
Inserting the expressions for $J(t)$, ${\cal {P}}(t)$ and $N(t)$
in the rate equations and ignoring terms of higher order in the deviation
from steady state we get the linear equation
\begin{equation}
\frac{d}{dt}\left[\begin{array}{c}
P_{1}(t)\\
N_{1}(t)
\end{array}\right]=\boldsymbol{M}\left[\begin{array}{c}
P_{1}(t)\\
N_{1}(t)
\end{array}\right]+\left[\begin{array}{c}
0\\
J_{1}(t)
\end{array}\right]\label{Rate dP dN}
\end{equation}
where
\begin{equation}
\boldsymbol{M}=\left[\begin{array}{cc}
-\frac{R_{sp}}{P_{s}} & aP_{s}\\
\frac{R_{sp}}{P_{s}}-\frac{1}{\tau_{p}} & -\Gamma_{N}
\end{array}\right]\label{M matrix}
\end{equation}
with
\begin{equation}
\Gamma_{N}=\frac{1}{\tau_{e}}+aP_{s}\label{Gamma N}
\end{equation}
and where we have used (\ref{P vs N}). We can solve (\ref{Rate dP dN})
by first Fourier transforming the equation. This gives
\begin{equation}
\left[\begin{array}{cc}
j\omega+\frac{R_{sp}}{P_{s}} & -aP_{s}\\
\frac{1}{\tau_{p}}-\frac{R_{sp}}{P_{s}} & j\omega+\Gamma_{N}
\end{array}\right]\left[\begin{array}{c}
\tilde{P}_{1}(f)\\
\tilde{N}_{1}(f)
\end{array}\right]=\left[\begin{array}{c}
0\\
\widetilde{J}_{1}(f)
\end{array}\right]
\end{equation}
and hence
\begin{equation}
\left[\begin{array}{c}
\tilde{P}_{1}(f)\\
\tilde{N}_{1}(f)
\end{array}\right]=\boldsymbol{H}(\omega)\left[\begin{array}{c}
0\\
\tilde{J}_{1}(f)
\end{array}\right]\label{dP dN vs dI }
\end{equation}
where $2\pi f=\omega$ and $\boldsymbol{H}(\omega)$ is the transfer
matrix
\begin{equation}
\boldsymbol{H}(\omega)=\left[\begin{array}{cc}
H_{PP} & H_{PN}\\
H_{NP} & H_{NN}
\end{array}\right]=\frac{1}{D}\left[\begin{array}{cc}
j\omega+\Gamma_{N} & aP_{s}\\
\frac{R_{sp}}{P_{s}}-\frac{1}{\tau_{p}} & j\omega+\frac{R_{sp}}{P_{s}}
\end{array}\right]\label{Laser mod transfer}
\end{equation}
and $D(\omega)$ is the determinant
\begin{equation}
D(\omega)=-\omega^{2}+\left(\Gamma_{N}+\frac{R_{sp}}{P_{s}}\right)j\omega+\frac{aP_{s}}{\tau_{p}}+\frac{R_{sp}}{P_{s}\tau_{e}}\,.\label{eq:Deno of mod trans}
\end{equation}
In the linear regime above threshold, where (\ref{P0 vs I0}) is a
good approximation to $P_{s}$, we can neglect the ratio $R_{sp}/P_{s}$
in (\ref{Laser mod transfer}) and (\ref{eq:Deno of mod trans}) and
obtain the simpler expressions
\begin{equation}
\boldsymbol{H}(\omega)\simeq\frac{1}{D}\left[\begin{array}{cc}
j\omega+\Gamma_{N} & aP_{s}\\
-\frac{1}{\tau_{p}} & j\omega
\end{array}\right]\label{Laser mod transfer-1}
\end{equation}
and
\begin{equation}
D(\omega)\simeq-\omega^{2}+j\Gamma_{N}\omega+\Omega_{R}^{2}\,.\label{Deno of mod trans-1}
\end{equation}
The parameter $\Omega_{R}$ is the relaxation resonance angular frequency
of the photon-carrier oscillation given by
\begin{equation}
\Omega_{R}^{2}=\frac{aP_{s}}{\tau_{p}}\,.\label{Relax freq}
\end{equation}

Eq.(\ref{dP dN vs dI }) gives the relation
\begin{equation}
\tilde{P}_{1}(f)=H_{PN}(\omega)\tilde{J}_{1}(f)\,.
\end{equation}
where

\begin{equation}
H_{PN}(\omega)=\frac{aP_{s}}{D}\simeq\frac{\tau_{p}\Omega_{R}^{2}}{-\omega^{2}+j\Gamma_{N}\omega+\Omega_{R}^{2}}\,.\label{H_PN(omega)}
\end{equation}
If $J_{1}(t)$ is the sinusoidal modulation $J_{1}(t)=\tfrac{\eta}{q}I_{m}\cos(\omega_{0}t)$
with current amplitude $I_{m}$, the Fourier transform is $\tilde{J}_{1}(f)=\tfrac{\eta}{2q}I_{m}\delta(f-f_{0})$
for $f>0$ and $2\pi f_{0}=\omega_{0}$. In this case
\begin{equation}
\tilde{P}_{1}(f)=\frac{\eta I_{m}}{2q}H_{PN}(\omega_{0})\delta(f-f_{0})
\end{equation}
for $f>0$ and therefore
\begin{equation}
P_{1}(t)=P_{m}\cos(\omega_{0}t+\theta)
\end{equation}
with amplitude
\begin{equation}
P_{m}=\frac{\eta I_{m}}{q}|H_{PN}(\omega_{0})|
\end{equation}
and phase shift $\theta=\arg H_{PN}(\omega_{0})$. The ratio $P_{m}/I_{m}$
is the modulation response. Since the output power from the laser
is proportional to the photon number ${\cal {P}}(t)$, the ratio is
a measure of the transfer of amplitude modulation of the current to
amplitude modulation of the output power at frequency $f_{0}$.

The transfer function $H_{PN}(\omega)$ is formally similar to the
transfer function (7.43)-[1] of the electronic circuit in
Figure 7.2-[1]. Using the simple form (\ref{H_PN(omega)})
\begin{equation}
H_{PN}(\omega)=-\frac{\tau_{p}\Omega_{R}^{2}}{(\omega-\omega_{+})(\omega-\omega_{-})}\label{eq:HPN trans func}
\end{equation}
where
\begin{equation}
\omega_{\pm}=\pm\Omega+j\Gamma_{N}/2\label{Relax freq 2}
\end{equation}
and $\Omega^{2}=\Omega_{R}^{2}-\Gamma_{N}^{2}/4$. The function $|H_{PN}(\omega)|/|H_{PN}(0)|$
 is flat for $\omega\ll\Omega_{R}$, it peaks close to $\omega=\Omega_{R}$
and it decreases as $1/(\omega^{2}-\Omega_{R}^{2})$ for $\omega\gg\Omega_{R}$.
At the frequency $f_{B}=\omega_{B}/2\pi$ for which $|H_{PN}(\omega_{B})|=\tfrac{1}{2}|H_{PN}(0)|$
the modulation response $P_{m}/I_{m}$ is half of its value at low
frequencies and it decreases as $1/(\omega^{2}-\Omega_{R}^{2})$ for
$f\gg f_{B}$. The frequency $f_{B}$ is therefore a measure of the
modulation bandwidth. From (\ref{eq:HPN trans func}) we find for
$\Gamma_{N}\ll\Omega_{R}$
\begin{equation}
2\pi f_{B}\approx\sqrt{3}\Omega_{R}
\end{equation}
so in our simple model both the relaxation frequency $f_{R}=\Omega_{R}/2\pi$
and the modulation bandwidth $f_{B}$ increase as the square root
of the photon number $P_{s}$ and hence as $\sqrt{I_{s}-I_{th}}$.
This agrees with experiments for moderate $P_{s}$ but for higher
photon numbers we have to take into account that the parameter $a$
introduced in (\ref{Linear Rst}) also depends on the photon number.
The dependence is often assumed to be of the form $a=a_{0}/(1+\epsilon{\cal {P}})$,
where $a_{0}$ and $\epsilon$ are constant parameters. It implies
that $\Omega_{R}$ tends to saturate as $P_{s}$ approaches $1/\epsilon$.
If we redo the small-signal analysis including the photon number dependence
of $a$ we get a modified loss parameter $\Gamma_{N}=1/\tau_{e}+\Omega_{R}^{2}(\tau_{p}+\epsilon/a_{0})$.

\begin{figure}
\begin{centering}
\includegraphics[scale=0.5]{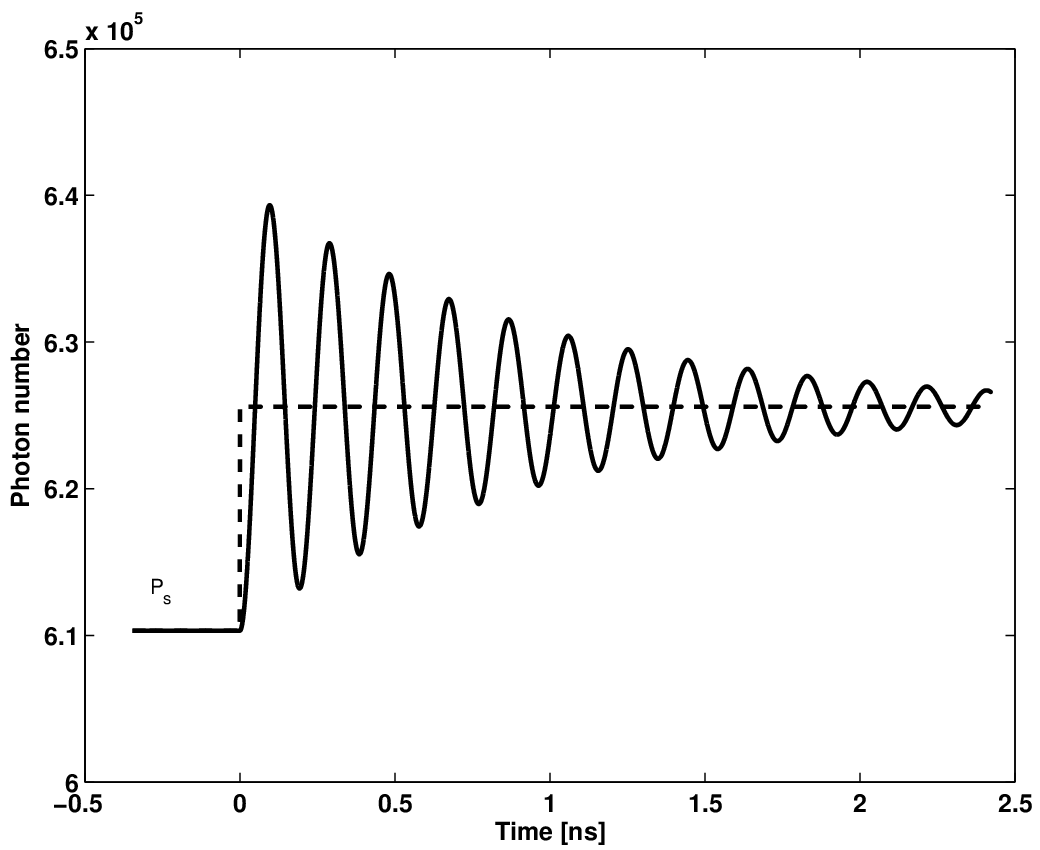}
\par\end{centering}

\caption{\selectlanguage{american}%
Transient response to a step current.\selectlanguage{english}%
}

\selectlanguage{american}%
\label{Transient response} \selectlanguage{english}%
\end{figure}

The inverse Fourier transform of $H_{PN}$ can be derived by using Eq.\
(7.59)-\cite{Jeppesen-2023} and gives the impulse response
\begin{equation}
h_{PN}(t)=\frac{aP_{s}}{\Omega}e^{-\tfrac{1}{2}\Gamma_{N}t}\sin(\Omega t)u(t)\label{Res PN}
\end{equation}
when the loss rate $\Gamma_{N}/2$ is less than $\Omega_{R}$. For
a general input current $I(t)=I_{s}+I_{1}(t)$ we then derive the
photon number $P_{1}(t)$ from the convolution
\begin{equation}
P_{1}(t)=\frac{\eta}{q}\int_{-\infty}^{t}h_{PN}(t-t')I_{1}(t')dt'\,.
\end{equation}
The case where the current is suddenly increased by $\Delta I$ can
be represented by $I_{1}(t)=\Delta Iu(t)$. The corresponding photon
number $P_{1}(t)$ is then
\begin{equation}
P_{1}(t)=\frac{\eta\Delta I}{q}\int_{0}^{t}h_{PN}(t')dt'u(t)=\tau_{p}\frac{\eta\Delta I}{q}\left(1-\frac{\Omega_{R}}{\Omega}e^{-\tfrac{1}{2}\Gamma_{N}t}\cos(\Omega t-\theta)\right)u(t)\label{Transient of P(t)}
\end{equation}
where $\tan(\theta)=\Gamma_{N}/(2\Omega)$. The transient behavior
after the current step is a damped oscillation that for large $t$
approaches $\Delta P=\tau_{p}\eta\Delta I/q$ as expected from (\ref{P0 vs I0}).
An numerical example is shown in Figure \ref{Transient response}.

\section{Guided wave solutions}

In order to determine the spectral properties of the laser diode it
is necessary to solve the wave equation for the electric field in
the laser diode waveguide. We consider the typical waveguide example
in Figure 10.1-[1], consisting of a thin stripe
of active material of InGaAsP of width $w=2$\foreignlanguage{american}{
\textmu{}m}, thickness $d=0.2$ \foreignlanguage{american}{\textmu{}m},
and length $L=300$ \foreignlanguage{american}{\textmu{}m} embedded
in cladding material of InP. The waveguide properties are determined
by the relative permittivity $\epsilon_{r}(x,y)=1+\tilde{\chi}(x,y)$
where $x$ and $y$ are the lateral and transverse coordinates in
Figure 10.1-[1].

For a waveguide that is uniform in the longitudinal z-direction the
Fourier transformed electric field is of the form   (see Eqs.\ (10.56)-[1] and (10.53)-[1] ) 
\begin{equation}
\tilde{\boldsymbol{E}}({\boldsymbol{r}},\omega)=\tilde{E}(\omega)\tilde{{\boldsymbol{F}}}(x,y,\omega)e^{-j\beta(\omega)z}\label{eq:Exyz}
\end{equation}
where the transverse function $\tilde{{\boldsymbol{F}}}_{t}(x,y,\omega)$
and the squared propagation constant $\beta^{2}(\omega)$ are eigenfunction
and eigenvalue solutions to the eigenvalue equation  (10.53)-[1].
For our example, where $d/w\ll1$ the fundamental solution is predominantly
a TE (transverse electric) mode, so we approximate (\ref{eq:Exyz})
by the simpler expression
\begin{equation}
\tilde{\boldsymbol{E}}({\boldsymbol{r}},\omega)=\hat{\boldsymbol{x}}\tilde{E}(\omega)U(x,y,\omega)e^{-j\beta(\omega)z}\label{eq:EUvsz}
\end{equation}
where $\tilde{{\boldsymbol{F}}}(x,y,\omega)$ is replaced by $\hat{\boldsymbol{x}}U(x,y,\omega)$
and $\hat{\boldsymbol{x}}$ is a unit vector along the lateral x-axis
in Figure (10.2)-[1]. We assume $\epsilon_{r}$
to be uniform within each of the active and cladding regions. This
implies that $\nabla_{t}\ln\epsilon_{r}=0$ in  (10.45)-[1] except
at the interfaces between the active and cladding material. Inserting
$\tilde{{\boldsymbol{F}}}_{t}=\hat{\boldsymbol{x}}U(x,y)$ in (10.45)-[1],
we get the eigenvalue equation
\begin{equation}
\left(\frac{\partial^{2}}{\partial x^{2}}+\frac{\partial^{2}}{\partial y^{2}}+k_{0}^{2}\epsilon_{r}\right)U(x,y)=\beta^{2}U(x,y)\label{U beta eqn}
\end{equation}
except at the active region boundaries where boundary conditions must
be imposed. The argument $\omega$ is here suppressed for notational
convenience. At the boundaries at $y=\pm d/2$ the tangential part
of $\tilde{\boldsymbol{E}}$ is continuous which means that $U(x,y)$
is continuous. The Fourier transforms of (10.4)-[1]
and (10.6)-[1] lead to the Maxwell equation $\nabla\times\tilde{\boldsymbol{E}}=-j\omega\mu_{0}\tilde{\boldsymbol{H}}$,
which has the z-component $j\omega\mu_{0}\tilde{H}_{z}=-\hat{\boldsymbol{z}}\cdot\nabla\times\tilde{\boldsymbol{E}}\propto\frac{\partial}{\partial y}U(x,y)$.
The magnetic field is continuous in the dielectric material so the
equation implies that $\frac{\partial}{\partial y}U(x,y)$ is continuous
at $y=\pm d/2$.

The susceptibility $\tilde{\chi}$ can be written as the sum $\tilde{\chi}=\tilde{\chi}_{b}+\tilde{\chi}_{a}$
of a background susceptibility $\tilde{\chi}_{b}$ and a susceptibility
$\tilde{\chi}_{a}$ due to the carriers in the active layer. The latter
depends on both carrier density and photon density at the local position
in the active layer. However, in this simple model we assume that
$\tilde{\chi}_{a}$ only depends on the average carrier density $N/V_{a}$
where $V_{a}$ is the volume of the active region. This means that
$\tilde{\chi}_{a}$ is assumed to be zero outside the active layer
and to be uniform within the active layer where it is only a function
$\tilde{\chi}_{a}(N)$ of the carrier number. The relative permittivity
is then of the form
\begin{equation}
\epsilon_{r}=1+\tilde{\chi}_{b}+\tilde{\chi}_{a}(N)=n_{b}^{2}+\Delta\epsilon\label{Delta eps}
\end{equation}
where
\begin{equation}
n_{b}^{2}=1+\textrm{Re}\{\tilde{\chi}_{b}\}\ \textrm{and}\ \Delta\epsilon=\tilde{\chi}_{a}(N)+j\textrm{Im}\{\tilde{\chi}_{b}\}\,.\label{nb index}
\end{equation}

For given carrier number and optical angular frequency there may be
more than one guided mode solution. In the present example with $\epsilon_{r}$
given by (\ref{Delta eps}) we may determine the solutions by a perturbation
method.  
We get
\begin{equation}
2\bar{\beta}\Delta\beta\simeq k_{0}^{2}\frac{\int\Delta\epsilon|U(x,y)|^{2}dxdy}{\int|U(x,y)|^{2}dxdy}\,.\label{Delta beta-1}
\end{equation}
where $\bar{\beta}$ is the unperturbed propagation constant. The
confinement factor $\Gamma$ is defined as the ratio between the power
in the part of the guided wave contained in the active region with
cross-sectional $A$ and the power in the entire optical mode, i.e.
\begin{equation}
\Gamma=\frac{\int_{A}|U(x,y)|^{2}dxdy}{\int|U(x,y)|^{2}dxdy}\,.\label{eq:Confinement factor}
\end{equation}
If we introduce the cross-sectional area $A_{a}$ associated with
the power in the active region by

\begin{equation}
\int_{A}|U(x,y)|^{2}dxdy=|U(0,0)|^{2}A_{a}\label{eq:Area Aa}
\end{equation}
and the cross-sectional area $A_{p}$ associated with the power in
the entire optical mode by

\begin{equation}
\int|U(x,y)|^{2}dxdy=|U(0,0)|^{2}A_{p}\label{eq:Area Ap}
\end{equation}
the confinement factor can simply be written $\Gamma=A_{a}/A_{p}$.
Since $\tilde{\chi}_{a}(N)$ is uniform within and zero outside the
active region we have

\begin{equation}
\int\tilde{\chi}_{a}(N)|U(x,y)|^{2}dxdy=\tilde{\chi}_{a}(N)\int_{A}|U(x,y)|^{2}dxdy=\tilde{\chi}_{a}(N)|U(0,0)|^{2}A_{a}\,.
\end{equation}
We also introduce the unperturbed modal index $\bar{n}$ defined by
$\bar{\beta}=k_{0}\bar{n}$. With this notation we can use (\ref{nb index})
and (\ref{Delta beta-1}) to write $\beta$ as
\begin{equation}
\beta\simeq\bar{\beta}+\Delta\beta=k_{0}\bar{n}+\frac{k_{0}}{2\bar{n}}\Gamma\tilde{\chi}_{a}(N)-j\tfrac{1}{2}\alpha_{i}=k_{0}n+j\tfrac{1}{2}(\Gamma g-\alpha_{i})\label{beta vs n&g}
\end{equation}
where $n$ is the modal index
\begin{equation}
n=\bar{n}+\frac{1}{2\bar{n}}\Gamma\,\textrm{Re}\{\tilde{\chi}_{a}(N)\}\label{n index}
\end{equation}
$g$ is the material gain
\begin{equation}
g(N)=\frac{k_{0}}{\bar{n}}\textrm{Im}\{\tilde{\chi}_{a}(N)\}\label{Mat gain g}
\end{equation}
and
\begin{equation}
\alpha_{i}=-\frac{k_{0}}{2\bar{n}}\frac{\int\textrm{Im}\{\tilde{\chi}_{b}\}|U(x,y)|^{2}dxdy}{\int|U(x,y)|^{2}dxdy}\label{Internal
abs}
\end{equation}
is the internal absorption coefficient. The gain parameter $g(N)$
can be positive or negative depending on $N$. The internal absorption
coefficient $\alpha_{i}$ is a material parameter that is always positive.
The functions $\tilde{\chi}_{a}(\omega,N)$, $n(\omega,N)$ and $g(\omega,N)$
can be derived from the theory of the energy distribution of electrons
and holes in the band diagram in Figure \ref{pn junction in laser diode},
see e.g. \cite{Agrawal-1993}. The derivation is outside the scope
of this discussion so we use the simple approximation where $n(\omega,N)$
is assumed to be linear in $\omega$ and $N$, and $g(\omega,N)$
to be of the form \cite{Agrawal-1993}
\begin{equation}
g(\omega,N)=g_{m}(N)\left[1-\left(\frac{\omega-\omega_{m}(N)}{\Delta\omega_{g}}\right)^{2}\right]\label{eq:g(omega,N)}
\end{equation}
for $\omega$ close to the maximum $\omega_{m}$. The gain decreases
quadratically in $\omega$ away from the maximum and has a width $\Delta\omega_{g}$.
$g_{m}(N)$ and $\omega_{m}(N)$ are assumed to be linear functions
of $N$. The photon energy $\hbar\omega_{m}$ lies between the band
gap energies of the active and the host material, see Figure \ref{pn junction in laser diode}.

We assume that $\tilde{E}(\omega)$ in (\ref{eq:EUvsz}) is centered
in a narrow band around a frequency $\omega_{0}$. The transverse
solution $U(x,y)$ will usually be slowly varying with frequency so
we take $U(x,y)$ to represent the solution at $\omega_{0}$. According
to (10.67)-[1] the $z$-component of the averaged Poyntings
vector for the field (\ref{eq:EUvsz}), i.e. the intensity, is
\begin{equation}
S_{z}({\boldsymbol{r}},t)=\frac{\beta_{r}}{2\mu_{0}\omega_{0}}|U(x,y)|^{2}|A_{e}(z,t)|^{2}\label{Wavegide I-1}
\end{equation}
for $\boldsymbol{F}=\hat{\boldsymbol{x}}U(x,y)$ and $\beta_{r}=\textrm{Re}\beta(\omega_{0})=k_{0}n$.
The envelope $A_{e}(z,t)$ is 
\begin{equation}
A_{e}(z,t)=\frac{1}{\pi}\int_{0}^{\infty}\tilde{E}(\omega)e^{j\{(\omega-\omega_{0})t-(\beta(\omega)-\beta_{r})z\}}d\omega\,.\label{Ae-laser}
\end{equation}
Ignoring dispersion to second and higher order
\begin{gather}
\beta(\omega)-\beta_{r}\simeq\frac{1}{v_{g}}(\omega-\omega_{0})+j\gamma/2\label{beta expansion}
\end{gather}
where
\begin{gather}
\gamma=\Gamma g-\alpha_{i}\label{Def gamma}
\end{gather}
is the net gain and
\begin{gather}
\frac{1}{v_{g}}=\textrm{Re}\frac{\partial\beta}{\partial\omega}=\frac{\partial(k_{0}n)}{\partial\omega}=\frac{1}{c}\frac{\partial(\omega n)}{\partial\omega}=\frac{n_{g}}{c}\label{Def vg}
\end{gather}
for $\omega=\omega_{0}$. Here, $v_{g}=c/n_{g}$ is the group velocity and $n_{g}=\frac{\partial(\omega n)}{\partial\omega}$
is the group index. Inserting (\ref{beta expansion}) in (\ref{Ae-laser})
\begin{gather}
A_{e}(z,t)=\frac{1}{\pi}\int_{0}^{\infty}\tilde{E}(\omega)e^{j\{(\omega-\omega_{0})(t-z/v_{g})\}}d\omega e^{\frac{\gamma z}{2}}=E(t-z/v_{g})e^{\frac{\gamma z}{2}}\label{Ae-laser-1}
\end{gather}
where $E(t)=2\int_{0}^{\infty}\tilde{E}(\omega)e^{j\{(\omega-\omega_{0})t\}}df$.
The intensity is then
\begin{equation}
S_{z}({\boldsymbol{r}},t)=\frac{1}{2}\epsilon_{0}nc|U(x,y)|^{2}|E(t-z/v_{g})|^{2}e^{\gamma z}\label{Wavegide I-1b}
\end{equation}
using $\epsilon_{0}\mu_{0}=1/c^{2}$ and $\beta_{r}=k_{0}n=\omega_{0}cn$.
We will in the following normalize $U(x,y)$ such that
\begin{equation}
\left|U\left(0,0\right)\right|=1\,.\label{Norm U}
\end{equation}
The power flow in the z-direction is then simply
\begin{equation}
P(z,t)=\int S_{z}({\boldsymbol{r}},t)dxdy=\frac{1}{2}\epsilon_{0}cnA_{p}|E(t-z/v_{g})|^{2}e^{\gamma z}\,.\label{Power vs z}
\end{equation}
It shows that the optical power increases exponentially with $z$
if $\gamma>0$ ($\Gamma g>\alpha_{i}$), and it decreases exponentially
if $\gamma<0$. For $\gamma=0$ the power flow satisfies $P(z,t)=P(z-v_{g}t,0)$,
which means that the power distribution moves undistorted along the
$z$-axis with velocity $v_{g}$.

For a harmonic solution where $E(t)=E_{0}e^{j\omega_{0}t}$ the power
flow is stationary and $P(z)=\tfrac{1}{2}\epsilon_{0}cnA_{p}|E_{0}|^{2}e^{\gamma z}$.
We can use the expression to establish a relation between the stimulated
emission rate $R_{st}$ in (\ref{Linear Rst}) and the material gain
parameter $g$. Let $W(z)$ be the number of photons per unit length.
The photons move with group velocity $v_{g}$ so during a time $dt$
a cross section of the waveguide at $z$ is passed by $W(z)v_{g}dt$
photons, i.e. by $v_{g}W(z)$ photons per unit time. Since each photon
has energy $\hbar\omega_{0}$ they carry the power $P(z)=\hbar\omega_{0}v_{g}W(z)$.
The change of power over a transverse slice of the waveguide of width
$dz$ is $dP=\frac{dP}{dz}dz=\gamma P(z)dz=(\Gamma g-\alpha_{i})P(z)dz$.
The term $\Gamma gP(z)dz$ is the energy generated in the slice per
unit time, and $\Gamma gP(z)dz/\hbar\omega_{0}$ is therefore the
rate of stimulated emissions $dR_{st}$ from the slice. Hence $dR_{st}=\Gamma gP(z)dz/\hbar\omega_{0}=v_{g}\Gamma gW(z)dz$.
Since $W(z)dz$ is the number of photons in the slice, the assumption
(\ref{Linear Rst}) implies that here $dR_{st}=a(N-N_{tr})W(z)dz$.
The two expressions for $dR_{st}$ require
\begin{equation}
v_{g}\Gamma g(\omega_{0},N)=a(N-N_{tr})\,.\label{eq:glinear1}
\end{equation}
The simple relation is only consistent with the parametrization of
gain in (\ref{eq:g(omega,N)}) for $\omega_{0}$ close to the gain
maximum at $\omega_{m}$ where $g(\omega,N)$ is a flat function of
$\omega$ and close to $g_{m}(N)$. The relation (\ref{eq:glinear1})
furthermore requires that $g_{m}(N)$ is linear and $a=v_{g}\Gamma\frac{\partial g}{\partial N}$.
Instead of the linear approximation to $g_{m}(N)$ one may use the
more realistic parametrization
\begin{equation}
g_{m}(N)=g_{0}\ln\left(\frac{N-N_{g}}{N_{tr}-N_{g}}\right)\label{Alt gm}
\end{equation}
where $g_{0}$, $N_{g}$ and $N_{tr}$ are constant fitting parameters
\cite{Coldren-1995}. However, to keep the discussion simple we stick
to the linear approximation (\ref{eq:glinear1}).

When the power is given by (\ref{Power vs z}) we still have the relation
\begin{equation}
P(z,t)=\hbar\omega_{0}v_{g}W(z,t)
\end{equation}
between optical power $P(z,t)$ and photon density per unit length
of the waveguide $W(z,t)$, i.e.
\begin{equation}
W(z,t)=\frac{\epsilon_{0}nn_{g}}{2\hbar\omega_{0}}A_{p}|E(t-z/v_{g})|^{2}e^{\gamma z}\,.\label{eq:Photon density RT new}
\end{equation}
However, this only holds when the field propagates in the $z$-direction.

\section{Laser oscillator model}


\begin{figure}
\begin{centering}
\includegraphics[scale=0.5]{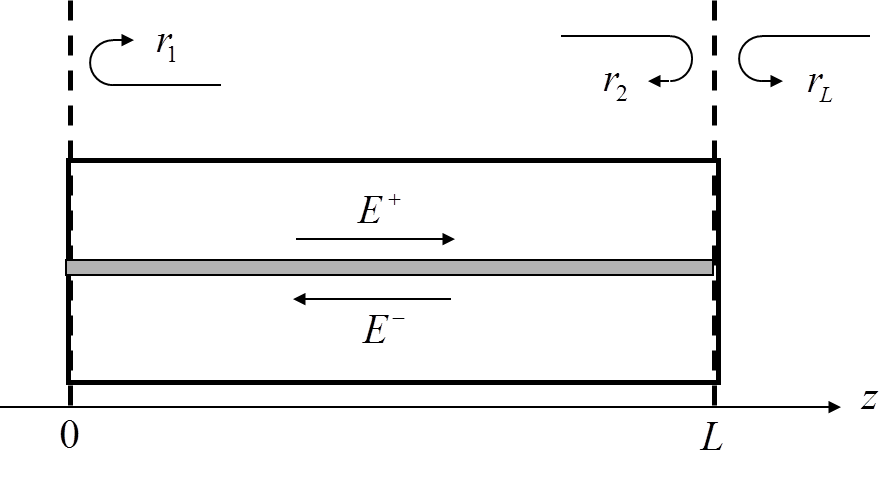}
\par\end{centering}

\centering{}\caption{\selectlanguage{american}%
Longitudinal cross section of the laser diode. Dashed vertical lines
are reference planes inside and at the laser facets. $r_{1}$ and
$r_{2}$ are the facet reflectivities seen from inside, and $r_{L}$
is the reflectivity of the laser waveguide seen from the right laser
facet.\selectlanguage{english}%
}

\selectlanguage{american}%
\label{Longitudinal cross section of laser diode} \selectlanguage{english}%
\end{figure}

Figure \ref{Longitudinal cross section of laser diode} shows
a longitudinal cross section of the laser diode. According to (\ref{eq:EUvsz})
the right and left travelling waves in the laser waveguide are
\begin{equation}
\tilde{\boldsymbol{E}}^{\pm}({\boldsymbol{r}})=\hat{\boldsymbol{x}}U(x,y)\tilde{E}^{\pm}(z)\label{E+/-xyz}
\end{equation}
where
\begin{equation}
\tilde{E}^{\pm}(z)=\tilde{E}^{\pm}(0)e^{\mp j\beta z}\,.\label{E+/- vs z}
\end{equation}
$\tilde{E}^{+}(z)$ is the right travelling wave and $\tilde{E}^{-}(z)$
is the left travelling. The tildes indicate that the fields are functions
of angular frequency $\omega$. It is implied that $\beta$ is a function
$\beta(\omega,N)$ of $\omega$ and carrier number $N$. The fields
have to satisfy the boundary conditions
\begin{equation}
\tilde{E}^{+}(0)=r_{1}\tilde{E}^{-}(0)\label{E+(0)}
\end{equation}
\begin{equation}
\tilde{E}^{-}(L)=r_{2}\tilde{E}^{+}(L)\label{E-(L)}
\end{equation}
where $r_{1}$ and $r_{2}$ are the internal reflection coefficients
at the left and right laser facets. For cleaved facets $r_{1}=r_{2}=0.32$
but the reflectivities can be changed to practically any value between
zero and one by coating the facets.

It follows from (\ref{E+/- vs z}), (\ref{E+(0)}) and $\tilde{E}^{-}(L)=\tilde{E}^{-}(0)e^{j\beta L}$
that
\begin{equation}
\tilde{E}^{+}(L)=\tilde{E}^{+}(0)e^{-j\beta L}=r_{1}\tilde{E}^{-}(0)e^{-j\beta L}=r_{1}e^{-j2\beta L}\tilde{E}^{-}(L)\,.\label{Bound. cond. at L }
\end{equation}
The equation means that the effective reflectivity of the laser waveguide
seen from a reference plane at and inside the laser facet in Figure
\ref{Longitudinal cross section of laser diode} is
\begin{equation}
r_{L}=r_{1}e^{-j2\beta L}\,.\label{rL at L}
\end{equation}
The field receives contributions from spontaneous emission during
a round-trip in the laser waveguide, so in order to include this we
have to add en extra term and modify (\ref{Bound. cond. at L } )
to
\begin{equation}
\tilde{E}^{+}(L)=r_{1}e^{-j2\beta L}\tilde{E}^{-}(L)+\tilde{F}^{+}\,.\label{E+E-F}
\end{equation}
We will see below how to deal with the function $\tilde{F}^{+}$ such
that it accounts for spontaneous emission. Inserting (\ref{E-(L)})
in (\ref{E+E-F}) we get the field equation
\begin{equation}
\left(1-G(\omega,N)\right)\tilde{E}^{+}(L)=\tilde{F}^{+}\label{Field eq at L}
\end{equation}
where
\begin{equation}
G(\omega,N)=r_{1}r_{2}e^{-j2\beta L}\,.\label{Loop gain G}
\end{equation}

The field power spectral densities $S_{E}(\omega)$ and $S_{F}^{+}(\omega)$
for $E^{+}(L)$ and $F^{+}$ are defined by (6.57)-[1].
From the field 
equation (\ref{Field eq at L}) the corresponding 
spectra are related by
\begin{equation}
S_{E^{+}}(\omega)=\frac{S_{F^{+}}(\omega)}{|1-G(\omega,N)|^{2}}\label{SE below th}
\end{equation}
provided we can assume the carrier number $N$ to be constant in time.
The relation is not applicable when $1-G(\omega,N)$ has a zero in
the lower half of the complex $\omega$-plane. 
The system becomes unstable when the
amplification, in the present case determined by the carrier number
$N$, increases to the point where a zero of $1-G(\omega,N)$ passes
from the upper half complex $\omega$-plane to the lower half. For
$s=j\omega$ this is when a zero passes from the left half complex
$s$-plane to the right half. The system will then start to oscillate
with increasing amplitude until it is clamped by nonlinear effects.
The oscillation frequency will be close to the frequency where the
zero of $1-G(\omega,N)$ passes the real axis in the complex $\omega$-plane
for increasing $N$. The equation
\begin{equation}
G(\omega,N)=1\label{Laser osc cond}
\end{equation}
for real $\omega$ is therefore called the oscillation condition.
It implies a gain condition $|G|=1$ and a phase condition $\arg(G)=0\,(\textrm{mod}\,2\pi)$.
Using the expression (\ref{beta vs n&g}) for $\beta$ the gain condition
becomes $|G|=r_{1}r_{2}\exp((\Gamma g-\alpha_{i})L)=1$ and hence
\begin{equation}
\Gamma g=\alpha_{i}-\frac{\ln(r_{1}r_{2})}{L}=\alpha_{i}+\alpha_{m}\label{Gain cond}
\end{equation}
where $\alpha_{m}=-\ln(r_{1}r_{2})/L$ is due to mirror loss at the
laser facets. The phase condition becomes
\begin{equation}
-\arg(G)=2\textrm{Re}\{\beta\}L=2k_{0}n(\omega,N)L=\frac{2L}{c}\omega n(\omega,N)=2\pi m\label{Phase cond}
\end{equation}
for integer $m$. The electric field solutions satisfying the phase
condition are called longitudinal modes.

\begin{figure}
\begin{centering}
\includegraphics[scale=0.65]{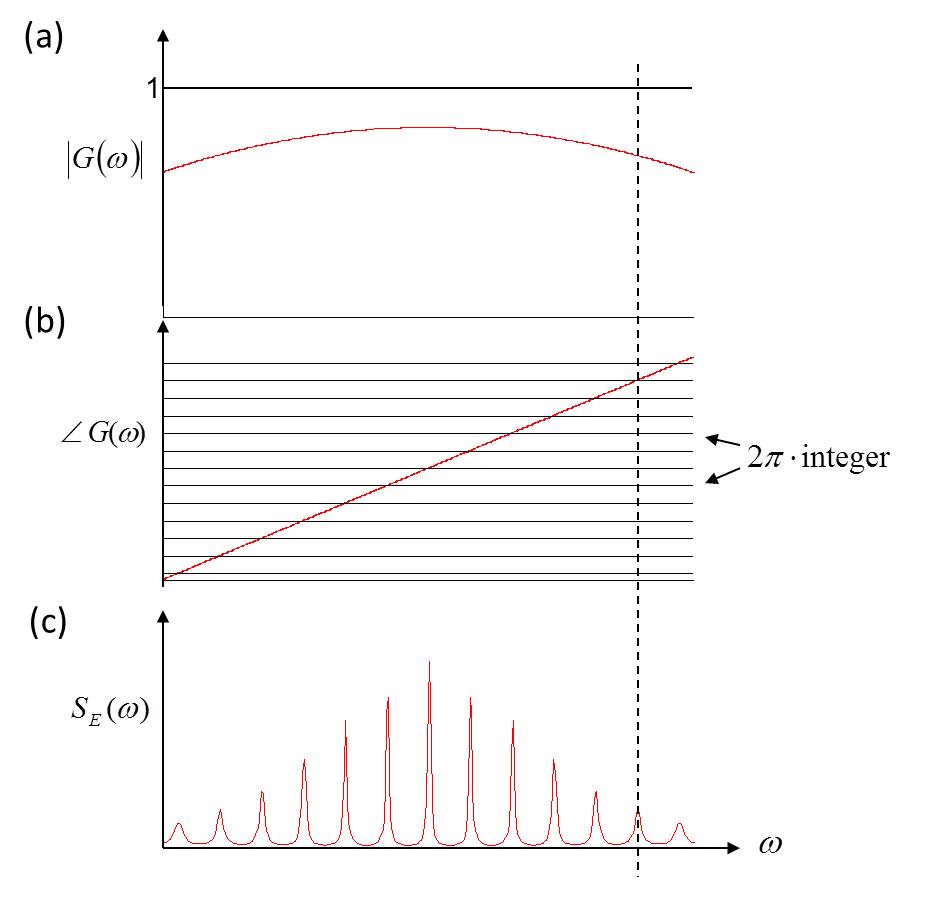}
\par\end{centering}

\centering{}\caption{\textit{\emph{Graphical}}\emph{ }\textit{\emph{solution of the oscillation
condition (\ref{Laser osc cond}). (a) Norm of the loop gain. (b)
Graphical solution of the phase condition (\ref{Phase cond}). The
spacing between horizontal lines is $2\pi$. (c) Spectrum (\ref{SE below th})
on a logarithmic scale.}}}

\label{Graphical solution of oscilllation condition}
\end{figure}

Figure \ref{Graphical solution of oscilllation condition}(c) shows
a model example of the spectrum we get if the norm of the loop gain
as a function of $\omega$ has the form in Figure \ref{Graphical solution of oscilllation condition}(a)
and $\arg(G)$ is the linear function of $\omega$ in Figure \ref{Graphical solution of oscilllation condition}(b).
We assume $S_{F^{+}}(\omega)$ to be constant over the considered
frequency range. The spectrum has sharp spikes at the frequencies
where the phase condition (\ref{Phase cond}) is satisfied. The norm
$|G(\omega,N)|$ increases with $N$ when $N$ is above the transparency
level $N_{tr}$ introduced in (\ref{Linear Rst}). The hight of a
spike is proportional to $(1-|G|)^{-2}$ at the spike frequency, so
the dominant spike will be at the mode frequency with smallest $1-|G|$
and its hight diverges as $|G|\rightarrow1$ unless $S_{F^{+}}(\omega)=0$.

The group velocity $v_{g}$ of an optical pulse in a waveguide is
given by (\ref{Def vg}). It therefore follows from (\ref{Phase cond})
that the angular frequency spacing $\Delta\omega$ between adjacent
modes is given by
\begin{equation}
2L\textrm{Re}\left\{ \frac{\partial\beta}{\partial\omega}\right\} \Delta\omega=\frac{2L}{v_{g}}\Delta\omega=2\pi\label{omega spacing}
\end{equation}
and the frequency spacing by
\begin{equation}
\Delta f=\frac{\Delta\omega}{2\pi}=\frac{v_{g}}{2L}=\frac{1}{\tau_{L}}\,.\label{frequency spacing}
\end{equation}
$\tau_{L}=2L/v_{g}$ is the round-trip time for an optical pulse in
the laser cavity. The spectrum in Figure \ref{Graphical solution of oscilllation condition}(c)
is a realistic picture of an experimental laser diode spectrum for
currents below the threshold current $I_{th}$.

The field equation (\ref{Field eq at L}) is derived for constant
carrier number $N$. We saw in Section \ref{Stationary solutions}
that with spontaneous emission $R_{sp}=0$, the carrier number is
clamped at $N=N_{0}$ for $I>I_{th}$, where $N_{0}$ is given by
(\ref{Clamped N0}). Since $R_{sp}=0$ implies $\tilde{F}^{+}=0$,
a non-zero field solution to the field equation (\ref{Field eq at L})
implies that the factor $1-G$ must be zero, i.e there is a solution
$(\omega_{0},N_{0})$ to the oscillation condition (\ref{Laser osc cond}).
The corresponding field solution is the harmonic solution
\begin{equation}
\boldsymbol{E}({\boldsymbol{r}},t)=\hat{\boldsymbol{x}}U(x,y)[E^{+}(z)+E^{-}(z)]e^{j\omega_{0}t}\label{eq:Complex harmonic sol new}
\end{equation}
where $E^{\pm}(z)=E^{\pm}(0)e^{\mp j\beta(\omega_{0},N_{0})z}$ and
$E^{+}(0)=r_{1}E^{-}(0)$. When $R_{sp}\neq0$ the amplitudes $E^{\pm}(z)$
will be time dependent and the field solution is then
\begin{equation}
\boldsymbol{E}({\boldsymbol{r}},t)=\hat{\boldsymbol{x}}U(x,y)[E^{+}(z,t)+E^{-}(z,t)]\label{Complex gen sol}
\end{equation}
where
\begin{equation}
E^{\pm}(z,t)=2\int_{0}^{\infty}\tilde{E}^{\pm}(z,\omega)e^{j\omega t}df\label{Epm(z,t)}
\end{equation}
and $\tilde{E}^{\pm}(z,\omega)=\tilde{E}^{\pm}(0,\omega)e^{\mp j\beta(\omega,N)z}$.
In the next section we derive a field equation for the case where
$R_{sp}\neq0$ and where the laser frequency and carrier number is
close to a solution $(\omega_{0},N_{0})$ to the oscillation condition.
But first we derive expressions for the total photon number and the
relative output power.

\subsubsection{Total photon number and relative output power}

The photon density (\ref{eq:Photon density RT new}) applies to the
case where photons are all right travelling. When the light is travelling
in both directions the photon density per unit length is
\begin{equation}
W(z,t)=\frac{\epsilon_{0}nn_{g}}{2\hbar\omega_{0}}A_{p}\left|E^{+}(z,t)+E^{-}(z,t)\right|^{2}\,.\label{eq:Photon density RLT new}
\end{equation}
The fields $E^{+}(z,t)$ and $E^{-}(z,t)$ are right and left travelling
waves and according to (\ref{eq:Photon density RT new}) they contribute
with photon densities
\begin{equation}
W^{\pm}(z,t)=\frac{\epsilon_{0}nn_{g}}{2\hbar\omega_{0}}A_{p}\left|E^{\pm}(z,t)\right|^{2}=\frac{\epsilon_{0}nn_{g}}{2\hbar\omega_{0}}A_{p}\left|E^{\pm}(0,t-z/v_{g})\right|^{2}e^{\pm\gamma z}\,.\label{Photondensitypm}
\end{equation}
The interference between right and left travelling waves gives rise
to a term proportional to $Re\{(E^{+}(z,t))^{*}E^{-}(z,t)\}$ in (\ref{eq:Photon density RLT new}).
For the harmonic solution (\ref{eq:Complex harmonic sol new})
\begin{equation}
(E^{+}(z,t))^{*}E^{-}(z,t)=(E^{+}(0))^{*}E^{-}(0)e^{j(\beta^{*}+\beta)z}=(E^{+}(0))^{*}E^{-}(0)e^{j2k_{0}nz}
\end{equation}
the real part of which is a standing wave that oscillates sinusoidally
with constant amplitude as a function of $z$. The period is $\pi/(k_{0}n)$
which is half the wavelength of light in the material at angular frequency
$\omega_{0}$.

The total photon number in the laser cavity is the integral
\begin{equation}
{\cal {P}}(t)=\int_{0}^{L}W(z,t)dz\simeq\int_{0}^{L}(W^{+}(z,t)+W^{-}(z,t))dz\,.\label{P integral}
\end{equation}
The integral over the oscillating interference term can be neglected
for the edge emitting laser where its relative contribution is of
the order $1/(2k_{0}nL)\ll1$. For other types of lasers, as for example
the vertical cavity surface emitting laser, it may matter how the
standing wave pattern is located compared to active region. Using
(\ref{Photondensitypm}) and $E^{+}(0,t)=r_{1}E^{-}(0,t)$ the photon
number becomes
\begin{gather}
{\cal {P}}(t)\simeq\frac{\epsilon_{0}nn_{g}}{2\hbar\omega_{0}}A_{p}\int_{0}^{L}\left(\left|E^{+}(0,t-z/v_{g})\right|^{2}e^{\gamma z}+\left|E^{-}(0,t-z/v_{g})\right|^{2}e^{-\gamma z}\right)dz\nonumber \\
=\frac{\epsilon_{0}nn_{g}}{2\hbar\omega_{0}}A_{p}\int_{0}^{L}\left(e^{\gamma z}+\frac{1}{r_{1}^{2}}e^{-\gamma z}\right)\left|E^{+}(0,t-z/v_{g})\right|^{2}dz\nonumber \\
\simeq\frac{\epsilon_{0}nn_{g}}{2\hbar\omega_{0}}A_{p}\int_{0}^{L}\left(e^{\gamma z}+\frac{1}{r_{1}^{2}}e^{-\gamma z}\right)dz\left|E^{+}(0,t)\right|^{2}\label{P integral-1}
\end{gather}
where we have assumed that $\left|E^{+}(0,t)\right|^{2}$ is slowly
varying compared to the round trip time in the cavity. With this approximation
we get
\begin{equation}
{\cal {P}}(t)\simeq K^{2}(N)e^{\gamma L}\left|E^{+}(0,t)\right|^{2}=K^{2}(N)\left|E^{+}(L,t)\right|^{2}\label{eq:P integral-2 new}
\end{equation}
where
\begin{equation}
K^{2}(N)=\frac{\epsilon_{0}nn_{g}}{2\hbar\omega\gamma r_{1}^{2}}(1-e^{-\gamma L})(r_{1}^{2}+e^{-\gamma L})A_{p}\,.\label{K2}
\end{equation}
Notice that $e^{-\gamma L}=r_{1}r_{2}$ when the gain condition $|G|=1$
is satisfied.

The output power $P_{out}(L)$ at the facet at $z=L$ is proportional
to $t_{2}^{2}|\tilde{E}^{+}(L,t)|^{2}$, where the transmission coefficient
$t_{2}$ is given by $t_{2}^{2}=1-r_{2}^{2}$ for a lossless facet.
At the facet at $z=0$ the output power $P_{out}(0)$ is similarly
proportional to $t_{1}^{2}|\tilde{E}^{-}(0,t)|^{2}$. The ratio between
the two powers is
\begin{equation}
\frac{P_{out}(L)}{P_{out}(0)}=\frac{t_{2}^{2}|\tilde{E}^{+}(L,t)|^{2}}{t_{1}^{2}|\tilde{E}^{-}(0,t)|^{2}}=\frac{t_{2}^{2}r_{1}^{2}e^{\gamma L}}{t_{1}^{2}}=\frac{t_{2}^{2}r_{1}}{t_{1}^{2}r_{2}}
\end{equation}
when the gain condition $e^{-\gamma L}=r_{1}r_{2}$ is satisfied.
The fraction $R_{2}$ of power emitted from the facet at $z=L$ is
then
\begin{equation}
R_{2}=\frac{P_{out}(L)}{P_{out}(0)+P_{out}(L)}=\frac{r_{1}t_{2}^{2}}{r_{2}t_{1}^{2}+r_{1}t_{2}^{2}}=\frac{r_{1}(1-r_{2}^{2})}{(r_{1}+r_{2})(1-r_{1}r_{2})}\,.\label{R2}
\end{equation}
We get the fraction $R_{1}$ of loss through the facet at $z=0$ by
interchanging the indices $"1"$ and $"2"$ in (\ref{R2}) or by using
$R_{1}+R_{2}=1$.

\section{The laser field equation}

In order to derive a field equation that applies for $R_{sp}\neq0$
and a time dependent carrier number we expand the loop gain $G(\omega,N)$
around the solution $(\omega_{0},N_{0})$ to the oscillation condition.
Thus $G(\omega_{0},N_{0})=1$ and
\begin{equation}
1-G(\omega,N)=1-G(\omega,N)/G(\omega_{0},N_{0})=1-e^{-j2L\Delta\beta}\simeq j2L\Delta\beta\label{G exp 1}
\end{equation}
where
\begin{equation}
\Delta\beta=\beta(\omega,N)-\beta(\omega_{0},N_{0})\simeq\textrm{Re}\left\{ \frac{\partial\beta}{\partial\omega}\right\} (\omega-\omega_{0})+\frac{\partial\beta}{\partial N}(N-N_{0})\,.\label{Delta beta exp}
\end{equation}
We ignore the derivative $\frac{\partial g}{\partial\omega}$ because
the solution $(\omega_{0},N_{0})$ to the oscillation condition must
be close to the maximum of $|G(\omega,N_{0})|$ where $\frac{\partial g}{\partial\omega}=0$.
From (\ref{Mat gain g}) we see that
\begin{equation}
\frac{\partial\beta}{\partial N}=\frac{k_{0}}{2\bar{n}}\Gamma\frac{\partial\tilde{\chi}_{a}}{\partial N}=\frac{k_{0}}{2\bar{n}}\Gamma\left(\textrm{Re}\left\{ \frac{\partial\tilde{\chi}_{a}}{\partial N}\right\} +j\textrm{Im}\left\{ \frac{\partial\tilde{\chi}_{a}}{\partial N}\right\} \right)=\frac{j}{2}(1+j\alpha)\Gamma\frac{\partial g}{\partial N}\label{dbeta/dN}
\end{equation}
where we have introduced the dimensionless parameter
\begin{equation}
\alpha=-\frac{\textrm{Re}\left\{ \frac{\partial\tilde{\chi}_{a}}{\partial N}\right\} }{\textrm{Im}\left\{ \frac{\partial\tilde{\chi}_{a}}{\partial N}\right\} }\label{alpha def}
\end{equation}
for $\omega=\omega_{0}$. It depends on the semiconductor material
and the spatial dimension of the active layer; for our example we
assume $\alpha=5$. It is usually referred to as the linewidth enhancement
factor; the latter name is justified in a later section. Introducing
the group velocity $v_{g}$ and the round-trip time $\tau_{L}=2L/v_{g}$
the expansions (\ref{G exp 1}) and (\ref{Delta beta exp}) lead to
\begin{equation}
1-G(\omega,N)\simeq\left[j(\omega-\omega_{0})-\frac{1}{2}(1+j\alpha)v_{g}\Gamma\frac{\partial g}{\partial N}(N-N_{0})\right]\tau_{L}\,.\label{G exp 2}
\end{equation}
The linear approximation (\ref{eq:glinear1}) implies $a=v_{g}\Gamma\frac{\partial g}{\partial N}$.
With the expansion (\ref{G exp 2}) the field equation (\ref{Field eq at L})
then becomes
\begin{equation}
[j(\omega-\omega_{0})-\frac{1}{2}(1+j\alpha)a(N-N_{0})]\tilde{E}^{+}(L)=\frac{1}{\tau_{L}}\tilde{F}^{+}\,.\label{eq:Eplus field new}
\end{equation}
In order to relate the equation to the photon number rate equation
(\ref{P rate 2}) we introduce the scaled envelope fields
\begin{equation}
E(t)=K\int_{0}^{\infty}\tilde{E}^{+}(L,\omega)e^{j(\omega-\omega_{0})t}df\label{Envelope E}
\end{equation}
\begin{equation}
F(t)=K\frac{1}{\tau_{L}}\int_{0}^{\infty}\tilde{F}^{+}(\omega)e^{j(\omega-\omega_{0})t}df\label{Envelope F}
\end{equation}
where $K^{2}$ is the factor (\ref{K2}). The definition $E(t)$ is
chosen such that by (\ref{eq:P integral-2 new}) we have the relation
\begin{equation}
{\cal {P}}(t)=|E(t)|^{2}\label{NormE}
\end{equation}
between total photon number and the envelope field. The approximation
implies that the photon density distribution scales with the output
power at the right facet. By taking the Fourier transform (\ref{Envelope E})
of the field equation (\ref{eq:Eplus field new}) for constant carrier
number $N$ we get the rate equation
\begin{equation}
\frac{d}{dt}E(t)=\frac{1}{2}(1+j\alpha)a(N-N_{0})E(t)+F(t)\label{Field rate eq}
\end{equation}
for the envelope field. For the photon number it gives the rate equation
\begin{equation}
\frac{d}{dt}{\cal {P}}(t)=E^{*}(t)\frac{d}{dt}E(t)+E(t)\frac{d}{dt}E^{*}(t)=a(N-N_{0}){\cal {P}}(t)+R_{sp}+F_{P}(t)\label{P rate 3}
\end{equation}
where $F_{P}(t)$ is defined by
\begin{equation}
F_{P}=E^{*}F+EF^{*}-R_{sp}\,.\label{Langevin-Fp}
\end{equation}
In order to compare the equation with (\ref{P rate 2}) we notice
that
\begin{equation}
v_{g}\Gamma g(\omega_{0},N_{0})=a(N_{0}-N_{tr})=\frac{1}{\tau_{p}}\label{Photon lt 1}
\end{equation}
according to (\ref{Clamped N0}) and (\ref{eq:glinear1}) for $\omega=\omega_{0}$
and $N=N_{0}$. Combining (\ref{Photon lt 1}) with the gain condition
(\ref{Gain cond}) we see that
\begin{equation}
\frac{1}{\tau_{p}}=v_{g}(\alpha_{i}+\alpha_{m})\label{Photon lt 2}
\end{equation}
which is an explicit relation between photon lifetime $\tau_{p}$
and the internal absorption $\alpha_{i}$ and mirror loss $\alpha_{m}$.
Inserting
\begin{equation}
a(N-N_{0})=a(N-N_{tr})-a(N_{0}-N_{tr})=a(N-N_{tr})-\frac{1}{\tau_{p}}\label{a(N-N0)andtaup}
\end{equation}
in (\ref{P rate 3}) it takes the form
\begin{equation}
\frac{d}{dt}{\cal {P}}(t)=\left[a(N-N_{tr})-\frac{1}{\tau_{p}}\right]{\cal {P}}(t)+R_{sp}+F_{P}(t)\label{eq:P rate 4 new}
\end{equation}
which is identical to (\ref{P rate 2}) except for the function $F_{P}(t)$.
As we shall see below, it is a fluctuating function with average $\langle F_{P}(t)\rangle=0$,
so (\ref{eq:P rate 4 new}) is a correction to (\ref{P rate 2}) that
includes noise contributions.

The field equation (\ref{Field rate eq}) was derived assuming $N$
to be constant. However, since it reproduces the photon number rate
equation (\ref{eq:P rate 4 new}) where $N$ may be time dependent,
we will also use (\ref{Field rate eq}) for time dependent $N$.

\section{Multimode field equation and multimode spectrum}

In the expansion of the loop gain $G(\omega,N)$ in (\ref{G exp 1})
we assumed that $(\omega,N)$ is close to the solution $(\omega_{0},N_{0})$
to the oscillation condition, $G(\omega,N)=1$, with lowest carrier
number $N_{0}$. We will now relax this assumption and study the field
equation (\ref{Loop gain G}) in a broader range of angular frequencies
around $\omega_{0}$. This leads to the field power spectral density
that applies for currents above the threshold current $I_{th}$. Remember,
the formula (\ref{SE below th}) for the field power spectral density
only applies for constant carrier number $N$ below $N_{0}$, i.e.
for currents below the threshold. We follow the method used in a number
of papers 
\cite{Olesen-1986,Tromborg-1987,Mork-1992,Tromborg-1994,Detoma-2005}
where the modes and sidemodes of a laser system are identified as
solutions to the oscillation condition, $G(\omega,N)=1$, for angular
frequencies $\omega$ in the complex plane.

\begin{figure}
\begin{centering}
\includegraphics[scale=0.5]{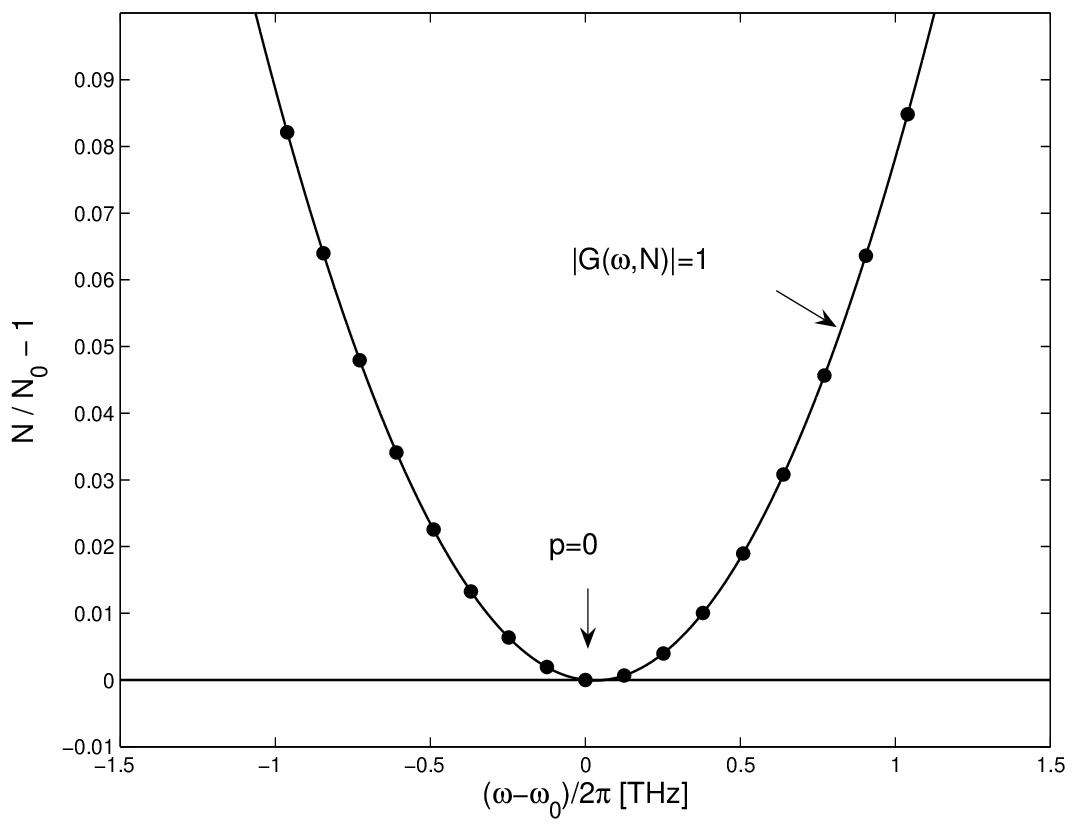}
\par\end{centering}

\caption{The bullets mark the solutions $(\omega_{p},N_{p})$ to the oscillation
condition $G(\omega,N)=1$. They represent the modes of the laser.
The solid curve is the parabola (\ref{Im-del-beta}) where the gain
condition $|G(\omega,N)|=1$ is satisfied.}

\label{N-vs-frequency}
\end{figure}

In the expansions (\ref{G exp 1}) and (\ref{Delta beta exp}) we
ignored the dependence of the gain on frequency. With $\beta=k_{0}n+j\tfrac{1}{2}(\Gamma g-\alpha_{i})$
as given by (\ref{beta vs n&g}) the frequency dependence is included
by adding the term $j\tfrac{1}{2}\Gamma(g(\omega,N_{0})-g(\omega_{0},N_{0}))$
to $\Delta\beta$ in (\ref{Delta beta exp}), so $2L\Delta\beta$
becomes
\begin{equation}
2L\Delta\beta=\tau_{L}[\omega-\omega_{0}+j\tfrac{1}{2}(1+j\alpha)a(N-N_{0})+j\tfrac{1}{2}v_{g}\Gamma(g(\omega,N_{0})-g(\omega_{0},N_{0}))]\,.\label{Del-beta-gain}
\end{equation}
It contains the r.h.s. of (\ref{G exp 2}) times $(-j)$. The approximation
(\ref{eq:g(omega,N)}) gives
\begin{gather}
g(\omega,N_{0})-g(\omega_{0},N_{0})\simeq-g_{m}(N_{0})\left[\left(\frac{\omega-\omega_{m}}{\Delta\omega_{g}}\right)^{2}-\left(\frac{\omega_{0}-\omega_{m}}{\Delta\omega_{g}}\right)^{2}\right]\nonumber \\
=\frac{2g_{m}(N_{0})(\omega_{m}-\omega_{0})}{(\Delta\omega_{g})^{2}}(\omega-\omega_{0})-\frac{g_{m}(N_{0})}{(\Delta\omega_{g})^{2}}(\omega-\omega_{0})^{2}
\end{gather}
and thus
\begin{equation}
2L\Delta\beta=(\tau_{L}+j\tau_{1})(\omega-\omega_{0})-j\tau_{2}^{2}(\omega-\omega_{0})^{2}+j\tfrac{1}{2}(1+j\alpha)\tau_{L}a(N-N_{0})\label{del-beta-N}
\end{equation}
where by (\ref{Photon lt 1})
\begin{equation}
\tau_{2}^{2}=\frac{\tau_{L}v_{g}\Gamma g_{m}(N_{0})}{2(\Delta\omega_{g})^{2}}\simeq\frac{\tau_{L}}{2\tau_{p}(\Delta\omega_{g})^{2}}\ \ \textrm{and}\ \ \tau_{1}=2\tau_{2}^{2}(\omega_{m}-\omega_{0})\,.\label{tau-1-2}
\end{equation}
We denote the solutions $(\omega_{p},N_{p})$ to the oscillation condition
\begin{equation}
G(\omega,N)=e^{-j2L\Delta\beta(\omega,N)}=1\label{Osc-con-mm}
\end{equation}
as the modes of the laser. With the approximation (\ref{del-beta-N})
we find the modes as solutions to the gain condition $|G|=1$, i.e.
$\textrm{Im}\{\Delta\beta\}=0$ or
\begin{equation}
\tfrac{1}{2}\tau_{L}a(N-N_{0})=-\tau_{1}(\omega-\omega_{0})+\tau_{2}^{2}(\omega-\omega_{0})^{2}\label{Im-del-beta}
\end{equation}
and to the phase condition $\arg G=-2\pi p$, i.e. $2L\textrm{Re}\{\Delta\beta\}=2\pi p$
or
\begin{equation}
\tau_{L}(\omega-\omega_{0})-\tfrac{1}{2}\alpha\tau_{L}a(N-N_{0})=2\pi p\label{Re-del-beta}
\end{equation}
for integer p. Inserting (\ref{Im-del-beta}) in (\ref{Re-del-beta})
we get the equation
\begin{equation}
(\tau_{L}+\alpha\tau_{1})(\omega-\omega_{0})-\alpha\tau_{2}^{2}(\omega-\omega_{0})^{2}=2\pi p\label{Freq-space}
\end{equation}
for the frequencies of the modes. It gives a mode spacing that is
slightly modified compared to the simple mode spacing $\Delta\omega=2\pi/\tau_{L}$
in (\ref{frequency spacing}). We show in Figure \ref{N-vs-frequency}
an example of the solutions to the oscillation condition as bullets
in the $(\omega,N)$-plane for the parameters of Table \ref{Table laser diode}
and for $\tau_{1}=0.0039$ ps, $\tau_{2}=0.1$ ps and $\tau_{L}=8$
ps. The solid curve through the bullets is the parabola (\ref{Im-del-beta}).

\subsubsection{Multimode field equation}

We can use the approximation $G=e^{-j2L\Delta\beta}$ to derive a
field equation that applies for a broader range of frequencies than
(\ref{Field rate eq}). By taking the Fourier transform of (\ref{Field eq at L})
and introducing the envelopes (\ref{Envelope E}) and (\ref{Envelope F})
we get
\begin{equation}
\int_{0}^{\infty}\left(1-e^{-j2L\Delta\beta}\right)E^{+}(\omega)e^{j(\omega-\omega_{0})t}df=\tau_{L}F(t)\,.\label{Field-eq-1}
\end{equation}
For constant $N$ it becomes
\begin{equation}
E(t)-e^{\tfrac{1}{2}(1+j\alpha)\tau_{L}a(N-N_{0})}h(t)\otimes E(t)=\tau_{L}F(t)\label{Field-eq-2}
\end{equation}
where $h(t)$ is the impulse response
\begin{equation}
h(t)={\cal {F}}^{-1}\left[e^{-j(\tau_{L}+j\tau_{1})\omega-\tau_{2}^{2}\omega^{2}}\right]=\frac{1}{\tau_{2}\sqrt{4\pi}}e^{-\frac{(t-\tau_{L}-j\tau_{1})^{2}}{4\tau_{2}^{2}}}\,.\label{Res-fun-lg}
\end{equation}
Notice, the parameter $\omega$ in (\ref{Res-fun-lg}) is a baseband
angular frequency. When we deal with the field envelope we will for
simplicity write the Fourier transform as $\tilde{E}(\omega)$ instead
of $\tilde{E}(\omega-\omega_{0})$. For $N=N_{0}$ the impulse response
$h(t)$ gives the output field after one roundtrip of an input delta
pulse in the laser cavity. It has a non-zero tail for $t<0$, so it
is only approximately causal. In order to ensure that (\ref{Res-fun-lg})
satisfies causality we can replace $h(t)$ by $h(t)u(t)$ where $u(t)$
is the step function. The transfer function
\begin{equation}
H(\omega)=e^{-j(\tau_{L}+j\tau_{1})\omega-\tau_{2}^{2}\omega^{2}}\label{Trans-fun-lg}
\end{equation}
is analytic in the lower half $\omega$-plane but it does not satisfy
the condition (B.51)-[1]
that would have been a sufficient condition for $H(\omega)$ to be
a causal transfer function.

Using the representation (1.24)-[1] for the delta function
with $\sigma=\sqrt{2}\tau_{2}$ the impulse response for $\tau_{1}$
is seen to become $\delta(t-\tau_{L})$ in the limit $\tau_{2}\rightarrow0$
and the field equation (\ref{Field-eq-2}) becomes
\begin{equation}
E(t)-e^{\tfrac{1}{2}(1+j\alpha)\tau_{L}a(N-N_{0})}E(t-\tau_{L})=\tau_{L}F(t)\,.\label{Field-eq-3}
\end{equation}
With $N\simeq N_{0}$ and $E(t-\tau_{L})\simeq E(t)-\tau_{L}\frac{d}{dt}E(t)$
the equation reproduces the field equation (\ref{Field rate eq}).

The field equation (\ref{Field-eq-1}) was derived under the assumption
that the carrier number is constant. However, we will assume that
it also holds for a time dependent carrier number $N(t)$ that satisfies
the rate equation (\ref{N rate 2}). Using (\ref{a(N-N0)andtaup})
it becomes
\begin{equation}
\frac{d}{dt}N=J_{s}-\frac{N}{\tau_{e}}-\left(a(N-N_{0})+\frac{1}{\tau_{p}}\right){\cal {P}}\label{Mod-rate-N}
\end{equation}
for constant carrier injection rate $J_{s}$. For $F(t)=0$, i.e.
in the limit of no spontaneous emission, the photon number $P_{0}$
satisfies the steady state equation (\ref{P0 vs I0}). The equations
(\ref{Field-eq-2}) and (\ref{Mod-rate-N}) can be used for a simple
simulation of the time-domain behaviour of the field envelope $E(t)$
and the carrier number $N(t)$ with $F(t)$ as a fluctuating driving
force.

\subsubsection{Multimode spectrum}

In order to derive the field power spectral density from (\ref{Field-eq-2})
for time varying $N(t)$ we assume
\begin{equation}
N(t)=N_{0}+\delta N(t)\label{delta-N}
\end{equation}
and
\begin{equation}
E(t)=E_{0}+\delta E(t)\label{delta-E}
\end{equation}
where $\delta N(t)$ and $\delta E(t)$ are small deviations from
the steady state values $N_{0}$ and $E_{0}$ for $F(t)=0$ and constant
carrier injection rate $J_{s}$. The photon number ${\cal {P}}(t)=|E(t)|^{2}$
is to 1st order
\begin{equation}
{\cal {P}}(t)\simeq E_{0}^{2}+E_{0}(\delta E(t)+\delta E^{*}(t))\label{delta-P}
\end{equation}
where $P_{0}=E_{0}^{2}$ is given by (\ref{P0 vs I0}). The convolution
$h(t)\otimes E_{0}$ is
\begin{equation}
h(t)\otimes E_{0}=E_{0}\int_{-\infty}^{\infty}h(t)dt=E_{0}H(0)=E_{0}\label{hxE0}
\end{equation}
so inserting (\ref{delta-P}) and (\ref{hxE0}) in the field equation
(\ref{Field-eq-2}) we get
\begin{gather}
E(t)-e^{\tfrac{1}{2}(1+j\alpha)\tau_{L}a\delta N(t)}h(t)\otimes E(t)\nonumber \\
=E_{0}(1-e^{\tfrac{1}{2}(1+j\alpha)\tau_{L}a\delta N(t)})+\delta E(t)-e^{\tfrac{1}{2}(1+j\alpha)\tau_{L}a\delta N(t)}h(t)\otimes\delta E(t)
=\tau_{L}F(t)\,.
\end{gather}
To 1st order it reduces to
\begin{equation}
\delta E(t)-h(t)\otimes\delta E(t)-E_{0}\frac{1}{2}(1+j\alpha)\tau_{L}a\delta N(t)=\tau_{L}F(t)\,.\label{delE-delN}
\end{equation}
For the rate equation (\ref{Mod-rate-N}) the insertion of (\ref{delta-N})
and (\ref{delta-P}) leads to
\begin{equation}
\frac{d}{dt}\delta N(t)=-\Gamma_{N}\delta N(t)-\frac{E_{0}}{\tau_{p}}(\delta E(t)+\delta E^{*}(t))\label{rate-del-N}
\end{equation}
to 1st order. The zero order terms cancel because of (\ref{P0 vs I0}).
$\Gamma_{N}$ is the damping rate $\Gamma_{N}=\frac{1}{\tau_{e}}+aP_{0}$
introduced in (\ref{Gamma N}). The Fourier transform of (\ref{rate-del-N})
gives
\begin{equation}
\widetilde{\delta N}(\omega)=-\frac{E_{0}}{\tau_{p}}\frac{\widetilde{\delta E}(\omega)+\widetilde{\delta E}^{*}(-\omega)}{j\omega+\Gamma_{N}}\label{Ft-del-N}
\end{equation}
which by insertion in the Fourier transform of (\ref{delE-delN})
gives
\begin{equation}
(1-H(\omega)+C(\omega))\widetilde{\delta E}(\omega)+C(\omega)\widetilde{\delta E}^{*}(-\omega)=\tau_{L}\tilde{F}(\omega)\label{CdelEF}
\end{equation}
where
\begin{equation}
C(\omega)=\frac{\tau_{L}aP_{0}(1+j\alpha)}{2\tau_{p}(j\omega+\Gamma_{N})}=\frac{\tau_{L}\Omega_{R}^{2}(1+j\alpha)}{2(j\omega+\Gamma_{N})}\label{COmG}
\end{equation}
and $\Omega_{R}$ is the relaxation resonance angular frequency (\ref{Relax freq})
given by $\Omega_{R}^{2}=aP_{0}/\tau_{p}$. By taking the complex
conjugate of (\ref{CdelEF}) and replacing $\omega$ with $-\omega$
we get
\begin{equation}
C^{*}(-\omega)\widetilde{\delta E}(\omega)+(1-H^{*}(-\omega)+C^{*}(-\omega))\widetilde{\delta E}^{*}(-\omega)=\tau_{L}\tilde{F}^{*}(-\omega)\label{CdelE*}
\end{equation}
The two equations (\ref{CdelEF}) and (\ref{CdelE*}) have the solution
\begin{equation}
\widetilde{\delta E}(\omega)=\tau_{L}\frac{\tilde{F}(\omega)(1-H^{*}(-\omega)+C^{*}(-\omega))-\tilde{F}^{*}(-\omega)C(\omega)}{D(\omega)}\,.\label{delEFC}
\end{equation}
The denominator is
\begin{gather}
D(\omega)=(1-H(\omega)+C(\omega))(1-H^{*}(-\omega)+C^{*}(-\omega))-C(\omega)C^{*}(-\omega)\nonumber \\
=(1-H^{*}(-\omega))(1-G_{0}(\omega))\label{DHCC*}
\end{gather}
where
\begin{equation}
G_{0}(\omega)=H(\omega)-C(\omega)-C^{*}(-\omega)(1-H(\omega))/(1-H^{*}(-\omega))\,.\label{LoopG0}
\end{equation}
Since $H(\omega)=G(\omega_{0}+\omega,N_{0})$ the function $G_{0}(\omega)$
is the loop gain when fluctuations of the carrier number around $N_{0}$
are taken into account.

The field power spectral density of $E_{c}(t)=E(t)e^{j\omega_{0}t}$
is a spectrum at optical angular frequencies around $\omega_{0}=2\pi f_{0}$
while the spectrum of the envelope $E(t)$ is a function of baseband
frequency $\omega=2\pi f$. Thus
\begin{gather}
S_{E_{c}}(f+f_{0})=\int_{-\infty}^{\infty}\langle E_{c}^{*}(t)E_{c}(t+\tau)\rangle e^{-j(\omega+\omega_{0})\tau}d\tau=\int_{-\infty}^{\infty}\langle E^{*}(t)E(t+\tau)\rangle e^{-j\omega\tau}d\tau\nonumber \\
=S_{E}(f)=E_{0}^{2}\delta(f)+S_{\delta E}(f).\label{SEcdelE}
\end{gather}
The arbitrariness of the phase of $\delta E(t)$ implies that $\langle\delta E(t)\rangle=0$
and there is therefore no mixed terms from $\langle E_{0}\delta E(t+\tau)\rangle$
and $\langle\delta E^{*}(t)E_{0}\rangle$.

According to (6.78)-[1], the linear relation
(\ref{delEFC}) gives a linear relation between the power spectral
density $S_{\delta E}(f)$ and the power spectral density of $F(t)$.
The function $F(t)$ is the envelope function (\ref{Envelope F})
derived from $\tilde{F}^{+}$ where $S_{F^{+}}(\omega)$ is the power
spectral density of spontaneous emission in (\ref{SE below th}) for
$\omega\simeq\omega_{0}$. The next chapter discusses in some detail
the stochastic properties of $F(t)$. For now we will simply quote
that $F(t)$ is a noise function with ensemble averages
\begin{equation}
\langle F(t)\rangle=\langle F(t)F(t')\rangle=\langle F^{*}(t)F^{*}(t')\rangle=0\label{Av-of-FF}
\end{equation}
and
\begin{equation}
\langle F^{*}(t)F(t')\rangle=R_{sp}\delta(t-t')\,.\label{Av-of-F*F}
\end{equation}
For $F(t)=u(t)+jv(t)$ the relations imply that $\langle u(t)\rangle=\langle v(t)\rangle=\langle u(t)v(t')\rangle=0$
and $\langle u(t)u(t')\rangle=\langle v(t)v(t')\rangle=\tfrac{1}{2}R_{sp}\delta(t-t')$.
The property $\langle\delta E(t)\rangle=0$ stated above follows from
(\ref{delEFC}) and $\langle F(t)\rangle=0$. The power spectral density
of $F(t)$ is
\begin{equation}
S_{F}(f)=\int_{-\infty}^{\infty}\langle F^{*}(t)F(t+\tau)\rangle e^{-j\omega\tau}d\tau=R_{sp}\,.\label{SF-Rsp}
\end{equation}
Similarly $S_{F^{*}}(f)=R_{sp}$ and the cross spectral densities
involving $\langle F(t)F(t')\rangle$ and $\langle F^{*}(t)F^{*}(t')\rangle$
are zero. By applying the theorem (6.73)-[1],
the power spectral density $S_{\delta E}(f)$ obtained from (\ref{delEFC})
then becomes
\begin{equation}
S_{\delta E}(f)=\tau_{L}^{2}R_{sp}\frac{|1-H^{*}(-\omega)+C^{*}(-\omega)|^{2}+|C(\omega)|^{2}}{|D(\omega)|^{2}}\,.\label{SdelE}
\end{equation}
The spectrum diverges at $\omega=0$ because $|D(\omega)|^{2}\propto\omega^{2}$
for $\omega\rightarrow0$. The Taylor expansion of $\omega^{2}S_{\delta E}(f)$
shows that
\begin{gather}
\omega^{2}S_{\delta E}(f)=\frac{\tau_{L}^{2}R_{sp}}{(\tau_{L}+\alpha\tau_{1})^{2}}\left(\tfrac{1}{2}(1+\alpha^{2})-\left(\alpha-\frac{\tau_{1}}{\tau_{L}}\right)\frac{\Gamma_{N}}{\Omega_{R}^{2}}\omega+\cdots\right)\nonumber \\
\simeq R_{sp}\left(\tfrac{1}{2}(1+\alpha^{2})-\frac{\alpha\Gamma_{N}}{\Omega_{R}^{2}}\omega+\cdots\right)\label{ExpSdelE}
\end{gather}
for $\tau_{1}\ll\tau_{L}$. The spectra $S_{E}(f)$ and $S_{\delta E}(f)$
are the result of an expansion in terms of $R_{sp}$. The delta function
in $S_{E}(f)$ is a zero order term and $S_{\delta E}(f)$ is a first
order term in $R_{sp}$. The leading singularities of $S_{E}(f)$
at $\omega=0$ come from the delta function in (\ref{SEcdelE}) and
the $1/\omega^{2}$ term of $S_{\delta E}(f)$ in (\ref{ExpSdelE}).
Both the delta function and the $1/\omega^{2}$ singularity are included
in a formal expansion of a Lorenzian $2\gamma/(\omega^{2}+\gamma^{2})$.
It represents by (1.23)-[1] the delta function in
the limit $\gamma\rightarrow0$, and for $\omega\neq0$ the expansion
in $\gamma$ is a geometric series. Thus
\begin{equation}
\frac{2\gamma}{\omega^{2}+\gamma^{2}}=\delta(f)+\frac{2\gamma}{\omega^{2}}\sum_{n=0}^{\infty}(-1)^{n}\left(\frac{\gamma}{\omega}\right)^{2n}=\delta(f)+\frac{2\gamma}{\omega^{2}}-\frac{2\gamma^{3}}{\omega^{4}}+\cdots\,.\label{Lorenz-exp}
\end{equation}
By replacing $P_{0}\delta(f)+R_{sp}(1+\alpha^{2})/(2\omega^{2})$
by $2P_{0}\gamma/(\omega^{2}+\gamma^{2})$ for
\begin{equation}
\gamma=\frac{R_{sp}(1+\alpha^{2})}{4P_{0}}\label{gam-Rsp}
\end{equation}
we get rid of the leading singularities in $S_{E}(f)$. If we furthermore
replace $\gamma_{1}/\omega$ by $\gamma_{1}\omega/(\omega^{2}+\gamma^{2})$
where $\gamma_{1}=-R_{sp}\alpha\Gamma_{N}/\Omega_{R}^{2}$, we also
get rid of the $1/\omega$ singularity. This results in the field
power spectral density
\begin{equation}
S_{E}(f)=\frac{2P_{0}\gamma}{\omega^{2}+\gamma^{2}}+\frac{\gamma_{1}\omega}{\omega^{2}+\gamma^{2}}+S_{\delta E}(f)-\frac{2P_{0}\gamma}{\omega^{2}}-\frac{\gamma_{1}}{\omega}\,.\label{FiniteS-E}
\end{equation}
The spectrum is the same as (\ref{SEcdelE}) to 1st order in $R_{sp}$
but it is finite at $\omega=0$ if we use $\tau_{1}=0$. For $\tau_{1}\neq0$
we have to use the precise expansion coefficients in (\ref{ExpSdelE})
to modify (\ref{SEcdelE}) to cancel the singularities at $f=0$.
We will later on confirm the Lorenzian shape of the spectrum near
$\omega=0$ by including higher order terms.
\begin{figure}
\begin{centering}
\includegraphics[scale=0.5]{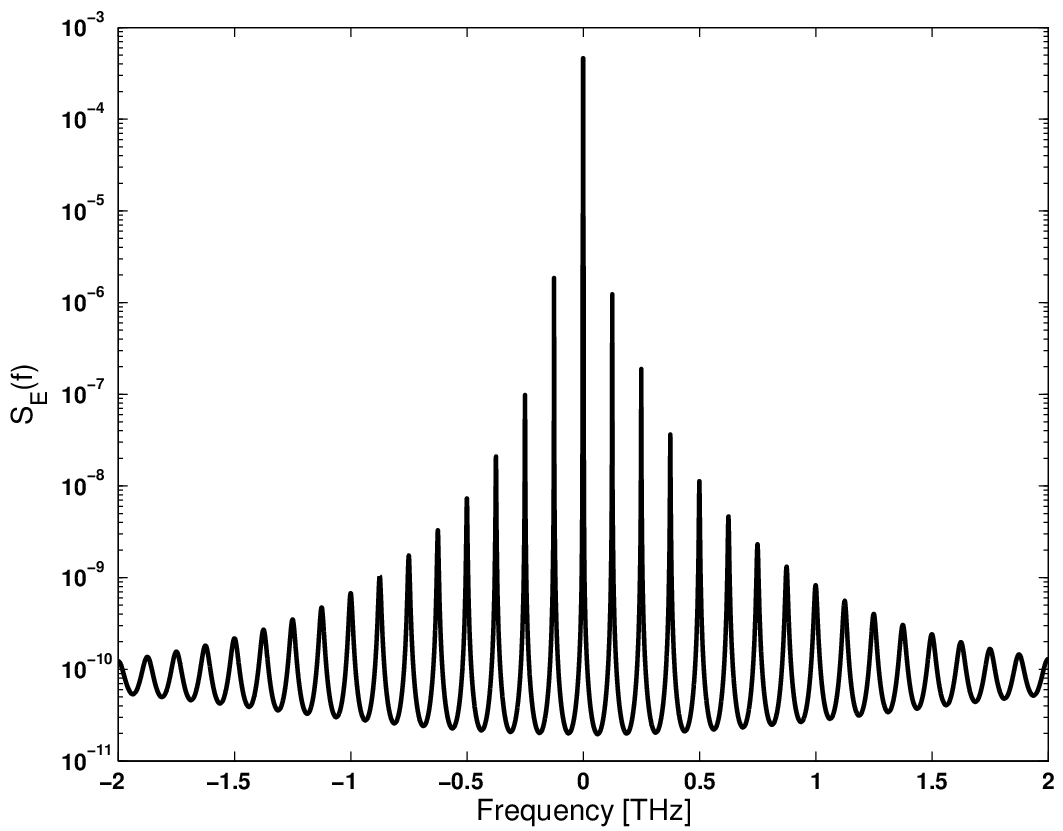}
\par\end{centering}

\caption{Power spectral density $S_{E}(f)$ calculated from (\ref{FiniteS-E}).}

\label{MultiMode-spectrum}
\end{figure}

\begin{figure}
\begin{centering}
\includegraphics[scale=0.5]{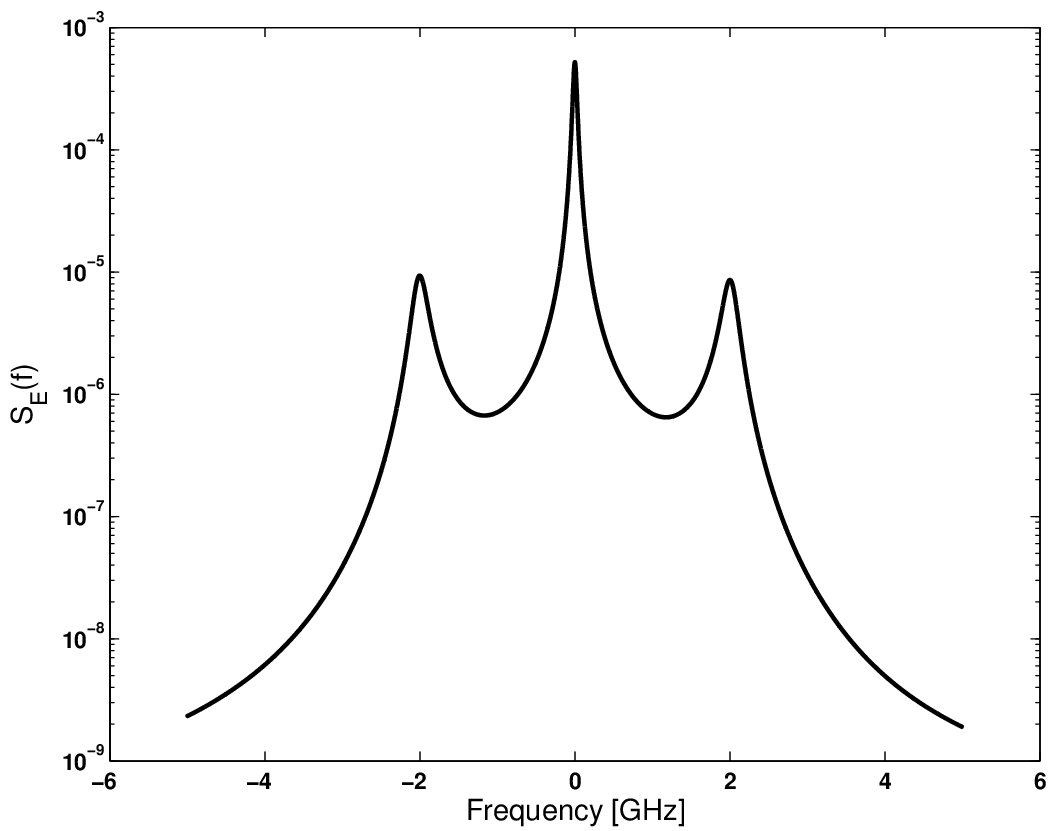}
\par\end{centering}

\caption{Details of the power spectral density $S_{E}(f)$ showing the central
Lorenzian shape and the adjacent relaxation resonances.}

\label{MultiMode-spectrum-a}
\end{figure}

The spectrum (\ref{FiniteS-E}) is shown in Figure \ref{MultiMode-spectrum}
for the parameters used in Figure \ref{N-vs-frequency} and for $J_{s}=1.3J_{th}$.
The central spike at $f=0$ is the Lorenzian and the other peaks are
sidemodes with approximate frequency spacing $1/\tau_{L}=125$ GHz.
The spectrum is calculated for constant $R_{sp}$. To be more realistic
we could have taken into account that $R_{sp}$ depends on frequency
and has the shape of the amplified spontaneous emission spectrum.
The Lorenzian is shown on an extended scale in Figure \ref{MultiMode-spectrum-a}
. It has two satellite peaks that we will identify below as relaxation
resonances at angular frequencies $\pm\Omega/2\pi$, where $\Omega=\sqrt{\Omega_{R}^{2}-(\Gamma_{N}/2)^{2}}$.
With no photons in the laser cavity the maximum carrier number is
given by $N/N_{0}-1=(J_{s}\tau_{e})/(J_{th}\tau_{e})-1=J_{s}/J_{th}-1=0.3$
for $J_{s}=1.3J_{th}$. This is around three times the vertical scale
in Figure \ref{N-vs-frequency}.

\subsubsection{Stability of the mode}

In the derivation of the spectrum (\ref{FiniteS-E}) we did not examine
whether the mode $(\omega_{0},N_{0})$ is actually a stable mode.
It will not be stable, and the spectrum will not be given by (\ref{FiniteS-E}),
if $1-G_{0}(\omega)$ has one or more zeros in the lower half complex
$\omega$-plane. If the mode is stable, the zeros of $1-G_{0}(\omega)$
in the upper half complex $\omega$-plane give rise to the spikes
in the spectrum and they therefore give the location and width of
the sidemodes. So let us show how to determine the complex solutions
to $G_{0}(\omega)=1$.

\begin{figure}
\begin{centering}
\includegraphics[scale=0.5]{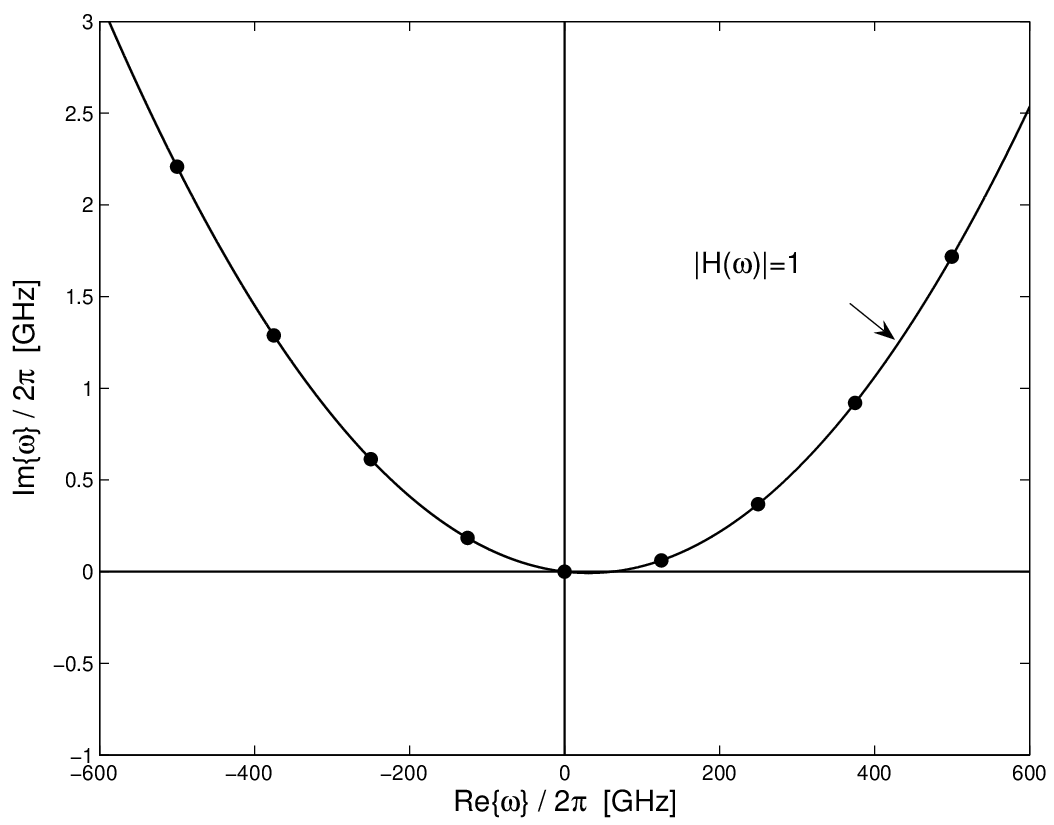}
\par\end{centering}

\caption{The bullets show the location of the solutions to $H(\omega)=1$ in
the complex $\omega$-plane. The solid curve shows the solution to
$|H(\omega)|=1$ }

\label{Complex zeros G0}
\end{figure}

When we extend $G_{0}(\omega)$ to complex $\omega$ we have to take
into account that $H^{*}(-\omega)$ is defined from $H(\omega)$ in
(\ref{Trans-fun-lg}) for real $\omega$ and is here $\exp(-j(\tau_{L}-j\tau_{1})\omega-\tau_{2}^{2}\omega^{2})$.
The continuation of $H^{*}(-\omega)$ to complex $\omega$ is therefore
the function $H_{1}(\omega)=\exp(-j(\tau_{L}-j\tau_{1})\omega-\tau_{2}^{2}\omega^{2})$,
which is the same as $H(\omega)$ except for a change of the sign
of $\tau_{1}$. In the same manner we see that the continuation of
$C^{*}(-\omega)$ to complex values is the function $C_{1}(\omega)=\frac{\tau_{L}\Omega_{R}^{2}(1-j\alpha)}{2(j\omega+\Gamma_{N})}$,
i.e. the same as $C(\omega)$ except for the change of sign of $\alpha$.
Thus
\begin{equation}
G_{0}(\omega)=H(\omega)-C(\omega)-C_{1}(\omega)(1-H(\omega))/(1-H_{1}(\omega))
\end{equation}
where all functions are defined for all complex $\omega$.

As a first step we will derive the zeros of $1-G_{0}(\omega)$ for
$C(\omega)=C_{1}(\omega)=0$, i.e. for $G_{0}(\omega)=H(\omega)$.
For $H(\omega)$ given by (\ref{Trans-fun-lg}) the zeros are the
complex solutions to
\begin{equation}
j(\tau_{L}+j\tau_{1})\omega+\tau_{2}^{2}(\omega)^{2}=j2\pi p\label{G-Isol}
\end{equation}
for integer $p$. Using the substitution $\omega=x+jy$ the real part
of (\ref{G-Isol}) gives the hyperbola
\begin{equation}
\left(y+\frac{\tau_{L}}{2\tau_{2}^{2}}\right)^{2}-\left(x-\frac{\tau_{1}}{2\tau_{2}^{2}}\right)^{2}=\frac{\tau_{L}^{2}-\tau_{1}^{2}}{(2\tau_{2}^{2})^{2}}\label{Hyperb}
\end{equation}
where $|H(\omega)|=1$. The imaginary part of (\ref{G-Isol}) gives
\begin{equation}
y=\frac{2\pi p-\tau_{L}x}{2\tau_{2}^{2}x-\tau_{1}}\,.\label{Phase-G0}
\end{equation}
The zeros on the upper branch of the hyperbola are shown as bullets
in the complex $\omega$-plane in Figure \ref{Complex zeros G0} for
the same parameters as in Figure \ref{N-vs-frequency}. The location
of the real part of the zeros for $p\neq0$ gives the approximate
position of the sidemodes of the mode. The spacing between the sidemodes
is close to the mode spacing given by (\ref{Freq-space}) but is not
exactly the same. The hyperbola has its center at $(\tau_{1}-j\tau_{L})/(2\tau_{2}^{2})$
in the complex $\omega$-plane and its lower branch vertex at $\textrm{Im}(\omega)\simeq-\tau_{L}/\tau_{2}^{2}\simeq-2\Delta\omega_{g}^{2}\tau_{p}$.
For our example $\tau_{L}/(2\pi\tau_{2}^{2})=127$ THz, i.e. the lower
branch is in the lower half complex $\omega$-plane far below the
real axis on the scale of Figure \ref{Complex zeros G0}. There are
also zeros on the lower branch and they should indicate that the mode
$(\omega_{0},N_{0})$ is unstable. However, laying more than $2\Delta\omega_{g}^{2}\tau_{p}$
below the real axis, the zeros rather indicate that the approximation
(\ref{eq:g(omega,N)}) is not realistic for large $|\textrm{Im}(\omega)|$%
\footnote{The $\Delta\beta$ in (\ref{del-beta-N}) does not satisfy a Kramers-Kronig
relations as required by causality%
.}. The zeros on the upper branch all lie above the real axis and do
not predict the mode to be unstable.

\begin{figure}
\begin{centering}
\includegraphics[scale=0.5]{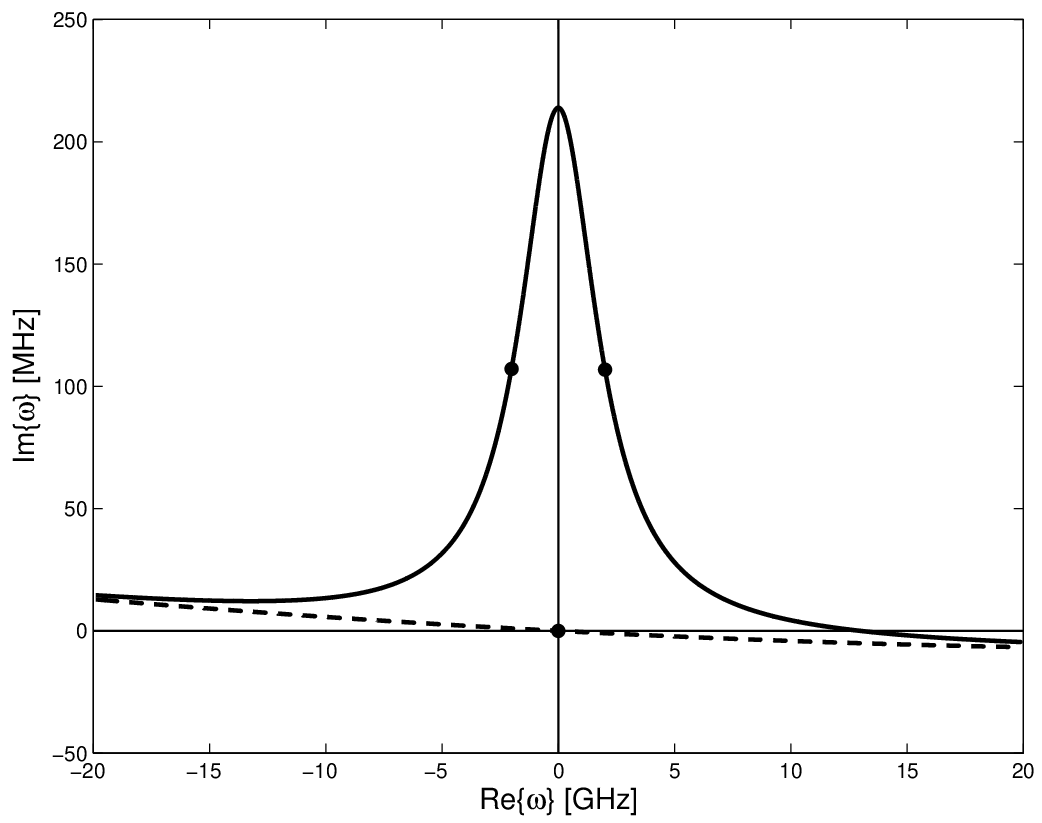}
\par\end{centering}

\caption{The solid curve shows where $|H(\omega)|=|1-G_{0}(\omega)+H(\omega)|$
for complex $\omega$ in the region $|\omega|\ll1/\tau_{L}$. The
bullets on the curve show the position of the solutions to $G_{0}(\omega)=1$.
The dashed curve is where $|H(\omega)|=1$ and the bullet on the curve
is where $H(\omega)=1$, i.e. at $\omega=0$. }

\label{Complex zeros G0-2a}
\end{figure}

\begin{figure}
\begin{centering}
\includegraphics[scale=0.33]{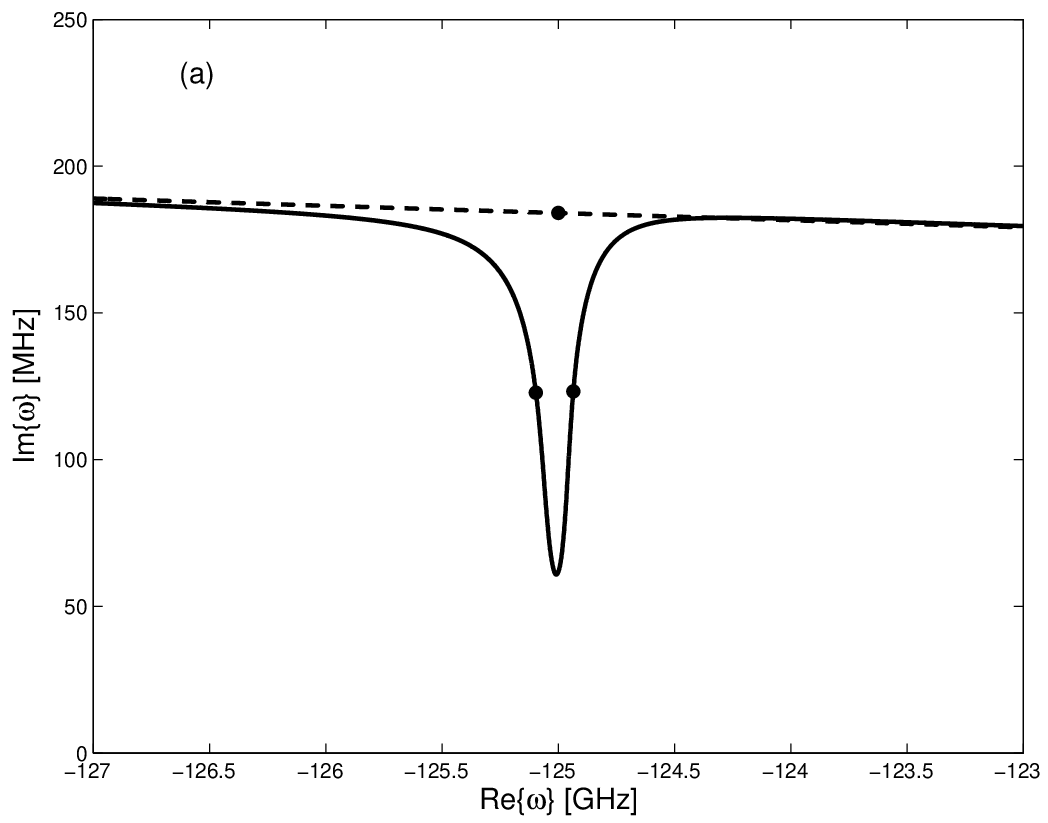}\includegraphics[scale=0.33]{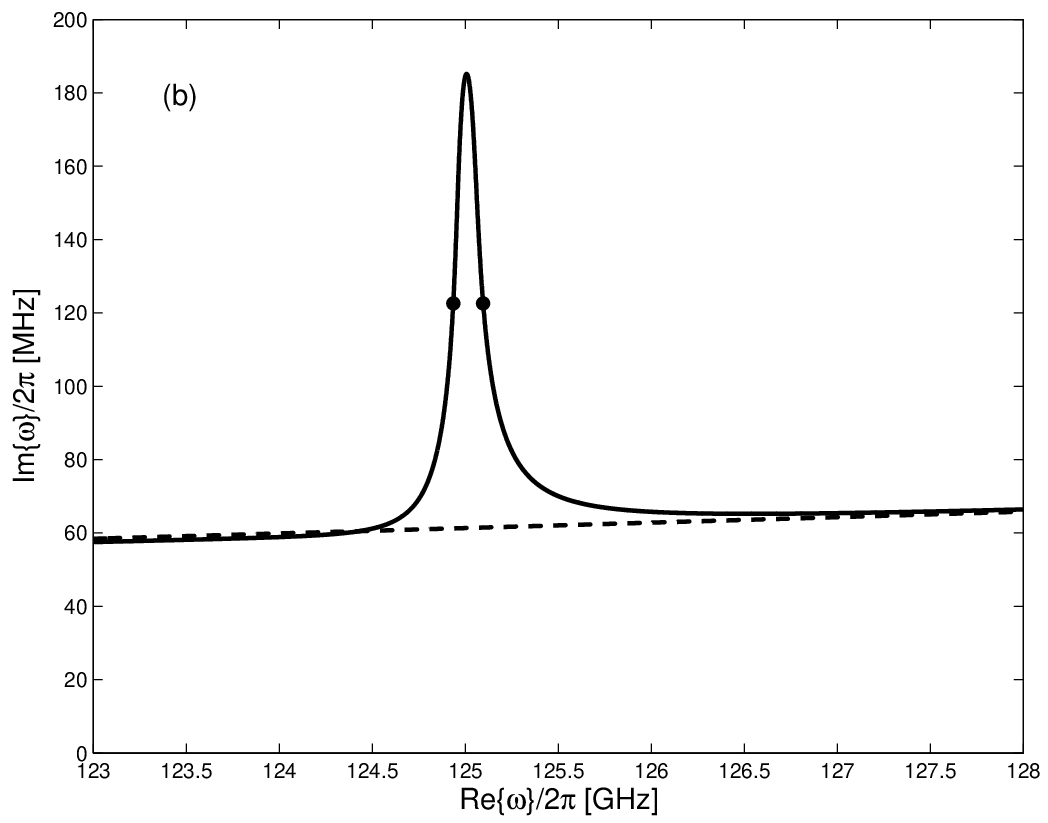}

\par\end{centering}

\caption{The solid curves show where $|H(\omega)|=|1-G_{0}(\omega)+H(\omega)|$
for complex $\omega$ near the solution to $H(\omega)=1$ for $\left.(a\right)$
$\textrm{Re}\{\omega\}\simeq-2\pi/\tau_{L}$ and for $\left.(b\right)$
$\textrm{Re}\{\omega\}\simeq2\pi/\tau_{L}$. The bullets on the curve
show the position of the solutions to $G_{0}(\omega)=1$. The dashed
curves are where $|H(\omega)|=1$ and the bullet on the curves is
where $H(\omega)=1$. }

\label{Complex zeros G0-2b}
\end{figure}

When carrier fluctuations are included the solutions to $G_{0}(\omega)=1$
are the solutions to
\begin{equation}
H(\omega)=1-G_{0}(\omega)+H(\omega)=1+C(\omega)+C_{1}(\omega)(1-H(\omega))/(1-H_{1}(\omega))\,.\label{Trans-w-C}
\end{equation}
The solutions must lie on the curve given by $|H(\omega)|=|1-G_{0}(\omega)+H(\omega)|$.
Since $C(\omega)$ and $C_{1}(\omega)$ decay as $1/|\omega|$ for
$|\omega|\gg\Gamma_{N}$ the curve will be close to the parabola $|H(\omega)|=1$
except near solutions to $H(\omega)=1$. The effect of carrier fluctuations
on solutions to $G_{0}(\omega)=1$ compared to solutions to $H(\omega)=1$
is illustrated in Figure \ref{Complex zeros G0-2a} and Figures \ref{Complex zeros G0-2b}(a)
and (b) for the regions around $\omega=0$ and the two adjacent sidemodes
at $Re\{\omega\}\simeq\mp2\pi/\tau_{L}$. The solid curves show the
solution to $|H(\omega)|=|1-G_{0}(\omega)+H(\omega)|$ and the bullets
on the curves show the location of the solutions to $G_{0}=1$. The
dashed curves show the solution to $|H(\omega)|=1$ and the bullet
on each curve marks the solution to $H(\omega)=1$.

The bullets on the solid curve in Figure \ref{Complex zeros G0-2a}
show the two solutions to (\ref{Trans-w-C}) that give rise to the
relaxation peaks in the spectrum in Figure \ref{MultiMode-spectrum-a}.
We can get an analytic estimate of the solutions to (\ref{Trans-w-C})
by considering the approximation where $\tau_{1}=\tau_{2}=0$. In
that case $H=H_{1}$ and (\ref{Trans-w-C}) becomes
\begin{equation}
H(\omega)=e^{-j\omega\tau_{L}}=1+\frac{\Omega_{R}^{2}\tau_{L}}{j\omega+\Gamma_{N}}\,.\label{Trans-H-simp}
\end{equation}
If we use the expansion $H(\omega)\simeq1-j\omega\tau_{L}$, then
(\ref{Trans-H-simp}) becomes $-j\omega=\Omega_{R}^{2}/(j\omega+\Gamma_{N})$
or $\omega^{2}-j\Gamma\omega-\Omega_{R}^{2}\simeq0$, which has the
standard solutions $\omega\simeq j\tfrac{1}{2}\Gamma_{N}\pm\Omega$.
However, this is a poor estimate of the solutions to (\ref{Trans-H-simp}).
In the numerical example in Figure \ref{Complex zeros G0-2a} the
imaginary part of the solution is larger than $\Gamma_{N}$. For $\omega=x+jy$
we find
\begin{equation}
|H(\omega)|^{2}=e^{2\tau_{L}y}=1+\frac{2\tau_{L}\Omega_{R}^{2}[\tfrac{1}{2}\Omega_{R}^{2}\tau_{L}+\Gamma_{N}-y]}{(\Gamma_{N}-y)^{2}+x^{2}}\,.\label{NormH-simp}
\end{equation}
If we now use the expansion $\exp(2\tau_{L}y)\simeq1+2\tau_{L}y$
and $x=\Omega_{R}$ we get the estimate
\begin{equation}
y\simeq\frac{1}{4}\Omega_{R}^{2}\tau_{L}+\tfrac{1}{2}\Gamma_{N}\,.\label{Est-of-loss}
\end{equation}
For our example this estimate is within $0.3\%$ of the numerical
value. It shows that using the expansion $e^{-j\omega\tau_{L}}\simeq1-j\omega\tau_{L}$
in (\ref{Trans-H-simp}), and in the setup of the field rate equation
(\ref{Field rate eq}), can be problematic except for short laser
cavities where $\Omega_{R}\tau_{L}\ll1$.

Figures \ref{Complex zeros G0-2b}(a) and (b) show that the effect
of the carrier fluctuations is to push the solutions of $H(\omega)=1$
closer to the real axis for the sidemode on the low frequency side
and further away from the axis for the sidemode on the high frequency
side. This explains why the amplitude of the left sidemode in the
spectrum in Figure \ref{MultiMode-spectrum} is higher than the right
sidemode even though the figures show that the $H(\omega)=1$ solution
to the right is closer to the real axis than that to the left. This
also shows that sidemodes are more suppressed when the mode is tuned
to the low frequency side of the gain peak. If we change sign of $\tau_{1}$,
the mode $(\omega_{0},N_{0})$ will be at a frequency as much above
the gain maximum as it is now below. In that case a calculation shows
that the solution of $G_{0}(\omega)=1$ for the left sidemode is above
but close to the real axis such that the left sidemode has a higher
amplitude than even the central mode at $f=0$. The two zeros of $G_{0}(\omega)=1$
for each sidemode give rise to splitting of the sidemode into two
peaks with frequency spacing around $\Omega/\pi$ or to corresponding
broadening of the sidemode.

The stability of the mode $(\omega_{p},N_{p})$ for $p\neq0$ is determined
by the location of the complex angular frequencies where the loop
gain is one. For constant $N=N_{p}$ and real $\omega$ the loop gain
is
\begin{equation}
G_{p}(\omega)=G(\omega_{0}+\omega,N_{p})=G(\omega_{0}+\omega,N_{0})e^{\tfrac{1}{2}(1+j\alpha)\tau_{L}a(N_{p}-N_{0})}\,.\label{Gi-loop}
\end{equation}
Since $G(\omega_{0}+\omega,N_{0})=H(\omega)$ for real $\omega$,
the continuation of $G_{p}(\omega)$ to complex $\omega$ is
\begin{equation}
G_{p}(\omega)=H(\omega)e^{\tfrac{1}{2}(1+j\alpha)\tau_{L}a(N_{p}-N_{0})}\,.
\end{equation}
From $|G_{p}(\omega)|=1$ and $\omega=x+jy$ we get the equation
\begin{equation}
\left(y+\frac{\tau_{L}}{2\tau_{2}^{2}}\right)^{2}-\left(x-\frac{\tau_{1}}{2\tau_{2}^{2}}\right)^{2}=\frac{\tau_{L}^{2}-\tau_{1}^{2}}{(2\tau_{2}^{2})^{2}}-\frac{\tau_{L}}{2\tau_{2}^{2}}a(N_{p}-N_{0})\label{Hyperb-for-Np}
\end{equation}
which is a hyperbola with the same asymptotes as (\ref{Hyperb}).
As long as the r.h.s. is positive the minimum of the upper branch
is at $x=\tau_{1}/(2\tau_{2}^{2})$ as in Figure \ref{Complex zeros G0}
but it is shifted vertically downwards to make the hyperbola cut the
real axis at $x=\omega_{p}-\omega_{0}$. This means that the complex
solutions to $G_{p}(\omega)=1$ closer to the minimum will have negative
imaginary part indicating that the mode $(\omega_{p},N_{p})$ is unstable.
However, we have to take into account that carrier fluctuations in
general may have a stabilizing effect that requires a closer analysis
if the shift only brings the solutions slightly below the real axis.

\selectlanguage{american}%

\section{Langevin noise functions\label{sec:Langevin-noise-functions}}

The functions $F(t)$ and $F_{P}(t)$ in (\ref{Field rate eq}) and
(\ref{eq:P rate 4 new}) are introduced to take into account the fluctuating
effect of spontaneous emission of photons into the waveguide mode.
They are often called Langevin noise functions or Langevin noise sources
after the French physicist Paul Langevin (1872-1946), and the stochastic
differential equations (\ref{Field rate eq}) and (\ref{eq:P rate 4 new})
are correspondingly called Langevin equations.

We use a classical model where the $i$-th spontaneous emission event
occurring at time $t_{i}$ adds a term $e^{j\theta_{i}}$ of unit
modulus and arbitrary phase $\theta_{i}$ to the complex electric
field envelope $E(t)$. This is accomplished if we choose $F(t)$
to be given by
\begin{equation}
F(t)=\sum_{i}e^{j\theta_{i}}\delta(t-t_{i})\label{Langev-1}
\end{equation}
in analogy with the discussion of shot noise in section 6.5 of [1]. 
The contribution to $E(t)$ in the
time interval $I=[t_{1},t_{2}]$ is then
\[
\int_{t_{1}}^{t_{2}}F(t')dt'=\sum_{t_{i}\in I}e^{j\theta_{i}}\,.
\]
For mathematical convenience we replace $F(t)$ in (\ref{Langev-1})
by its time-average over the interval $I_{t}=[t-\Delta t,t]$ of width
$\Delta t$,
\begin{equation}
F(t)=\frac{1}{\Delta t}\sum_{t_{i}\in I_{t}}e^{j\theta_{i}}\label{TimeavF(t)}
\end{equation}
where $\Delta t$ is small compared to the time constants in the rate
equations (\ref{Field rate eq}) and (\ref{N rate 2}). The ensemble
average of $e^{j\theta_{i}}=\cos\theta_{i}+j\sin\theta_{i}$ is zero
because the phases $\theta_{i}$ are random and therefore
\begin{equation}
\langle F(t)\rangle=\frac{1}{\Delta t}\left\langle \sum_{t_{i}\in I_{t}}e^{j\theta_{i}}\right\rangle =0\,.\label{AvF(t)}
\end{equation}
The correlation functions $\langle F(t)F^{*}(t')\rangle$, $\langle F(t)F(t')\rangle$
and $\langle F^{*}(t)F^{*}(t')\rangle$ can be calculated by the same
method as used in section 6.5.1 of [1].
Thus by insertion
of (\ref{TimeavF(t)})
\begin{equation}
\langle F(t)F^{*}(t')\rangle=\frac{1}{{\Delta t}^{2}}\left\langle \sum_{t_{i},t_{k}}e^{j(\theta_{i}-\theta_{k})}\right\rangle \label{CorFF*-1}
\end{equation}
where $t_{i}\in I_{t}$ and $t_{k}\in I_{t'}$. The ensemble average
of the terms where $t_{i}\neq t_{k}$ is zero because the phases are
random, but from the terms where $t_{i}=t_{k}$ we get the contribution
\begin{equation}
\langle F(t)F^{*}(t')\rangle=\frac{1}{{\Delta t}^{2}}\left\langle \sum_{t_{i}\in I_{t}\cap I_{t'}}1\right\rangle \label{CorFF*-2}
\end{equation}
where $I_{t}\cap I_{t'}$ is the intersection of the intervals $I_{t}$
and $I_{t'}$. The rate of spontaneous emission is $R_{sp}$ so
\begin{equation}
\left\langle \sum_{t_{i}\in I_{t}\cap I_{t'}}1\right\rangle =R_{sp}|I_{t}\cap I_{t'}|\label{CorFF*-3}
\end{equation}
where $|I_{t}\cap I_{t'}|$ denotes the length of the interval $I_{t}\cap I_{t'}$.
Since $|I_{t}\cap I_{t'}|=\Delta t\Lambda((t'-t)/\Delta t)$, where
$\Lambda(t/\Delta t)$ is the triangular function (1.10)-[1],
the correlation relation (\ref{CorFF*-2}) becomes
\begin{equation}
\langle F(t)F^{*}(t')\rangle=R_{sp}\frac{1}{\Delta t}\Lambda((t'-t)/\Delta t)\label{CorFF*-4}
\end{equation}
and by (6.97)-[1] we get
\begin{equation}
\langle F(t)F^{*}(t')\rangle=R_{sp}\delta(t-t')\label{CorFF*-delta}
\end{equation}
in the limit $\Delta t\rightarrow0$.

When we use the same procedure to calculate $\langle F(t)F(t')\rangle$
and $\langle F^{*}(t)F^{*}(t')\rangle$ we find that instead of the
sum of terms $e^{j(\theta_{i}-\theta_{k})}$ we get sums of terms
$e^{j(\theta_{i}+\theta_{k})}$ and $e^{-j(\theta_{i}+\theta_{k})}$
which have random phases even for $t_{i}=t_{k}$. Thus
\begin{equation}
\langle F(t)F(t')\rangle=\langle F^{*}(t)F^{*}(t')\rangle=0\,.\label{CorFF}
\end{equation}

From the model of $F(t)$ we can determine the correlation properties
of $F_{P}(t)$ defined by (\ref{Langevin-Fp}). First we determine
the average $\langle E(t)F^{*}(t)\rangle$ in the limit where (\ref{CorFF*-delta})
applies. By integrating the field equation (\ref{Field rate eq})
we get
\begin{equation}
E(t)=\frac{1}{2}(1+j\alpha)a\int_{-\infty}^{t}(N(t')-N_{0})E(t')dt'+\int_{-\infty}^{t}F(t')dt'\,.
\end{equation}
The cross-correlation $\langle(N(t')-N_{0})E(t')F^{*}(t)\rangle$
is zero for $t'<t$ because $F^{*}(t)$ is uncorrelated to field and
carrier number at earlier times and it is finite at $t'=t$. Therefore
\begin{multline}
\langle E(t)F^{*}(t)\rangle=\frac{1}{2}(1+j\alpha)a\int_{-\infty}^{t}\langle(N(t')-N_{0})E(t')F^{*}(t)\rangle dt'\\
+\int_{-\infty}^{t}\langle F(t')F^{*}(t)\rangle dt'=R_{sp}\int_{-\infty}^{t}\delta(t-t')dt'=\tfrac{1}{2}R_{sp}\,.\label{Av E(t)F*(t)}
\end{multline}
The same argument shows that $\langle E(t)F(t)\rangle=0$. Since $R_{sp}$
is real
\begin{equation}
\langle E(t)F^{*}(t)+E^{*}(t)F(t)\rangle=2Re\{\langle E(t)F^{*}(t)\rangle\}=R_{sp}\label{AvEF*-1}
\end{equation}
so the definition (\ref{Langevin-Fp}) gives the average
\begin{equation}
\langle F_{P}(t)\rangle=0\label{Av Fp}
\end{equation}
as claimed in the previous subsection. Furthermore, by insertion of
(\ref{Langevin-Fp}) the autocorrelation of $F_{P}(t)$ is
\begin{equation}
\langle F_{P}(t)F_{P}(t')\rangle=\langle\{E(t)F^{*}(t)+E^{*}(t)F(t)\}\{E(t')F^{*}(t')+E^{*}(t')F(t')\}\rangle\\
-R_{sp}^{2}\,.\label{Auto Fp}
\end{equation}
From the average on the r.h.s. we get terms that are the product of
four functions such as f.ex. $\langle E(t)F^{*}(t)E^{*}(t')F(t')\rangle$.
We adopt the factorization approximation
\begin{multline}
\langle E(t)F^{*}(t)E^{*}(t')F(t')\rangle\simeq\langle E(t)F^{*}(t)\rangle\langle E^{*}(t')F(t')\rangle+\langle E(t)E^{*}(t')\rangle\langle F^{*}(t)F(t')\rangle\\
+\langle E(t)F(t')\rangle\langle F^{*}(t)E^{*}(t')\rangle
=\frac{1}{4}R_{sp}^{2}+\langle{\cal {P}}\rangle R_{sp}\delta(t-t')\label{Av EF*E*F}
\end{multline}
where we have inserted (\ref{Av E(t)F*(t)}) and (\ref{CorFF*-delta})
and used $\langle E(t)E^{*}(t)\rangle=\langle{\cal {P}}\rangle$ according
to (\ref{NormE}). Since $\langle E(t)F(t')\rangle=0$ for $t'\geq t$
the product 
 $\langle E(t)F(t')\rangle\langle F^{*}(t)E^{*}(t')\rangle$ is zero
for all $t$ and $t'$. Similarly
\begin{multline}
\langle E(t)F^{*}(t)E(t')F^{*}(t')\rangle=\langle E(t)F^{*}(t)\rangle\langle E(t')F^{*}(t')\rangle+\langle E(t)E(t')\rangle\langle F^{*}(t)F^{*}(t')\rangle\\
+\langle E(t)F^{*}(t')\rangle\langle F^{*}(t)E(t')\rangle=\frac{1}{4}R_{sp}^{2}+R_{sp}^{2}u(t-t')u(t'-t)\label{Av EF*EF*}
\end{multline}
using (\ref{Av E(t)F*(t)}) and (\ref{CorFF}). Since $\langle E(t)F^{*}(t')\rangle=0$
for $t'>t$ and $\langle F^{*}(t)E(t')\rangle=0$ for $t'<t$ the
product $\langle E(t)F^{*}(t')\rangle\langle F^{*}(t)E(t')\rangle$
is zero except for $t=t'$ where it is $\tfrac{1}{4}R_{sp}^{2}$.
The product can therefore be written as $R_{sp}^{2}u(t-t')u(t'-t)$
in terms of the step function $u(t)$. The average in (\ref{Auto Fp})
is the sum of (\ref{Av EF*E*F}) and (\ref{Av EF*EF*}) and their
complex conjugates, i.e.
\begin{equation}
\langle F_{P}(t)F_{P}(t')\rangle=2\langle{\cal {P}}\rangle R_{sp}\delta(t-t')\label{CorFpFp}
\end{equation}
where the term with step functions is ignored compared to the delta
function.

The carrier number is fluctuating due to current injection, non-radiative
recombination, spontaneous and stimulated emission and stimulated
absorption. In order to take into account the fluctuations we add
a Langevin noise function $F_{N}(t)$ to the rate equation in (\ref{N rate 2})
and thus obtain the corrected rate equation
\begin{equation}
\frac{d}{dt}N=J-\frac{N}{\tau_{e}}-a(N-N_{tr}){\cal {P}}+F_{N}(t)\,.\label{N rate FN}
\end{equation}
For convenience we have introduced the carrier injection rate $J=\eta I/q$.
Adding (\ref{eq:P rate 4 new}) and (\ref{N rate FN}) we get
\begin{equation}
\frac{d}{dt}(N+{\cal {P}})=J-\frac{N}{\tau_{e}}-\frac{{\cal {P}}}{\tau_{p}}+R_{sp}+F_{+}(t)\,.\label{N+P rate}
\end{equation}
where $F_{+}(t)=F_{N}(t)+F_{P}(t)$. Injection of carriers, non-radiative
recombination and photon decay due to material absorption and mirror
loss can be assumed to be independent Poisson processes.
Let us first consider the contribution
to $F_{+}(t)$ from current injection. In analogy with (\ref{TimeavF(t)})
we use the classical model
\begin{equation}
F_{+}(t)=\frac{k}{\Delta t}-\langle J\rangle\label{Def F+}
\end{equation}
where $k$ is the number of electrons injected in the time interval
$I_{t}=[t-\Delta t,t]$. It is a stochastic variable that satisfies
the Poisson probability distribution  with average
$\langle k\rangle=\langle J\rangle\Delta t$ and variance $\langle k^{2}\rangle=\langle k\rangle^{2}+\langle k\rangle$.

In order to calculate the correlation $\langle F_{+}(t)F_{+}(t')\rangle$
we assume $t'\geq t$ and introduce the set difference $I_{t}\setminus I_{t'}$
as the section of $I_{t}$ that is not in $I_{t'}$. Similarly $I_{t'}\setminus I_{t}$
is the section of $I_{t'}$ that is not in $I_{t}$. The correlation
can then be written as
\begin{flalign}
\langle F_{+}(t)F_{+}(t')\rangle & =\left\langle \left(\left(k_{1}+k_{2}\right)/\Delta t-\langle J\rangle\right)\left(\left(k_{2}+k_{3}\right)/\Delta t-\langle J\rangle\right)\right\rangle \nonumber \\
 & =\frac{1}{\Delta t^{2}}\langle(k_{1}+k_{2})(k_{2}+k_{3})\rangle-\langle J\rangle^{2}\label{CorF+F+ }
\end{flalign}
where $k_{1}$, $k_{2}$ and $k_{3}$ are the number of electrons
injected in the three intervals $I_{t}\setminus I_{t'}$, $I_{t}\cap I_{t'}$
and $I_{t'}\setminus I_{t}$. The three numbers are uncorrelated which
means for example $\left\langle k_{1}k_{2}\right\rangle =\left\langle k_{1}\right\rangle \left\langle k_{2}\right\rangle $
and they have averages $\langle k_{1}\rangle=\langle J\rangle|I_{t}\setminus I_{t'}|=\langle J\rangle|I_{t'}\setminus I_{t}|=\langle k_{3}\rangle$
and $\langle k_{2}\rangle=\langle J\rangle|I_{t}\cap I_{t'}|$. Therefore
\begin{equation}
\langle(k_{1}+k_{2})(k_{2}+k_{3})\rangle=(\langle k_{1}\rangle+\langle k_{2}\rangle)^{2}+\langle k_{2}\rangle=(\langle J\rangle\Delta t)^{2}+\langle J\rangle|I_{t}\cap I_{t'}|
\end{equation}
where we have used that $\langle k_{2}^{2}\rangle=\langle k_{2}\rangle^{2}+\langle k_{2}\rangle$
and $|I_{t}\setminus I_{t'}|+|I_{t}\cap I_{t'}|=\Delta t$. Inserted
in (\ref{CorF+F+ }) the correlation becomes
\begin{equation}
\langle F_{+}(t)F_{+}(t')\rangle=\tfrac{1}{\Delta t^{2}}\langle J\rangle|I_{t}\cap I_{t'}|=\langle J\rangle\tfrac{1}{\Delta t}\Lambda(\tfrac{t'-t}{\Delta t})\label{CorF+F+-J}
\end{equation}
similarly to (\ref{CorFF*-4}). The result is independent of the order
of $t$ and $t'$ and it therefore also applies for $t'\leq t$. It
gives the correlation
\begin{equation}
\langle F_{+}(t)F_{+}(t')\rangle=\langle J\rangle\delta(t'-t)\label{CorF+F+-delta}
\end{equation}
in the limit $\Delta t\rightarrow0$. 
We have assumed that the injection of electrons is Poisson distributed,
but that actually depends on the circuitry that provides the current
to the laser. This is sometimes taken into account by replacing $\langle J\rangle$
by $\xi\langle J\rangle$ where $\xi$ is a parameter in the interval
$[0,1]$.

The terms $N/\tau_{s}$ and $\mathcal{P}/\tau_{p}$ in (\ref{N+P rate})
are rates of single events that are mutually uncorrelated and uncorrelated
with the electron injection rate $J$. The full autocorrelation relation
for $F_{+}(t)$ is therefore
\begin{equation}
\langle F_{+}(t)F_{+}(t')\rangle=\left(\langle J\rangle+\frac{\langle N\rangle}{\tau_{e}}+\frac{\langle{\cal {P}}\rangle}{\tau_{p}}\right)\delta(t'-t)\,.\label{Cor F+F+}
\end{equation}
Notice, it is the averages of the rates $J$, $N/\tau_{s}$ and $\mathcal{P}/\tau_{p}$
that contribute to the correlation irrespective of the sign with which
the rates appear in (\ref{N+P rate}).

The photon decay is active in both (\ref{eq:P rate 4 new}) and (\ref{N+P rate})
and implies the cross-correlation
\begin{equation}
\langle F_{+}(t)F_{P}(t')\rangle=\frac{\langle{\cal {P}}\rangle}{\tau_{p}}\delta(t'-t)\label{CC-F+Fp}
\end{equation}
and hence
\begin{multline}
\langle F_{N}(t)F_{N}(t')\rangle=\langle F_{+}(t)F_{+}(t')\rangle-2\langle F_{+}(t)F_{P}(t')\rangle+\langle F_{P}(t)F_{P}(t')\rangle\\
=\left(\langle J\rangle+\frac{\langle N\rangle}{\tau_{e}}-\frac{\langle{\cal {P}}\rangle}{\tau_{p}}+2\langle{\cal {P}}\rangle R_{sp}\right)\delta(t'-t)=2\left(\frac{\langle N\rangle}{\tau_{e}}+\langle{\cal {P}}\rangle R_{sp}\right)\delta(t'-t)\,.\label{CorFNFN}
\end{multline}
We have here inserted the average of (\ref{N+P rate})
\begin{equation}
\langle J\rangle=\frac{\langle N\rangle}{\tau_{e}}+\frac{\langle{\cal {P}}\rangle}{\tau_{p}}-R_{sp}\label{Av
current}
\end{equation}
and assumed $\langle{\cal {P}}\rangle\gg1$. We finally also have
the cross-correlation
\begin{equation}
\langle F_{P}(t)F_{N}(t')\rangle=\langle F_{P}(t)F_{+}(t')\rangle-\langle F_{P}(t)F_{P}(t')\rangle=\langle{\cal {P}}\rangle\left(\frac{1}{\tau_{p}}-2R_{sp}\right)\delta(t'-t)\,.\label{CC-FpFN}
\end{equation}
The correlation relations (\ref{CorFpFp}), (\ref{CorFNFN}) and (\ref{CC-FpFN})
are with minor modifications equal to the corresponding relations
in \cite{Agrawal-1993} and \cite{Coldren-1995}.

The rate equations (\ref{N rate FN}) and (\ref{eq:P rate 4 new})
allow calculation of the carrier and photon numbers as a function
of time. From the field equation (\ref{Field rate eq}) we can also
get an equation for the phase $\phi(t)$ of the electric field envelope
$E(t)=|E(t)|e^{j\phi(t)}$. Multiplying (\ref{Field rate eq}) by
$E^{*}(t)$ and subtracting the complex conjugate gives
\begin{equation} 
E^{*}(t)\frac{d}{dt}E(t)-E(t)\frac{d}{dt}E^{*}(t)=2j|E(t)|^{2}\frac{d}{dt}\phi(t)
=j\alpha a(N-N_{0})|E(t)|^{2}+E^{*}(t)F(t)-E(t)F^{*}(t)
\end{equation}
and hence
\begin{equation}
\frac{d}{dt}\phi(t)=\frac{1}{2}\alpha a(N-N_{0})+F_{\phi}(t)\label{Phase rate}
\end{equation}
where
\begin{equation}
F_{\phi}(t)=\frac{1}{2j{\cal {P}}(t)}(E^{*}(t)F(t)-E(t)F^{*}(t))\,.\label{Fphi def.}
\end{equation}
$F_{\phi}(t)$ is a Langevin noise function that is the source of
part of the laser phase noise. By using (\ref{Av EF*E*F}), (\ref{Av EF*EF*})
and $\langle E(t)F^{*}(t)\rangle=\tfrac{1}{2}R_{sp}$ we derive the
correlation relations
\begin{eqnarray}
\langle F_{\phi}(t)\rangle & = & 0\label{Av-Fphi}\\
\langle F_{\phi}(t)F_{\phi}(t')\rangle & = & \frac{R_{sp}}{2\langle{\cal {P}}\rangle}\delta(t'-t)\label{Cor-Fphi}\\
\langle F_{\phi}(t)F_{P}(t')\rangle & = & 0\,.\label{CC-FphiFp}
\end{eqnarray}
The phase does not appear in the photon and carrier number rate equations
and therefore phase fluctuations cannot influence the photon and carrier
fluctuations. Hence also
\begin{equation}
\langle F_{\phi}(t)F_{N}(t')\rangle=0\,.\label{CC-FphiFN}
\end{equation}
This completes the list of auto- and cross-correlations of the noise
functions. For notational convenience we write the relations as
\begin{equation}
\langle F_{i}(t)F_{j}(t')\rangle=2D_{ij}\delta(t-t')\,.\label{diff-const}
\end{equation}
for $i,j=P,N,\phi$. The coefficients $D_{ij}$ are named diffusion
coefficients in \cite{Agrawal-1993} and $2D_{ij}$ are named correlation
strengths in \cite{Coldren-1995}. In summary

\begin{equation}
D_{PP}=R_{sp}\langle{\cal {P}}\rangle\ \ \ D_{\phi\phi}=R_{sp}/4\langle{\cal {P}}\rangle\ \ \ D_{P\phi}=0\label{diff-const-1}
\end{equation}
\begin{equation}
D_{NN}=R_{sp}\langle{\cal {P}}\rangle+\langle N\rangle/\tau_{e}\ \ \ D_{PN}=\langle{\cal {P}}\rangle(1/2\tau_{p}-R_{sp})\ \ \ \ D_{N\phi}=0\,.\label{diff-const-2}
\end{equation}
The cross power spectral densities of the noise functions are simply
\begin{equation}
S_{F_{i}F_{j}}(f)=\int_{-\infty}^{\infty}\langle F_{i}^{*}(t)F_{j}(t+\tau)\rangle e^{-j\omega\tau}d\tau=2D_{ij}\,.\label{cros-spec of F}
\end{equation}
According to the Wiener-Khinchine theorem,
the cross power spectral density is also given by
\begin{equation}
S_{F_{i}F_{j}}(f)=\lim_{T\rightarrow\infty}\frac{1}{T}\langle\tilde{F}_{i,T}^{*}(f)\tilde{F}_{j,T}(f)\rangle\label{cros-spec of F-1}
\end{equation}
where $\tilde{F}_{i,T}(f)$ is the Fourier transform of the truncated
noise function \\
 \mbox{$F_{i}(t)\Pi(t/T)$} for $i=P$, $N$ and $\phi$. $\Pi(t/T)$
is the rectangular function defined in (1.6)-[1]. One can
easily verify that the r.h.s.~of (\ref{cros-spec of F-1}) is $2D_{ij}$
in agreement with (\ref{cros-spec of F}).

With these relations we can now calculate a number of laser noise
spectra.

\subsection{Perturbation expansion of the laser rate equations}

The Langevin noise functions can be considered as small perturbations
to the solution of the laser rate equation without the noise functions.
Their influence on the solution can be calculated by introducing a
parameter $\lambda$ that scales the noise functions in the rate equations,
i.e. we get the equations
\begin{align}
\frac{d}{dt}{\cal {P}}(t) & =a(N-N_{0}){\cal {P}}(t)+R_{sp}+\lambda F_{P}(t)\label{P rate 00}\\
\frac{d}{dt}N\left(t\right) & =J-\frac{N}{\tau_{e}}-a(N-N_{tr}){\cal {P}}+\lambda F_{N}(t)\,.\label{N rate 00}
\end{align}
The solutions will then depend on $\lambda$ and we may assume that
they can be Taylor expanded as
\begin{align}
{\cal {P}} & =P_{s}+\lambda P_{1}+\lambda^{2}P_{2}+\cdots\label{P expansion}\\
N & =N_{s}+\lambda N_{1}+\lambda^{2}N_{2}+\cdots\,.\label{N expansion}
\end{align}
Inserting the expansions in (\ref{P rate 00}) and (\ref{N rate 00})
the equations become polynomials in $\lambda$ on both sides of the
equality signs. Each equation implies that terms of the same order
of $\lambda$ on the two sides of the equation are equal. To zero
order we reproduce the equations
\begin{equation}
a(N_{s}-N_{0})P_{s}+R_{sp}=0
\end{equation}
\begin{equation}
J_{s}=\frac{N_{s}}{\tau_{e}}+a(N_{s}-N_{tr})P_{s}
\end{equation}
for the steady state solution (\ref{P vs N}) and (\ref{I vs N})
for $J_{s}=\eta I/q$. To 1st order we get
\begin{equation}
\frac{d}{dt}\left[\begin{array}{c}
P_{1}\\
N_{1}
\end{array}\right]=\boldsymbol{M}\left[\begin{array}{c}
P_{1}\\
N_{1}
\end{array}\right]+\left[\begin{array}{c}
F_{P}\\
F_{N}
\end{array}\right]\label{Rate eq P1&N1}
\end{equation}
where $\boldsymbol{M}$ is the matrix in (\ref{M matrix}) and where
we have used $aN_{0}=aN_{tr}+\frac{1}{\tau_{p}}$ according to (\ref{Clamped N0}).
To 2nd order 
\begin{equation}
\frac{d}{dt}\left[\begin{array}{c}
P_{2}\\
N_{2}
\end{array}\right]=\boldsymbol{M}\left[\begin{array}{c}
P_{2}\\
N_{2}
\end{array}\right]+aN_{1}P_{1}\left[\begin{array}{c}
1\\
-1
\end{array}\right]\,.\label{second order rate}
\end{equation}
Since $\langle F_{P}\rangle=\langle F_{N}\rangle=0$ the average of
(\ref{Rate eq P1&N1}) gives
\begin{equation}
\boldsymbol{M}\left[\begin{array}{c}
\langle P_{1}\rangle\\
\langle N_{1}\rangle
\end{array}\right]=\left[\begin{array}{c}
0\\
0
\end{array}\right]
\end{equation}
and therefore $\langle P_{1}\rangle=\langle N_{1}\rangle=0$. From
the average of (\ref{second order rate}) we find
\begin{equation}
\left[\begin{array}{c}
\langle P_{2}\rangle\\
\langle N_{2}\rangle
\end{array}\right]=-a\langle N_{1}P_{1}\rangle\boldsymbol{M}^{-1}\left[\begin{array}{c}
1\\
-1
\end{array}\right]=-a\frac{\langle N_{1}P_{1}\rangle}{\Omega_{R}^{2}}\left[\begin{array}{c}
-\tau_{e}^{-1}\\
\tau_{p}^{-1}
\end{array}\right]\,.\label{<P2&N2>}
\end{equation}

If the noise functions $F_{P}(t)$ and $F_{N}(t)$ are replaced by
truncated functions $F_{P,T}(t)=F_{P}(t)\Pi(t/T)$ and $F_{N,T}(t)=F_{N}(t)\Pi(t/T)$
we can solve (\ref{Rate eq P1&N1}) by Fourier transformation of the
equation. This gives
\begin{equation}
\left[\begin{array}{cc}
j\omega+\frac{R_{sp}}{P_{s}} & -aP_{s}\\
\frac{1}{\tau_{p}}-\frac{R_{sp}}{P_{s}} & j\omega+\Gamma_{N}
\end{array}\right]\left[\begin{array}{c}
\tilde{P}_{1}\\
\tilde{N}_{1}
\end{array}\right]=\left[\begin{array}{c}
\tilde{F}_{P,T}\\
\tilde{F}_{N,T}
\end{array}\right]
\end{equation}
and hence
\begin{equation}
\left[\begin{array}{c}
\tilde{P}_{1}\\
\tilde{N}_{1}
\end{array}\right]=\boldsymbol{H}\left[\begin{array}{c}
\tilde{F}_{P,T}\\
\tilde{F}_{N,T}
\end{array}\right]\label{P&N vs FpFn}
\end{equation}
where $\boldsymbol{H}$ is the transfer matrix (\ref{Laser mod transfer}).

From (\ref{P&N vs FpFn}) we can now calculate the power spectral
density of $P_{1}$ and cross power spectral density of $P_{1}$ and
$N_{1}$. For notational convenience we drop the index $"1"$ and
write for example $S_{PN}(f)$ instead of $S_{P_{1}N_{1}}(f)$ when
there is no risk of confusion. 
The power spectral density $S_{ij}(f)$, where $i$ and $j$ are $P$
or $N$, is given by
\begin{align}
S_{ij}(f) & =\lim_{T\rightarrow\infty}\frac{1}{T}\left\langle \left(\sum_{k}H_{ik}\tilde{F}_{k,T}\right)^{*}\left(\sum_{m}H_{jm}\tilde{F}_{m,T}\right)\right\rangle =2\sum_{k,m}H_{ik}^{*}D_{km}H_{jm}\,.\label{Sij(f)}
\end{align}
The summations are over $P$ and $N$. We have here used that
\begin{equation}
2D_{ij}=\lim_{T\rightarrow\infty}\frac{1}{T}\left\langle \tilde{F}_{i,T}^{*}\tilde{F}_{j,T}\right\rangle \label{Dij FiFj spectrum}
\end{equation}
according to (\ref{cros-spec of F}) and (\ref{cros-spec of F-1}).
The relation (\ref{Sij(f)}) can be written in the compact form
\begin{equation}
\boldsymbol{S}=2\boldsymbol{H}^{*}\boldsymbol{D}\boldsymbol{H}^{T}\label{S compact matrix}
\end{equation}
where $\boldsymbol{S}$ is the matrix with elements $S_{ij}$ and
$\boldsymbol{D}$ is the matrix with elements $D_{ij}$. The superscript
$T$ means that the matrix is transposed.

\section{Relative intensity noise (RIN) spectrum }

The photon number ${\cal {P}}(t)$ is an internal variable in the
laser that is not directly measurable. However, in our simple model
it is proportional to the output power which from the right facet
at $z=L$ is
\begin{equation}
P_{out}(t)=\hbar\omega_{0}R_{2}v_{g}\alpha_{m}{\cal {P}}(t)\,.\label{Pout}
\end{equation}
The factor $v_{g}\alpha_{m}$ is the rate of loss of photons through
the laser facets. Together with $v_{g}\alpha_{i}$ it gives the total
loss rate $1/\tau_{p}$ of photons in the laser, see (\ref{Photon lt 2}).
$R_{2}$ is the fraction of loss through the facet at $z=L$ given
by (\ref{R2}). The product $R_{2}v_{g}\alpha_{m}{\cal {P}}(t)$ is
then the number of photons transmitted through the right facet per
second. Multiplied by the photon energy $\hbar\omega_{0}$ we get
the output power (\ref{Pout}).

The noise in the photon number is transferred to the output power
via the relation (\ref{Pout}). However, we have to take into account
that each photon transmitted through the facet means a decrease of
${\cal {P}}(t)$ by one photon. The effect is named partition noise
and can be included by adding a noise term $\hbar\omega_{0}F_{0}(t)$
to (\ref{Pout})
\begin{equation}
P_{out}(t)=\hbar\omega_{0}R_{2}v_{g}\alpha_{m}{\cal {P}}(t)+\hbar\omega_{0}F_{0}(t)\label{Pout+F0}
\end{equation}
where
\begin{equation}
F_{0}(t)=\frac{1}{\Delta t}\left(\sum_{t_{i}\in I_{t}}1\right)-\langle P_{out}\rangle/\hbar\omega_{0}\,.\label{Def F0}
\end{equation}
The summation is over the events where photons are transmitted in
the time interval $I_{t}$. The expression implies $\langle F_{0}(t)\rangle=0$
and since it is analogous to (\ref{Def F+}) it also implies that
$F_{0}(t)$ satisfies the correlation relation
\begin{equation}
\langle F_{0}(t)F_{0}(t')\rangle=\frac{\langle P_{out}\rangle}{\hbar\omega_{0}}\delta(t'-t)
\end{equation}
in the limit $\Delta t\rightarrow0$. Moreover, the function $F_{0}(t)$
is anticorrelated with the noise function $F_{P}(t)$ defined in (\ref{Langevin-Fp})
and (\ref{P rate 3}) since each transmission of a photon coincides
with loss of one internal photon. Hence
\begin{equation}
\langle F_{0}(t)F_{P}(t')\rangle=-\frac{\langle P_{out}\rangle}{\hbar\omega_{0}}\delta(t'-t)\label{<F0FP>}
\end{equation}
in the limit $\Delta t\rightarrow0$. In this description of partition
noise we have followed the semiclassical presentation by Coldren and
Corzine \cite{Coldren-1995}. It can also be considered as beat noise
between a vacuum field and a signal field as was done in a quantum
optics description by Yamamoto and Imoto \cite{Yamamoto-1986}.

For constant input current $J=J_{s}$ we have $\mathcal{P}\left(t\right)=P_{s}+P_{1}(t)$
to 1st order. The noise of the output power is therefore to 1st order
\begin{equation}
\delta P_{out}=P_{out}-\left\langle P_{out}\right\rangle \simeq\hbar\omega_{0}R_{2}v_{g}\alpha_{m}P_{1}+\hbar\omega_{0}F_{0}(t)\label{delta Pout}
\end{equation}
where $\left\langle \mathcal{P}\left(t\right)\right\rangle =P_{s}$,
$\langle F_{0}(t)\rangle=0$ and
\begin{equation}
\left\langle P_{out}\right\rangle \simeq\hbar\omega_{0}R_{2}v_{g}\alpha_{m}P_{s}\,.\label{<Pout>}
\end{equation}
The power spectral density $S_{\delta P}(f)$ of $\delta P_{out}$
is an important measure of the signal quality of the laser. The ratio
$S_{\delta P}(f)/\langle P_{out}\rangle^{2}$ is usually named the
relative intensity noise (RIN) spectrum and is denoted by RIN$(f)$.
From (\ref{delta Pout})and (\ref{<Pout>}) we find
\begin{equation}
\textrm{RIN}(f)=\frac{S_{\delta P}(f)}{\langle P_{out}\rangle^{2}}\simeq\frac{1}{P_{s}^{2}}S_{P_{1}}(f)+\frac{2\hbar\omega_{0}}{P_{s}\langle P_{out}\rangle}\textrm{Re}\left(S_{P_{1}F_{0}}(f)\right)+\frac{(\hbar\omega_{0})^{2}}{\langle P_{out}\rangle^{2}}S_{F_{0}F_{0}}(f)\,.\label{RIN}
\end{equation}
The power spectral density $S_{P_{1}}(f)$ is a short notation for
$S_{PP}(f)$ given in (\ref{Sij(f)}). By inserting the $\boldsymbol{H}$
matrix from (\ref{Laser mod transfer-1}) and the $\boldsymbol{D}$
matrix from (\ref{diff-const-1}) and (\ref{diff-const-2}) the spectrum
becomes
\begin{gather}
S_{PP}(f)=2\left[D_{PP}|H_{PP}|^{2}+4D_{PN}\textrm{Re}\{H_{PP}H_{PN}^{*}\}+2D_{NN}|H_{PN}|^{2}\right]\nonumber \\
=2P_{s}\frac{R_{sp}(\omega^{2}+\tau_{e}^{-2})+\Gamma_{N}\Omega_{R}^{2}+a^{2}N_{s}P_{s}/\tau_{e}}{(\omega^{2}-\Omega_{R}^{2})^{2}+(\Gamma_{N}\omega)^{2}}\,.
\end{gather}
Since
\begin{eqnarray}
S_{F_{P}F_{0}}(f) & = & \int_{-\infty}^{\infty}\left\langle F_{P}(t)F_{0}(t+\tau)\right\rangle e^{-j\omega\tau}d\tau=-\frac{\langle P_{out}\rangle}{\hbar\omega_{0}}\\
S_{F_{0}F_{0}}(f) & = & \int_{-\infty}^{\infty}\left\langle F_{0}(t)F_{0}(t+\tau)\right\rangle e^{-j\omega\tau}d\tau=\frac{\langle P_{out}\rangle}{\hbar\omega_{0}}
\end{eqnarray}
and $S_{F_{N}F_{0}}\left(f\right)=0$ because $N_{1}(t)$ and $F_{0}\left(t\right)$
are uncorrelated we have
\begin{equation}
S_{P_{1}F_{0}}(f)=H_{PP}^{*}S_{F_{P}F_{0}}(f)=\frac{j\omega-\Gamma_{N}}{D^{*}}\frac{\langle P_{out}\rangle}{\hbar\omega_{0}}\,.
\end{equation}
The RIN spectrum is therefore
\begin{equation}
\textrm{RIN}(f)=\frac{2}{P_{s}}\frac{R_{sp}(\omega^{2}+\tau_{e}^{-2})+a^{2}N_{s}P_{s}/\tau_{e}}{(\omega^{2}-\Omega_{R}^{2})^{2}+(\Gamma_{N}\omega)^{2}}+\frac{\hbar\omega_{0}}{\langle P_{out}\rangle}\,.\label{eq:RIN(f)}
\end{equation}
The constant term $\hbar\omega_{0}/\langle P_{out}\rangle$ is a RIN
contribution stemming from shot noise. This can be understood as follows.
To this contribution corresponds a power spectral density $S_{\delta P}=\langle P_{out}\rangle^{2}\cdot\hbar\omega_{0}/\langle P_{out}\rangle=\langle P_{out}\rangle\hbar\omega_{0}$.
If the output power is detected by an ideal photodetector the output
current from the detector will be $I=qP_{out}/(\hbar\omega_{0})$
and the corresponding power spectral density of the current fluctuations
$\Delta I$ is therefore $S_{\triangle I}=\left(q/(\hbar\omega_{0})\right)^{2}S_{\delta P}=\left(q/(\hbar\omega_{0})\right)^{2}\cdot\langle P_{out}\rangle\hbar\omega_{0}=q\left\langle I\right\rangle $, in agreement with (6.100)-[1].

\begin{figure}
\begin{centering}
\centering \includegraphics[scale=0.5]{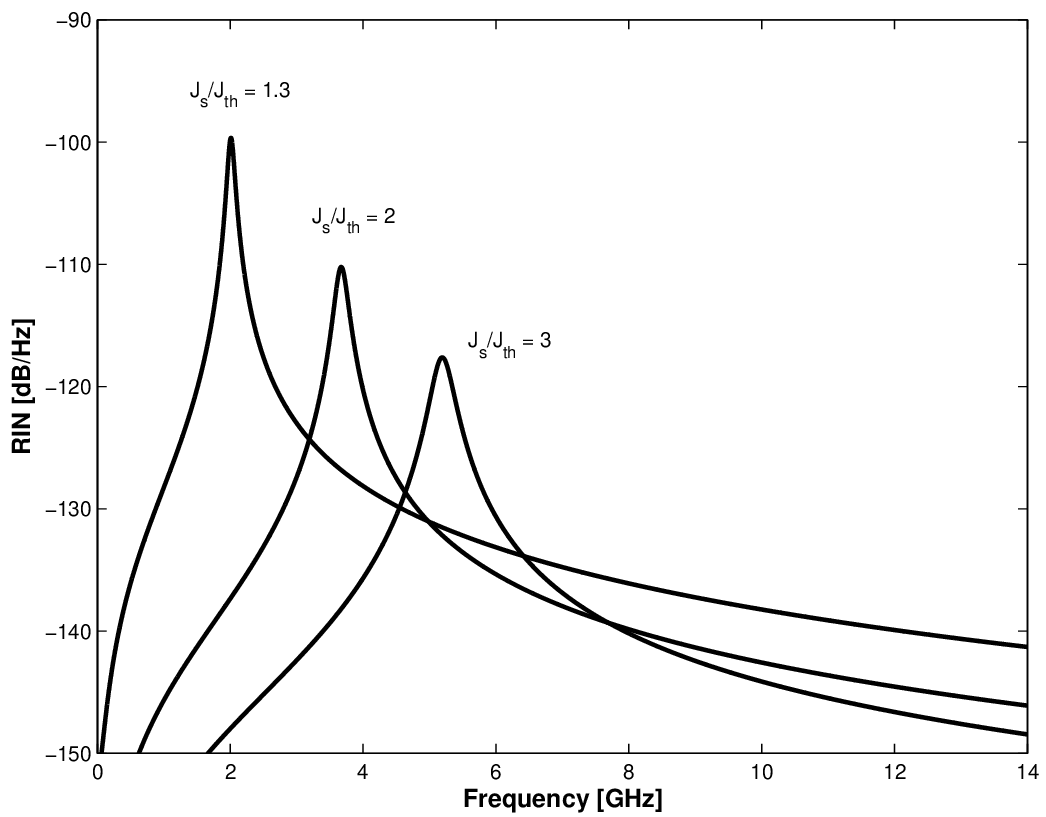}
\par\end{centering}

\caption{RIN spectrum for $J_{s}/J_{th}=$1.3, 2 and 3. The relaxation frequency
is 2 GHz for $J_{s}=1.3J_{th}$ and it increases proportional to $\sqrt{J_{s}-J_{th}}$.}

\label{RIN spectrum}
\end{figure}

\begin{figure}
\begin{centering}
\includegraphics[scale=0.6]{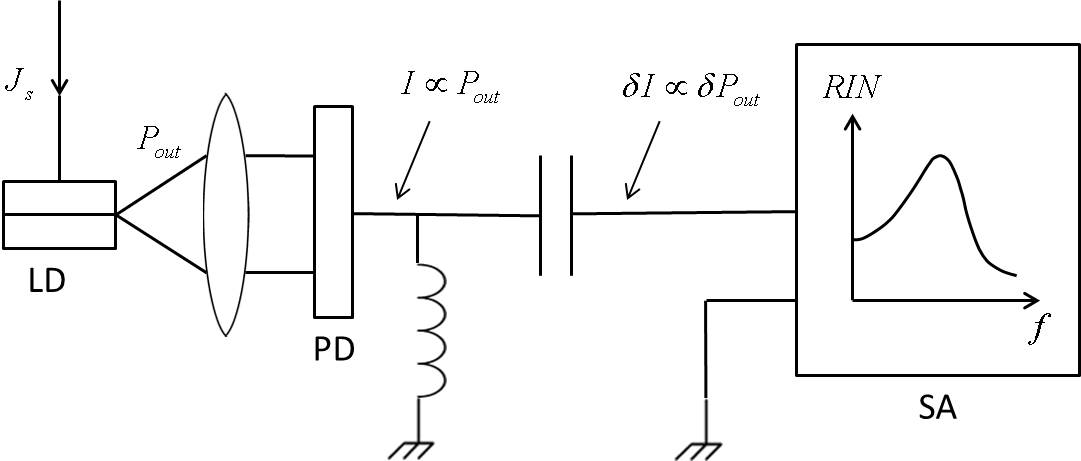}
\par\end{centering}

\caption{\selectlanguage{american}%
Experimental setup to measure RIN spectrum. LD: laser diode, PD: photodiode.
SA: electronic spectrum analyzer.%
}

\selectlanguage{american}%
\label{Experimental setup}
\end{figure}

The shape of the RIN$(f)$ spectrum is dominated by the denominator
$|D|^{2}$. It peaks close to the relaxation resonance frequency $f_{R}=\Omega_{R}/2\pi$
as shown in the example in Figure \ref{RIN spectrum}. The calculation
uses the parameters of Table \ref{Table laser diode} for bias currents
$J_{s}/J_{th}=$ 1.3, 2 and 3 corresponding to output powers $P_{out}=$
1.07 mW, 3.6 mW and 7.1 mW. Figure \ref{Experimental setup} shows
an experimental setup to measure the spectrum.

The variance $\langle(\delta P_{out}(t))^{2}\rangle$ of the output
power is 
the integral over the power spectral
density. Hence
\begin{equation}
\frac{\langle(\delta P_{out}(t))^{2}\rangle}{\langle P_{out}(t)\rangle^{2}}=\frac{2\int_{0}^{B}S_{\delta P}(f)df}{\langle P_{out}(t)\rangle^{2}}=2\int_{0}^{B}\textrm{RIN}(f)df\,.
\end{equation}
where $B$ is the bandwidth of the detection system. This means that
the signal-to-noise (S/N) ratio for the output power is
\begin{equation}
\frac{\langle P_{out}(t)\rangle}{\sqrt{\langle(\delta P_{out}(t))^{2}\rangle}}=\left(2\int_{0}^{B}\textrm{RIN}(f)df\right)^{-\tfrac{1}{2}}\,.
\end{equation}
The integral of RIN(f) without the shot noise term can be derived
from a contour integral around the poles $\omega_{\pm}=\pm\Omega+j\Gamma_{N}/2$
in the upper half of the complex $\omega$-plane. Using
\[
\textrm{RIN}(f)-\frac{\hbar\omega_{0}}{\langle P_{out}\rangle}=\frac{2}{P_{s}}\frac{R_{sp}(\omega^{2}+\tau_{e}^{-2})+a^{2}N_{s}P_{s}/\tau_{e}}{\left(\omega-\omega_{+}\right)\left(\omega-\omega_{+}^{*}\right)\left(\omega-\omega_{-}\right)\left(\omega-\omega_{-}^{*}\right)}
\]
the Cauchy integral formula for two poles, the
result becomes
\begin{equation}
2\int_{0}^{\infty}\left(\textrm{RIN}(f)-\frac{\hbar\omega_{0}}{\langle P_{out}\rangle}\right)df=\frac{R_{sp}(\Omega_{R}^{2}+1/\tau_{e}^{2})+a^{2}N_{s}P_{s}/\tau_{e}}{P_{s}\Gamma_{N}\Omega_{R}^{2}}\,.
\end{equation}
If the bandwidth $B$ is much greater than the resonance frequency
$f_{R}$ then
\begin{equation}
2\int_{0}^{B}\textrm{RIN}(f)df\simeq\frac{R_{sp}(\Omega_{R}^{2}+1/\tau_{e}^{2})+a^{2}N_{s}P_{s}/\tau_{e}}{P_{s}\Gamma_{N}\Omega_{R}^{2}}+\frac{2\hbar\omega_{0}B}{\langle P_{out}\rangle}\,.
\end{equation}
For $J_{s}=1.3J_{th}$ the numerical example in Figure \ref{RIN spectrum}
and for $B=10$ GHz, we get a S/N-ratio of -45 dB.

\section{Frequency noise}

The phase $\phi$ of the envelope field $E(t)=|E(t)|e^{j\phi(t)}$
is governed by the rate equation
\begin{equation}
\frac{d}{dt}\phi(t)=\frac{1}{2}\alpha a(N-N_{0})+\lambda F_{\phi}(t)\label{Phase rate-1}
\end{equation}
where we have introduced the scale factor $\lambda$ in (\ref{Phase rate}).
By taking the average we get a frequency shift
\begin{equation}
\hat{\omega}=\left\langle \frac{d\phi}{dt}\right\rangle =\tfrac{1}{2}\alpha a(\langle N\rangle-N_{0})\,.\label{eq:Freq shift}
\end{equation}
It is therefore convenient to introduce a modified phase shift
\begin{equation}
\varphi(t)=\phi(t)-\hat{\omega}t
\end{equation}
for which
\begin{equation}
\left\langle \frac{d\varphi}{dt}\right\rangle =\left\langle \frac{d\phi}{dt}\right\rangle -\hat{\omega}=0\,.\label{<dvarphi/dt>}
\end{equation}
The 1st order equation for $\varphi(t)$ is
\begin{equation}
\frac{d\varphi}{dt}=\frac{1}{2}\alpha aN_{1}(t)+F_{\phi}(t)\,.\label{dvarphi/dt}
\end{equation}
Using the notation $\dot{\varphi}=\frac{d\varphi}{dt}$ we write the
power spectral density of the frequency noise as $S_{\dot{\varphi}}(f)$.
The noise function $F_{\phi}(t)$ is not correlated with the noise
functions $F_{P}(t)$ and $F_{N}(t)$ and therefore
\begin{equation}
S_{\dot{\varphi}}(f)=\left(\frac{\alpha a}{2}\right)^{2}S_{N_{1}}(f)+S_{F_{\phi}}(f)
\end{equation}
where $S_{N_{1}}(f)$ is a short notation for $S_{NN}(f)$. From (\ref{Sij(f)}),
(\ref{Laser mod transfer-1}), (\ref{Deno of mod trans-1}) and the
diffusion constants in (\ref{diff-const-1}) and (\ref{diff-const-2})
we find
\begin{gather}
S_{N_{1}}(f)=|H_{NP}|^{2}2D_{PP}+4Re\left\{ H_{NP}^{*}H_{NN}\right\} D_{PN}+|H_{NN}|^{2}2D_{NN}\nonumber \\
=2\left[\frac{R_{sp}P_{s}}{\tau_{p}^{2}}+\omega^{2}\left(R_{sp}P_{s}+\frac{N_{s}}{\tau_{e}}\right)\right]/|D|^{2}\,.
\end{gather}
By (\ref{cros-spec of F}) and (\ref{diff-const-1}) we have $S_{F_{\phi}}(f)=2D_{\phi\phi}=R_{sp}/(2P_{s})$
so using $\Omega_{R}^{2}=aP_{s}/\tau_{p}$ the power spectral density
of the frequency noise becomes
\begin{equation}
S_{\dot{\varphi}}(f)=\frac{\alpha^{2}\Omega_{R}^{4}}{2P_{s}}\frac{R_{sp}(1+(\omega\tau_{p})^{2})+(\omega\tau_{p})^{2}N_{s}/(P_{s}\tau_{e})}{(\omega^{2}-\Omega_{R}^{2})^{2}+(\Gamma_{N}\omega)^{2}}+\frac{R_{sp}}{2P_{s}}\label{Freq noise spectrum}
\end{equation}
with the DC value
\begin{equation}
S_{\dot{\varphi}}(0)=\frac{R_{sp}(1+\alpha^{2})}{2P_{s}}\,.\label{Sphi(0)}
\end{equation}
The spectrum peaks close to the relaxation resonance frequency and
approaches $R_{sp}/(2P_{s)}$ for $f\rightarrow\infty$. A numerical
example of the spectrum is shown in Figure \ref{Frequency noise spectrum}.

\begin{center}
\begin{figure}
\begin{centering}
\includegraphics[scale=0.3]{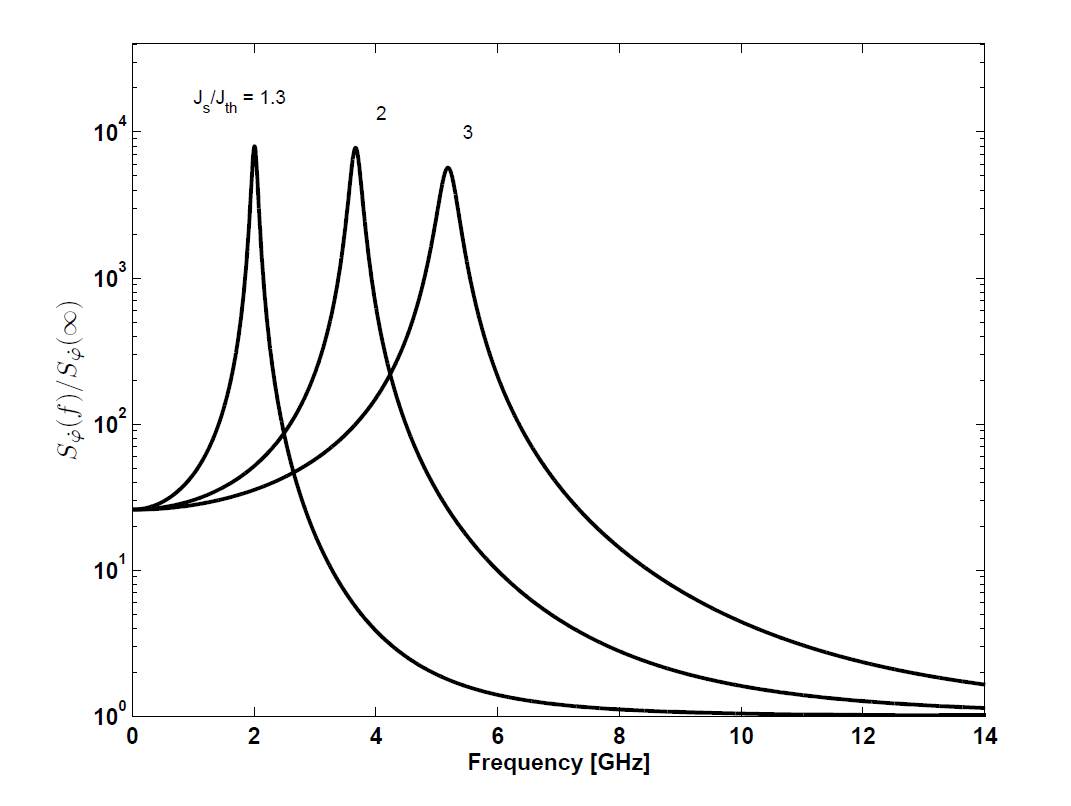}
\par\end{centering}

\caption{\selectlanguage{american}%
Power spectral density of frequency noise for $J_{s}/J_{th}=$1.3,
2 and 3. \selectlanguage{english}%
}

\selectlanguage{american}%
\label{Frequency noise spectrum}
\end{figure}

\par\end{center}

\section{Power spectral density of the laser field}

The power spectral density of the laser envelope field $E(t)$ is
the Fourier transform
\begin{equation}
S_{E}(f)=\int_{-\infty}^{\infty}\mathcal{R}_{E}(\tau)e^{-j\omega\tau}df\label{Laser field spectrum}
\end{equation}
of the autocorrelation
\begin{equation}
\mathcal{R}_{E}(\tau)=\left\langle E^{*}(t)E(t+\tau)\right\rangle \label{Laser field auto}
\end{equation}
for constant injection current $J(t)=J_{s}$. We have normalized the
electric field such that $|E(t)|^{2}={\cal {P}}$. Therefore $E(t)=\sqrt{{\cal {P}}}e^{j\phi(t)}$
and
\begin{equation}
\mathcal{R}_{E}(\tau)=\left\langle \sqrt{{\cal {P}}(t){\cal {P}}(t+\tau)}e^{j(\phi(t+\tau)-\phi(t)}\right\rangle \,.\label{Laser field auto-1}
\end{equation}
In terms of the modified phase shift $\varphi$
\begin{equation}
\phi(t+\tau)-\phi(t)=\hat{\omega}\tau+\Delta\varphi\label{Deltaphi}
\end{equation}
where
\begin{equation}
\Delta\varphi=\varphi(t+\tau)-\varphi(t)=\int_{t}^{t+\tau}\frac{d\varphi(t')}{dt'}dt'
\end{equation}
and
\begin{equation}
\left\langle \Delta\varphi\right\rangle =\int_{t}^{t+\tau}\left\langle \frac{d\varphi}{dt'}\right\rangle dt'=0\,.\label{<Delta phi>}
\end{equation}
Introducing the notation
\begin{equation}
x=\sqrt{{\cal {P}}(t){\cal {P}}(t+\tau)}
\end{equation}
we can rewrite (\ref{Laser field auto-1}) as
\begin{equation}
\mathcal{R}_{E}(\tau)=\left\langle xe^{j\Delta\varphi}\right\rangle e^{j\hat{\omega}\tau}=\left(\left\langle (\Delta xe^{j\Delta\varphi}\right\rangle +\langle x\rangle\left\langle e^{j\Delta\varphi}\right\rangle \right)e^{j\hat{\omega}\tau}\label{Laser field auto-2}
\end{equation}
where $\Delta x=x-\langle x\rangle$.The variables $\Delta x$ and
$\Delta\varphi$ are zero mean variables and in the joint Gaussian
approximation one has 
\begin{equation}
\left\langle e^{j\Delta\varphi}\right\rangle =e^{-\frac{1}{2}\langle(\Delta\varphi)^{2}\rangle}\label{eq:avr. ejdeltafi}
\end{equation}
and
\begin{equation}
\left\langle \Delta xe^{j\Delta\varphi}\right\rangle =j\left\langle \Delta x\Delta\varphi\right\rangle e^{-\frac{1}{2}\langle(\Delta\varphi)^{2}\rangle}\,.
\end{equation}
In this approximation we then get
\begin{equation}
\mathcal{R}_{E}(\tau)\simeq\left\{ \langle x\rangle+j\left\langle \Delta x\Delta\varphi\right\rangle \right\} e^{-\frac{1}{2}\langle(\Delta\varphi)^{2}\rangle+j\hat{\omega}\tau}\,.\label{Laser field auto app}
\end{equation}

From the expansion (\ref{P expansion}) of $P\left(t\right)$ we find
directly to 2nd order in $\lambda$

\begin{gather}
x=\sqrt{{\cal {P}}(t){\cal {P}}(t')}\nonumber \\
\simeq\left[P_{s}^{2}+\lambda P_{s}(P_{1}(t)+P_{1}(t'))+\lambda^{2}(P_{s}(P_{2}(t)+P_{2}(t'))+P_{1}(t)P_{1}(t'))\right]^{\tfrac{1}{2}}\,.
\end{gather}
This expression is then Taylor expanded also to 2nd order in $\lambda$

\begin{equation}
x\simeq P_{s}+\frac{\lambda}{2}(P_{1}(t)+P_{1}(t'))+\frac{\lambda^{2}}{2}\left(P_{2}(t)+P_{2}(t')-\frac{(P_{1}(t)-P_{1}(t'))^{2}}{4P_{s}}\right)\,.\label{eq:sq(PP)expansion}
\end{equation}
Approximating $x$ by (\ref{eq:sq(PP)expansion}) for $\lambda=1$
and $t'=t+\tau$ we obtain the 2nd order average
\begin{equation}
\langle x\rangle\simeq\hat{P}+\frac{\langle P_{1}(t)P_{1}(t+\tau)\rangle}{4P_{s}}\label{<x>}
\end{equation}
where we have used $\langle P_{1}\rangle=0$ and introduced the constant
\begin{equation}
\hat{P}=P_{s}+\langle P_{2}\rangle-\frac{\langle P_{1}^{2}\rangle}{4P_{s}}\,.\label{eq:P hat}
\end{equation}
Furthermore, to 1st order $\Delta x=x-\langle x\rangle=\tfrac{1}{2}(P_{1}(t)+P_{1}(t+\tau))$
so to 2nd order
\begin{align}
j\left\langle \Delta x\Delta\varphi\right\rangle  & \simeq\frac{j}{2}\left\langle (P_{1}(t)+P_{1}(t+\tau))(\varphi(t+\tau)-\varphi(t))\right\rangle \nonumber \\
 & =\frac{j}{2}\langle P_{1}(t)\varphi(t+\tau)-\varphi(t)P_{1}(t+\tau)\rangle\,.\label{<P1phi-phiP1>}
\end{align}
Notice, the averages are independent of $t$ for stationary stochastic
processes.

The Fourier transform of $\mathcal{R}_{E}(\tau)e^{-j\hat{\omega}\tau}$
is by (\ref{Laser field spectrum}) the power spectral density of
the envelope field $S_{E}(f+\hat{f})$ where $\hat{\omega}=2\pi\hat{f}$.
Using the approximations (\ref{Laser field auto app}), (\ref{<x>})
and (\ref{<P1phi-phiP1>}) it becomes the frequency domain convolution
\begin{equation}
S_{E}(f+\hat{f})=\left(\hat{P}\delta(f)+\frac{S_{P_{1}}(f)}{4P_{s}}+Im\{S_{\varphi P_{1}}(f)\}\right)\otimes\mathcal{L}(f)\label{S_E conv}
\end{equation}
where
\begin{equation}
\mathcal{L}(f)=\int_{-\infty}^{\infty}e^{-\tfrac{1}{2}\langle(\Delta\varphi)^{2}\rangle-j\omega\tau}d\tau\,.\label{Lineshape}
\end{equation}
is a lineshape function. The spectrum $S_{E}(f+\hat{f})$ is centered
at $f=0$, which means that $S_{E}(f)$ is centered at $f=\hat{f}$.

\subsubsection{Variance $\langle(\Delta\varphi)^{2}\rangle$ and the Lineshape function}

Let us first examine the lineshape function $\mathcal{L}(f)$. 
The variance of $\Delta\varphi$ can be calculated
from
\begin{equation}
\langle(\Delta\varphi)^{2}\rangle=\int_{-\infty}^{\infty}S_{\Delta\varphi}(f)df\,.\label{variance phi-1}
\end{equation}
where $S_{\Delta\varphi}(f)$ is the power spectral density of $\Delta\varphi$.
Since ${\cal {F}}[\Delta\varphi)](f)=(e^{j\omega\tau}-1)\tilde{\varphi}(f)$
it follows from (6.81)-[1] that $S_{\Delta\varphi}(f)=|e^{j\omega\tau}-1|^{2}S_{\varphi}(f)=4\sin^{2}(\omega\tau/2)S_{\varphi}(f)$
where $S_{\varphi}(f)$ is the power spectral density of $\varphi$.
Moreover, ${\cal {F}}[\dot{\varphi}](f)=j\omega\tilde{\varphi}$ implies
that the power spectral density of $\dot{\varphi}$ is $S_{\dot{\varphi}}(f)=\omega^{2}S_{\varphi}(f)$,
so
\begin{equation}
S_{\Delta\varphi}(f)=\frac{4\sin^{2}(\omega\tau/2)}{\omega^{2}}S_{\dot{\varphi}}(f)
\end{equation}
and hence the variance is
\begin{equation}
\langle(\Delta\varphi)^{2}\rangle=\int_{-\infty}^{\infty}S_{\dot{\varphi}}(f)\frac{\sin^{2}(\omega\tau/2)}{(\omega/2)^{2}}df\label{eq:phi-2}
\end{equation}
in terms of the power spectal density of the frequency noise. Using
the substitution $v=\omega|\tau|/2=\pi f|\tau|$ we can rewrite (\ref{eq:phi-2})
as
\begin{equation}
\langle(\Delta\varphi)^{2}\rangle=\frac{|\tau|}{\pi}\int_{-\infty}^{\infty}S_{\dot{\varphi}}\left(\frac{v}{\pi|\tau|}\right)\frac{\sin^{2}v}{v^{2}}dv\,.\label{variance phi-3}
\end{equation}
According to (B.8)-[1], for $|\tau|\rightarrow\infty$
the integral converges to
\begin{equation}
\lim_{\tau\rightarrow\infty}\int_{-\infty}^{\infty}S_{\dot{\varphi}}\left(\frac{v}{\pi|\tau|}\right)\frac{\sin^{2}v}{v^{2}}dv=S_{\dot{\varphi}}(0)\int_{-\infty}^{\infty}\frac{\sin^{2}v}{v^{2}}dv=\pi S_{\dot{\varphi}}(0)\label{variance phi-4}
\end{equation}
so for large $|\tau|$ the variance can be approximated by
\begin{equation}
\langle(\Delta\varphi)^{2}\rangle\simeq|\tau|S_{\dot{\varphi}}(0)=2\gamma|\tau|\label{variance phi-5}
\end{equation}
where the parameter $\gamma$ is indirectly defined and by (\ref{Sphi(0)})
given by
\begin{equation}
\gamma=\frac{1}{2}S_{\dot{\varphi}}(0)=\frac{R_{sp}(1+\alpha^{2})}{4P_{s}}\,.\label{Field decay rate}
\end{equation}
It can be shown that in a model where $\varphi(t)$ is performing
a random walk \cite{Henry-1986} that changes the phase by $\Delta\varphi$
during the time $\tau$, the increase of the variance is proportional
to $|\tau|$. In such a model we have $\mathcal{R}_{E}(\tau)\propto e^{-\gamma|\tau|}$,
i.e. the field autocorrelation decays with rate $\gamma$ and $1/\gamma$
is therefore called the coherence time. For the lineshape function
(\ref{Lineshape}) the approximation (\ref{variance phi-5}) gives
the simple Lorentzian
\begin{equation}
\mathcal{L}(f)\simeq\frac{2\gamma}{\omega^{2}+\gamma^{2}}=\frac{1}{2\pi}\frac{\Delta f}{f^{2}+(\tfrac{1}{2}\Delta f)^{2}}\label{Loretzian app}
\end{equation}
where $\Delta f=\gamma/\pi$ is the FWHM linewidth of (\ref{Loretzian app}).
The result is called the Lorentzian approximation. The factor $1+\alpha^{2}$
in (\ref{Field decay rate}) explains why the parameter $\alpha$
is called the linewidth enhancement factor. For a typical value $\alpha=5$,
it increases the linewidth by a factor 26.

\begin{figure}
\begin{centering}
\includegraphics[scale=0.4]{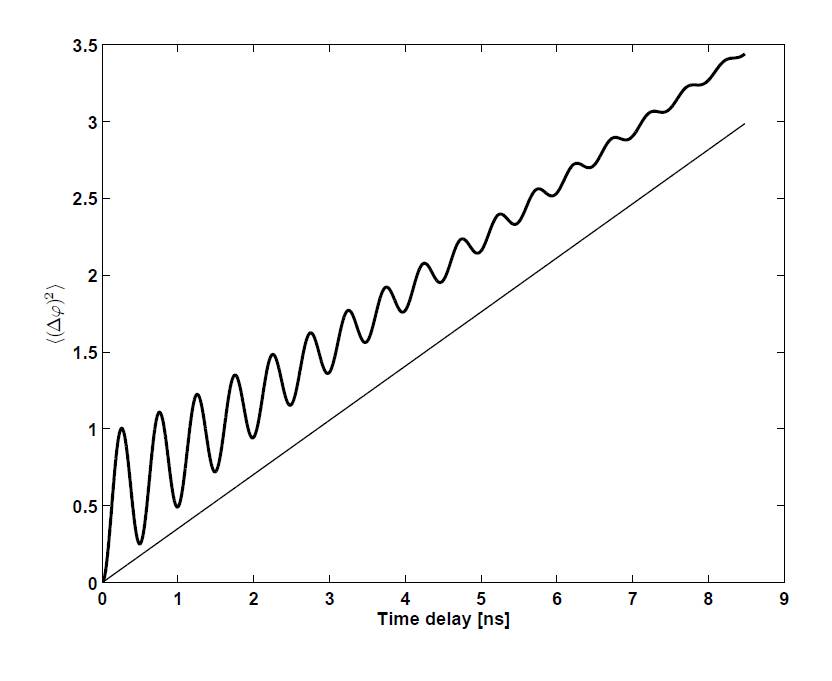}\centering
\par\end{centering}

\caption{Variance $\langle(\Delta\varphi)^{2}\rangle$ versus time delay for
$J_{s}=1.3J_{th}$. Thick solid curve shows (\ref{variance phi-7})
and the thin line is the approximation (\ref{variance phi-5}).}

\label{Variance versus phase delay}
\end{figure}

\begin{figure}
\begin{centering}
\includegraphics[scale=0.5]{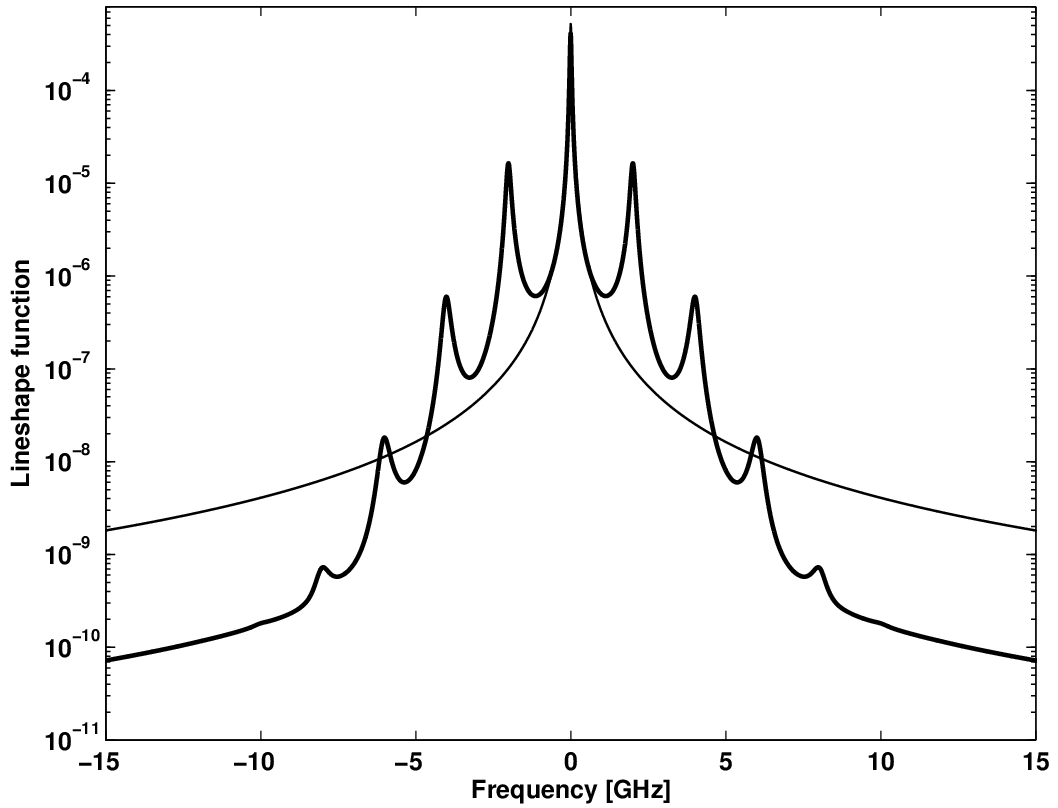}\centering
\par\end{centering}

\caption{Thick solid curve: lineshape function $\mathcal{L}(f)$ using (\ref{variance phi-7}).
Thin solid curve: Lorentzian approximation (\ref{Loretzian app})}

. \label{Lineshape function}
\end{figure}

We can get a more detailed picture of the variance by using the approximation
\begin{equation}
S_{\dot{\varphi}}(f)\simeq\frac{R_{sp}}{2P_{s}}\left(1+\frac{\alpha^{2}\Omega_{R}^{4}}{(\omega^{2}-\Omega_{R}^{2})^{2}+(\Gamma_{N}\omega)^{2}}\right)\label{Freq
noise spectrum-app}
\end{equation}
where we have neglected terms in (\ref{Freq noise spectrum}) proportional
to $\omega\tau_{p}$. With this expression the variance is
\begin{gather}
\langle(\Delta\varphi)^{2}\rangle=2\gamma|\tau|+2\int_{-\infty}^{\infty}(S_{\dot{\varphi}}(f)-S_{\dot{\varphi}}(0))\frac{1-\cos(\omega\tau)}{\omega^{2}}df\nonumber \\
=2\gamma|\tau|+\frac{\alpha^{2}R_{sp}}{2\pi P_{s}}Re\left[\int_{-\infty}^{\infty}\frac{2\Omega_{R}^{2}-\omega^{2}-\Gamma_{N}^{2}}{(\omega^{2}-\Omega_{R}^{2})^{2}+(\Gamma_{N}\omega)^{2}}(1-e^{j\omega\tau})d\omega\right]\,.\label{variance
phi-6}
\end{gather}
The integral can be calculated as a contour integral around the poles
$\omega_{\pm}=\pm\Omega+j\Gamma_{N}/2$ in the upper complex $\omega$-plane
\cite{Henry-1986,Agrawal-1993}. It gives
\begin{equation}
\langle(\Delta\varphi)^{2}\rangle=2\gamma|\tau|+\frac{\alpha^{2}R_{sp}\Omega_{R}}{2P_{s}\Gamma_{N}\Omega}\left(\cos(3\theta)-e^{-\tfrac{1}{2}\Gamma_{N}|\tau|}\cos(\Omega|\tau|-3\theta)\right)\label{variance phi-7}
\end{equation}
where $\tan(\theta)=\Gamma_{N}/(2\Omega)$ and hence $\cos(3\theta)=\Omega(1-(\Gamma_{N}/\Omega_{R})^{2})/\Omega_{R}$.
The variance is therefore a sum of the linear function $2\gamma|\tau|+\frac{\alpha^{2}R_{sp}}{2\Gamma_{N}P_{s}}(1-(\Gamma_{N}/\Omega_{R})^{2})$
and a damped relaxation oscillation. The variance versus $\tau$ is
shown in Figure \ref{Variance versus phase delay} for $J_{s}=1.3J_{th}$
and for the parameters of Table \ref{Table laser diode}. The corresponding
lineshape function $\mathcal{L}(f)$ from (\ref{Lineshape}) and (\ref{variance phi-7})
is shown in Figure \ref{Lineshape function} together with the Lorentzian
approximation (\ref{Loretzian app}). The linewidth is $\Delta f=$
56 MHz. For low frequencies the two curves are very close but at larger
frequencies the relaxation oscillations of (\ref{variance phi-7})
give rise to satellite peaks at multiples of the relaxation frequency.

For $|\tau|\rightarrow0$ the integral in (\ref{variance phi-3})
converges to $\pi S_{\dot{\varphi}}(\infty)=\pi R_{sp}/2P_{s}$ according
to (\ref{Freq noise spectrum}). Therefore
\[
\langle(\Delta\varphi)^{2}\rangle\simeq\frac{R_{sp}}{2P_{s}}|\tau|=\frac{2\gamma}{1+\alpha^{2}}|\tau|\,.
\]
for small $|\tau|$, which can also be shown to agree with (\ref{variance phi-7})
by expanding the variance to 1st order in $|\tau|$. This explains
why the tails of $\mathcal{L}(f)$ based on (\ref{variance phi-7})
is a factor $1+\alpha^{2}$ lower than the tails of the Lorentzian
approximation.

\subsubsection{Comparison of spectral contributions}

\begin{figure}
\begin{centering}
\centering \includegraphics[scale=0.5]{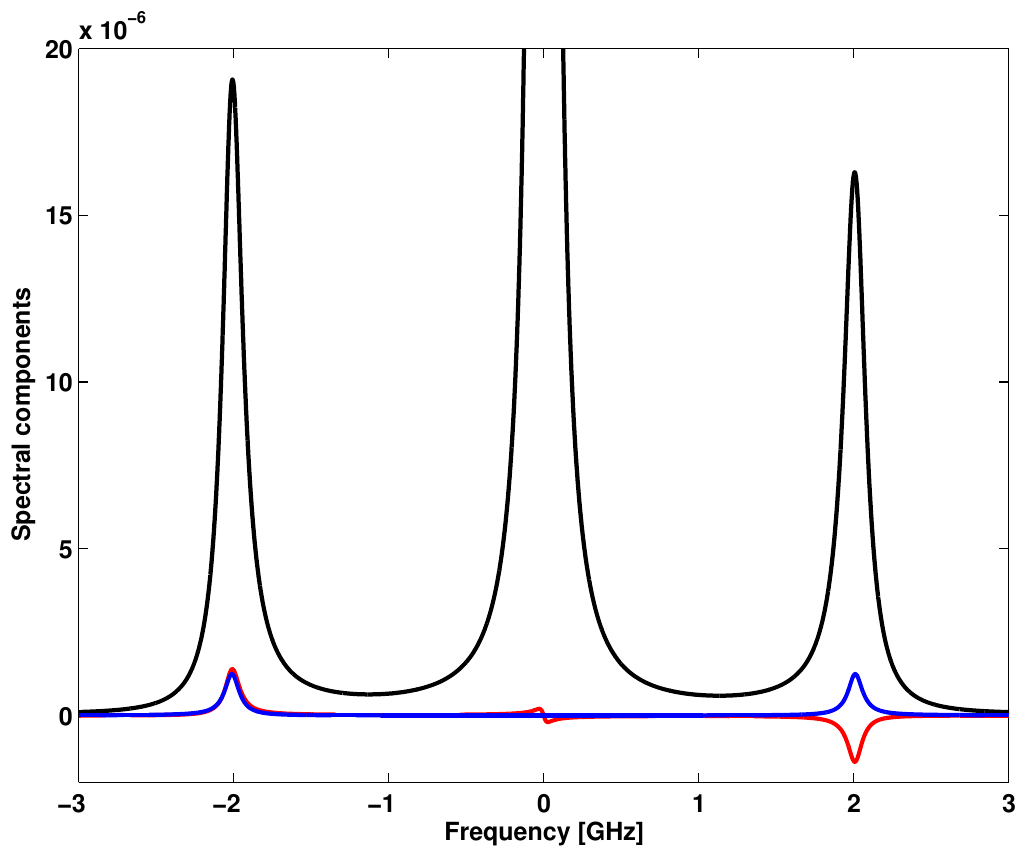}
\par\end{centering}

\caption{Contributions to $S_{E}(f+\hat{f})$. Total spectrum (black) is the
sum of $P_{s}\mathcal{L}(f)$, intensity noise (blue) and amplitude-phase
noise (red).}

\label{Comp of field spectrum}
\end{figure}

The dominant contribution to the field power spectrum $S_{E}(f+\hat{f})$
is the convolution $\hat{P}\delta(f)\otimes\mathcal{L}(f)=\hat{P}\mathcal{L}(f)$
in (\ref{S_E conv}). The black curve in Figure \ref{Comp of field spectrum}
shows a clipped spectrum on a linear scale for the same example parameters
as in Figure \ref{Lineshape function} and for the variance (\ref{variance phi-7}).
It is calculated as the sum of $P_{s}\mathcal{L}(f)$ and the contributions
from intensity noise $S_{P_{1}}(f)\otimes\mathcal{L}(f)/4P_{s}$,
shown as the blue curve, and the phase-amplitude coupling $\textrm{Im}\{S_{\varphi P_{1}}(f)\}\otimes\mathcal{L}(f)$
shown as the red curve. We ignore the small difference between $\hat{P}$
and $P_{s}$ in (\ref{eq:P hat}). The intensity noise and the phase-amplitude
noise have a minor influence on the total spectrum, the most visible
effect being the asymmetry of the satellite peaks. Below we explain
how the contributions are calculated.

It follows from (\ref{Lineshape}) that the integral over the lineshape
function is
\begin{equation}
\int_{-\infty}^{\infty}\mathcal{L}(f)df=e^{-\tfrac{1}{2}\langle(\Delta\varphi)^{2}\rangle}|_{\tau=0}=1\,.\label{Lineshape int}
\end{equation}
Furthermore, $\mathcal{L}(f)$ has a narrow spike at $f=0$. It can
therefore be approximated by $\delta(f)$ when it is convoluted with
functions like $S_{P_{1}}(f)$ (=$S_{PP}(f)$), where the spectral
structures are much broader than the width of the spike in $\mathcal{L}(f)$.
Thus
\begin{equation}
S_{P_{1}}(f)\otimes\mathcal{L}(f)\simeq S_{P_{1}}(f)\,.\label{S_PxL}
\end{equation}

The contribution to $S_{E}(f+\hat{f})$ due to phase-amplitude coupling
has to be treated with special care because of the divergence of $S_{\varphi P_{1}}(f)$
at $f=0$. The Fourier transform of the rate equation (\ref{dvarphi/dt})
with truncated noise sources is
\begin{equation}
\tilde{\varphi}(f)=\left(\frac{1}{2}\alpha a\tilde{N}_{1}+\tilde{F}_{\phi T}(f)\right)\left(P\frac{1}{j\omega}+\frac{1}{2}\delta(f)\right)
\end{equation}
according to (4.100)-[1]. The symbol $"P"$ means that we have
to take the principal value when integrating over frequency.
\foreignlanguage{english}{
Since $\langle F_{P}(t)F_{\phi}(t')\rangle=\langle F_{N}(t)F_{\phi}(t')\rangle=0$,
and hence $\langle P_{1}(t)F_{\phi}(t')\rangle=0$, the phase-amplitude
cross power spectral density is
\begin{equation}
S_{\varphi P_{1}}(f)=\frac{1}{2}\alpha aS_{NP}(f)\left(-P\frac{1}{j\omega}+\frac{1}{2}\delta(f)\right)
\end{equation}
with imaginary part
\begin{equation}
\textrm{Im}\{S_{\varphi P_{1}}(f)\}=\frac{1}{2}\alpha a\left(\textrm{Re}\{S_{NP}(f)\}P\frac{1}{\omega}+\frac{1}{2}\textrm{Im}\{S_{NP}(0)\}\delta(f)\right)\,.
\end{equation}
The spectrum $S_{NP}$ follows from (\ref{S compact matrix}) and
is
\begin{equation}
S_{NP}(f)=2[D_{PP}H_{NP}^{*}H_{PP}+D_{PN}(H_{NN}^{*}H_{PP}+H_{NP}^{*}H_{PN})+D_{NN}H_{NN}^{*}H_{PN}]\,.
\end{equation}
The matrix $\boldsymbol{D}$ is real, and so is the matrix $\boldsymbol{H}$
for $\omega=0$ as we see from (\ref{Laser mod transfer}). Thus $\textrm{Im}\{S_{NP}(0)\}=0$,
\begin{equation}
\textrm{Re}\{S_{NP}(f)\}=\frac{2P_{s}}{\tau_{p}|D|^{2}}\left[-R_{sp}\Gamma_{N}+(\tfrac{1}{2}-R_{sp}\tau_{p})(\omega^{2}-\Omega_{R}^{2})\right]
\end{equation}
and
\begin{equation}
\textrm{Im}\{S_{\varphi P_{1}}(f)\}=g(\omega)P\frac{1}{\omega}\label{Im S_Pphi}
\end{equation}
where
\begin{equation}
g(\omega)=-\alpha\Omega_{R}^{2}\frac{R_{sp}\Gamma_{N}+(R_{sp}\tau_{p}-\tfrac{1}{2})(\omega^{2}-\Omega_{R}^{2})}{(\omega^{2}-\Omega_{R}^{2})^{2}+(\Gamma_{N}\omega)^{2}}\label{g in S_Pphi}
\end{equation}
is an even function of $\omega$. The convolution in (\ref{S_E conv})
of $\textrm{Im}\{S_{\varphi P_{1}}(f)\}$ with $\mathcal{L}(f)$ is
the principal value integral
\begin{equation}
\textrm{Im}\{S_{\varphi P_{1}}(f)\}\otimes\mathcal{L}(f)=\frac{1}{2\pi}P\int_{-\infty}^{\infty}\mathcal{L}(f-f')\frac{g(\omega')}{\omega'}d\omega'\,.\label{Lxg/omega}
\end{equation}
We first calculate the integral over a finite symmetric interval $[-b,b]$,
and then we take the limit $b\rightarrow\infty$. The finite integral
is written as
\begin{gather}
\frac{1}{2\pi}P\int_{-b}^{b}\mathcal{L}(f-f')\frac{g(\omega')}{\omega'}d\omega'=\frac{1}{2\pi}\int_{-b}^{b}\mathcal{L}(f-f')\frac{g(\omega')-g(0)}{\omega'}d\omega'\nonumber \\
+g(0)\frac{1}{2\pi}P\int_{-b}^{b}\frac{\mathcal{L}(f-f')}{\omega'}d\omega'\,.\label{Finite Lxg/om}
\end{gather}
The integrant of the first integral on the r.h.s. is finite at $\omega'=0$
and the integral therefore converges as an ordinary integral. Moreover,
$(g(\omega)-g(0))/\omega$ is slowly varying compared to the central
spike of $\mathcal{L}(f)$ so we approximate $\mathcal{L}(f)$ by
a delta function and get
\begin{equation}
\frac{1}{2\pi}P\int_{-\infty}^{\infty}\mathcal{L}(f-f')\frac{g(\omega')-g(0)}{\omega'}d\omega'\simeq\frac{g(\omega)-g(0)}{\omega}\,.\label{Lx(g-g0)/om}
\end{equation}
For the Lorentzian approximation $\mathcal{L}(f)=\frac{2\gamma}{\omega^{2}+\gamma^{2}}$
the second integral on the r.h.s. of (\ref{Finite Lxg/om}) is
\begin{equation}
\frac{1}{2\pi}P\int_{-b}^{b}\frac{\mathcal{L}(f-f')}{\omega'}d\omega'=\frac{1}{2\pi}\int_{-b}^{b}\frac{\mathcal{L}(f-f')-\mathcal{L}(f)}{\omega'}d\omega'+\frac{\mathcal{L}(f)}{2\pi}P\int_{-b}^{b}\frac{1}{\omega'}d\omega'\,.
\end{equation}
The second integral on the r.h.s. is zero because the integrant is
an odd function. The first integral on the r.h.s. is
\begin{gather}
\frac{1}{2\pi}\int_{-b}^{b}\frac{\mathcal{L}(f-f')-\mathcal{L}(f)}{\omega'}d\omega'=\frac{1}{2\pi}\frac{2\gamma}{\omega^{2}+\gamma^{2}}\int_{-b}^{b}\frac{2\omega-\omega'}{(\omega'-\omega)^{2}+\gamma^{2}}d\omega'\nonumber \\
=\frac{1}{2\pi}\frac{2\gamma}{\omega^{2}+\gamma^{2}}\left(\frac{1}{2}\ln\frac{(\omega+b)^{2}+\gamma^{2}}{(\omega-b)^{2}+\gamma^{2}}+\omega\int_{-b}^{b}\frac{d\omega'}{(\omega'-\omega)^{2}+\gamma^{2}}\right)\,.
\end{gather}
The first term in the parenthesis is zero in the limit $b\rightarrow\infty$
and the second is $\omega\pi/\gamma$ for $b\rightarrow\infty$. Hence
\begin{equation}
\lim_{b\rightarrow\infty}\frac{1}{2\pi}P\int_{-b}^{b}\frac{\mathcal{L}(f-f')}{\omega'}d\omega'=\frac{\omega}{\omega^{2}+\gamma^{2}}\label{Lx1/om}
\end{equation}
for the Lorentzian lineshape function. For $\mathcal{L}(f)$ based
on the variance (\ref{variance phi-7}) the l.h.s. of (\ref{Lx1/om})
can be calculated numerically as the imaginary part of the Fourier
transform of $\exp(-\tfrac{1}{2}\langle(\Delta\varphi)^{2}\rangle)u(-\tau)$
where $u(\tau)$ is the step function.}

Combining the results of (\ref{Finite Lxg/om}), (\ref{Lx(g-g0)/om})
and (\ref{Lx1/om}) we finally have
\begin{equation}
\textrm{Im}\{S_{\varphi P_{1}}(f)\}\otimes\mathcal{L}(f)\simeq\frac{g(\omega)-g(0)}{\omega}+g(0)\frac{\omega}{\omega^{2}+\gamma^{2}}\,.\label{S_PphixL}
\end{equation}
The convolution with $\mathcal{L}(f)$ has turned a diverging function
$g(\omega)/\omega$ into a finite function of frequency. For the parameters
of Figure \ref{Lineshape function} the function $\textrm{Im}\{S_{\varphi P_{1}}(f)\}\otimes\mathcal{L}(f)$,
which is the term that contributes to $S_{E}(f+\hat{f})$, is the
red curve shown in Figure \ref{Comp of field spectrum}. It does not
matter whether we use the Lorentzian lineshape function or $\mathcal{L}(f)$
based on (\ref{variance phi-7}). The two methods give results that
agree within the line thickness in Figure \ref{Comp of field spectrum}.

From (\ref{g in S_Pphi}) we find

\begin{equation}
g(0)=\gamma_{1}+\alpha(R_{sp}\tau_{p}-\tfrac{1}{2})\label{g(0)}
\end{equation}
where $\gamma_{1}=-R_{sp}\alpha\Gamma_{N}/\Omega_{R}^{2}$ is the
parameter introduced in the field power spectrum in (\ref{FiniteS-E})
and which gave the contribution $\gamma_{1}\omega/(\omega^{2}+\gamma^{2})$
to the spectrum. It gives the same contribution to the spectrum in
(\ref{S_PphixL}) but in addition we get a contribution of the same
form from the second term in $g(0)$. The latter comes from the correlation
strength $D_{PN}$ in (\ref{diff-const-2}) but that was not included
in the calculation of the spectrum in (\ref{FiniteS-E}).

The solid black curve in Figure \ref{Comp of field spectrum} is the
total spectrum $S_{E}(f+\hat{f})$ composed of $P_{s}\mathcal{L}(f)$
plus the blue and the red curves. It is asymmetric due to the asymmetry
of the red curve. The asymmetry was demonstrated experimentally in
\cite{Vahala-1983} and also explained as being caused by the amplitude-phase
coupling.

\subsubsection{Noise induced frequency shift}

The noise induced frequency shift $\hat{\omega}$ in (\ref{eq:Freq shift})
is to 2nd order
\begin{equation}
\hat{\omega}=\frac{1}{2}a(N_{s}+\langle N_{2}\rangle-N_{0})\label{Freq shift 1}
\end{equation}
where
\begin{equation}
\langle N_{2}\rangle=-\frac{\tau_{e}}{\tau_{p}}\langle P_{2}\rangle=-\frac{\langle N_{1}P_{1}\rangle}{P_{s}}\label{<N2>}
\end{equation}
according to (\ref{<P2&N2>}). The average $\langle N_{1}(t)P_{1}(t)\rangle$
can be calculated from the cross power spectral density $S_{NP}(f)$,
which is
\begin{equation}
S_{NP}(f)=\int_{-\infty}^{\infty}\langle N_{1}(t)P_{1}(t+\tau)\rangle e^{-j\omega\tau}d\tau\,.
\end{equation}
The integral over frequency is then
\begin{equation}
\int_{-\infty}^{\infty}S_{NP}(f)df=\int_{-\infty}^{\infty}\langle N_{1}(t)P_{1}(t+\tau)\rangle\delta(\tau)d\tau=\langle N_{1}(t)P_{1}(t)\rangle\,.\label{<N1P1>}
\end{equation}
From (\ref{Sij(f)}) we find
\begin{gather}
S_{NP}(f)=2\left[H_{NP}^{*}H_{PP}D_{PP}+(H_{NP}^{*}H_{PN}+H_{NN}^{*}H_{PP})D_{PN}+H_{NN}^{*}H_{PN}D_{NN}\right]\nonumber \\
=\frac{2}{|D|^{2}}\left[-D_{PP}(j\omega+\Gamma_{N})/\tau_{p}-D_{PN}(\Omega_{R}^{2}+j\omega(j\omega+\Gamma_{N}))-j\omega aP_{s}D_{NN}\right]\,.\label{SNP(f)}
\end{gather}
When we integrate $S_{NP}(f)$ over frequency we get no contribution
from terms that are odd functions of $\omega$, i.e. we get no contribution
from the imaginary terms. This implies
\begin{equation}
\langle N_{1}(t)P_{1}(t)\rangle=\frac{1}{\pi}\int_{-\infty}^{\infty}\frac{-D_{PP}\Gamma_{N}/\tau_{p}+D_{PN}(\omega^{2}-\Omega_{R}^{2})}{(\omega^{2}-\Omega_{R}^{2})^{2}+(\Gamma_{N}\omega)^{2}}d\omega\,.\label{int(SNP)}
\end{equation}
The integration can be performed by contour integration around the
poles $\omega_{\pm}=\pm\Omega+j\Gamma_{N}/2$ in the upper half of
the complex $\omega$-plane. This gives the simple result
\begin{equation}
\langle N_{1}(t)P_{1}(t)\rangle=-\frac{D_{PP}}{\Omega_{R}^{2}\tau_{p}}=-\frac{R_{sp}}{a}\label{<N1P1> result}
\end{equation}
and hence
\begin{equation}
\langle N_{2}\rangle=-\frac{\tau_{e}}{\tau_{p}}\langle P_{2}\rangle=\frac{R_{sp}}{aP_{s}}\,.\label{<N2>result}
\end{equation}
By (\ref{P vs N})
\begin{equation}
N_{0}-N_{s}=\frac{R_{sp}}{aP_{s}}
\end{equation}
so the 2nd order result (\ref{Freq shift 1}) gives $\hat{\omega}=0$.
This means that the spontaneous emission does not give rise to a shift
in the lasing frequency.

The average photon number is to 2nd order given by
\begin{equation}
\langle{\cal {P}}\rangle=P_{s}+\langle P_{2}\rangle=P_{s}-\frac{R_{sp}\tau_{p}}{aP_{s}\tau_{e}}\label{Average P}
\end{equation}
according to (\ref{<N2>result}). The stationary solution $P_{s}$
is itself a function of $R_{sp}$. We derive a simple relation for
$P_{s}$ from (\ref{Steady N}), i.e. from
\begin{gather}
J_{s}=\frac{N_{s}}{\tau_{e}}+a(N_{s}-N_{tr})P_{s}\nonumber \\
=\frac{N_{s}-N_{0}}{\tau_{e}}+\frac{N_{0}}{\tau_{e}}+a(N_{s}-N_{0})P_{s}+a(N_{0}-N_{tr})P_{s}\nonumber \\
=-\frac{R_{sp}}{aP_{s}\tau_{e}}+\frac{N_{0}}{\tau_{e}}-R_{sp}+\frac{P_{s}}{\tau_{p}}\label{Alt Js}
\end{gather}
where we have used that (\ref{P vs N}) implies $a(N_{0}-N_{s})P_{s}=R_{sp}$
and (\ref{Clamped N0}) implies $a(N_{0}-N_{tr})=1/\tau_{p}$. From
(\ref{Alt Js}) we get
\begin{equation}
P_{s}=\tau_{p}\left(J_{s}-\frac{N_{0}}{\tau_{e}}\right)+R_{sp}\tau_{p}\left(1+\frac{1}{aP_{s}\tau_{e}}\right)=P_{0}+R_{sp}\tau_{p}\left(1+\frac{1}{aP_{s}\tau_{e}}\right)\label{Alt Ps}
\end{equation}
above threshold; here (\ref{P0 vs I0}) and $J_{th}=N_{0}/\tau_{e}$
were used. Combining (\ref{Average P}) and (\ref{Alt Ps}) gives
\begin{equation}
\langle{\cal {P}}\rangle=P_{0}+R_{sp}\tau_{p}\label{Average P-1}
\end{equation}
above threshold and to 2nd order. So after long calculations we have
obtained the almost trivial result that spontaneous emission increases
the average photon number by the number of spontaneously emitted photons
during the lifetime of photons in the cavity.

\section{Power spectral density of the external field}

The optical field envelope $E(t)$ for which we have derived the power
spectral density in the previous section is an internal field in the
laser. The output field envelope at the right laser facet is proportional
to
\begin{equation}
E_{out}(t)=\sqrt{P_{out}(t)}e^{j\psi(t)}\label{Eout}
\end{equation}
where $P_{out}(t)$ is the output power (\ref{Pout+F0}). The phase
$\psi(t)$ is modified compared to the phase $\varphi(t)$ of the
internal field envelope $E(t)$ by the partition noise or shot noise
added at the laser facet. In the work of Yamamoto et al.~\cite{Yamamoto-1986,Yamamoto-1986PRA}
(see also \cite{Tromborg-1994,Haus-2000}) the shot noise is included
as a quantum optics phenomenon due to beating between the output signal
and the vacuum field. It is shown that the power spectral density
of $\psi(t)$ is
\begin{equation}
S_{\psi}(f)=S_{\varphi}(f)+\frac{\hbar\omega_{0}}{4\langle P_{out}\rangle}\,.\label{Spectrum-psi}
\end{equation}
The ratio $\hbar\omega_{0}/\langle P_{out}\rangle$ is the familiar
shot noise term that also appeared in the $\textrm{RIN}$ spectrum
in (\ref{eq:RIN(f)}). The term was here introduced as a partition
noise caused by the exit laser facet that either transmit or reflect
the photons in the laser waveguide. Without going into quantum optics
arguments we can see how the factor $1/4$ comes about.

By (\ref{delta Pout}) we have $P_{out}(t)=\langle P_{out}\rangle+\delta P_{out}$
so for $E_{0}=\sqrt{\langle P_{out}\rangle}$ a 1st order expansion
gives
\begin{equation}
E_{out}(t)\simeq E_{0}+\frac{\delta P_{out}(t)}{2E_{0}}+jE_{0}\psi(t)=E_{0}+E_{1}+jE_{2}\label{Eout-exp}
\end{equation}
where $E_{1}=\tfrac{1}{2}\delta P_{out}(t)/E_{0}$ and $E_{2}=E_{0}\psi(t)$
are the in-phase part and the quadrature part of the noise contribution,
respectively. The power spectral density of $E_{1}$ is given by
\begin{equation}
S_{E_{1}}(f)=\left(\frac{1}{2E_{0}}\right)^{2}S_{\delta P_{out}}=\frac{1}{4\langle P_{out}\rangle}S_{\delta P_{out}}=\frac{1}{4}\langle P_{out}\rangle\textrm{RIN}(f)\label{Spectrum-SE1}
\end{equation}
where (\ref{RIN}) was used, and the power spectral density of $E_{2}$
is
\begin{equation}
S_{E_{2}}(f)=\langle P_{out}\rangle S_{\psi}(f)\,.\label{Spectrum-SE2}
\end{equation}
From (\ref{Spectrum-SE1}) and (\ref{eq:RIN(f)}) we see that $S_{E_{1}}(f)$
gets the contribution $\hbar\omega_{0}/4$ from shot noise. If we
assume that the in-phase and quadrature components of the envelope
field get the same contribution from shot noise it follows from (\ref{Spectrum-SE2})
that $S_{\psi}(f)$ gets the contribution $\frac{1}{4}\hbar\omega_{0}/\langle P_{out}\rangle$
in (\ref{Spectrum-psi}).

It is essential that the shot noise is limited by a filter. The expression
for the variance $\langle(\Delta\psi)^{2}\rangle$, which appears
in the expression for the power spectral density of the external field,
must similarly to $\langle(\Delta\varphi)^{2}\rangle$ in (\ref{eq:phi-2})
be
\begin{equation}
\langle(\Delta\psi)^{2}\rangle=2\int_{0}^{B}4S_{\psi}(f)\sin^{2}(\omega\tau/2)df=2\int_{0}^{B}S_{\dot{\psi}}(f)\frac{\sin^{2}(\omega\tau/2)}{\left(\omega/2\right)^{2}}df\label{Variance-psi}
\end{equation}
with the filter bandwidth $B$ as the upper integration limit. The
integral diverges due to the shot noise term in (\ref{Spectrum-psi})
for $B\rightarrow\infty$.

\begin{figure}
\begin{centering}
\includegraphics[scale=0.5]{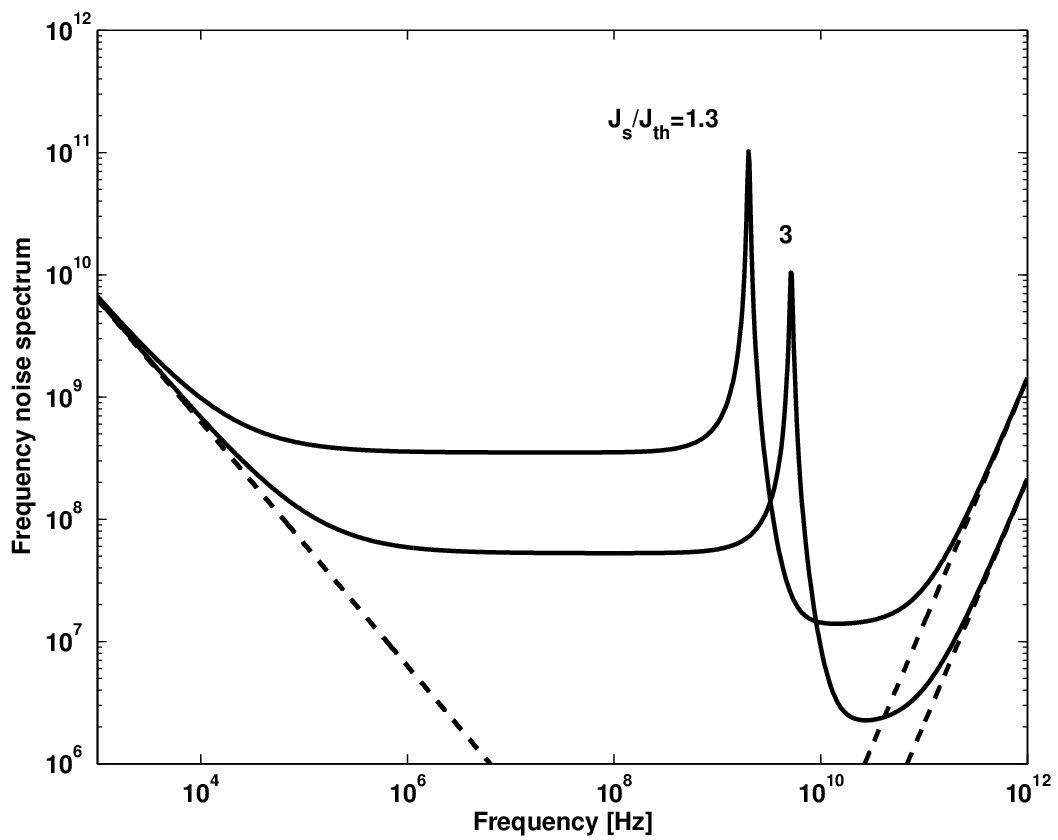}
\par\end{centering}

\caption{\selectlanguage{american}%
Frequency noise spectrum $S_{\dot{\psi}}(f)$ for $J_{s}/J_{th}=1.3$
and $3$ . The dashed curve at low frequencies is the $1/f$-noise
contribution for $\omega_{N}=6.3\cdot10^{6}\ s^{-1}$ and the dashed
curves at high frequencies are the shot noise contributions.\selectlanguage{english}%
}

\selectlanguage{american}%
\label{Ext-FM-noise-spectum}
\end{figure}

We also have to take into account that in practice there is a low-frequency
contribution to the FM-noise spectrum $S_{\dot{\psi}}(f)$ of the
form $\omega_{N}^{2}/|\omega|$ where $\omega_{N}$ is a constant
that depends on the type of laser \cite{Kikuchi-1985,Okoshi-1988,Petermann-1991,Kikuchi-2012}
but does not depend on the laser output power. The contribution is
named $1/f$-noise and it typically dominates the spectrum for $f<100\ \textrm{kHz}$
\cite{Kikuchi-2012}. The origin of the $1/f$-noise is not fully
understood and the topic is still a very active field of research.
The noise term makes the integral in (\ref{Variance-psi}) diverge
logarithmically at $f=0$. The divergence problem is dealt with by
noting that a measurement that takes the time $T$ does not involve
lower frequencies than $f=1/T$. The integral in (\ref{Variance-psi})
should therefore also have a lower cut-off at $f=1/T$ \cite{Okoshi-1988,Petermann-1991}.
Including $1/f$ and shot noise the FM-noise spectrum becomes
\begin{equation}
S_{\dot{\psi}}(f)=\omega^{2}S_{\psi}(f)=S_{\dot{\varphi}}(f)+\frac{\omega_{N}^{2}}{|\omega|}+\frac{\hbar\omega_{0}}{4\langle P_{out}\rangle}\omega^{2}\,.
\end{equation}
The example of $S_{\dot{\varphi}}(f)$ shown in Figure \ref{Frequency noise spectrum}
for $J_{s}/J_{th}=1.3$ and $3$ will now give the curves in Figure
\ref{Ext-FM-noise-spectum} showing $S_{\dot{\psi}}(f)$ as the solid
curves. The $1/f$-noise contribution is shown as the dashed line
at low frequencies. We use the value $\omega_{N}=6.3\cdot10^{6}\ \textrm{s}^{-1}$
from \cite{Petermann-1991}. The shot noise contributions are shown
as the dashed lines at high frequencies. For increasing output power
both the low and high frequency part of $S_{\dot{\varphi}}(f)$ and
the shot noise term scales as $1/P_{out}$ while the $1/f$-noise
term is unchanged. The latter will therefore influence the spectrum
at increasingly higher frequencies for increasing output power.

\subsubsection{Measurement of the variance $\langle(\Delta\psi)^{2}\rangle$}

\begin{figure}
\begin{centering}
\includegraphics[scale=0.55]{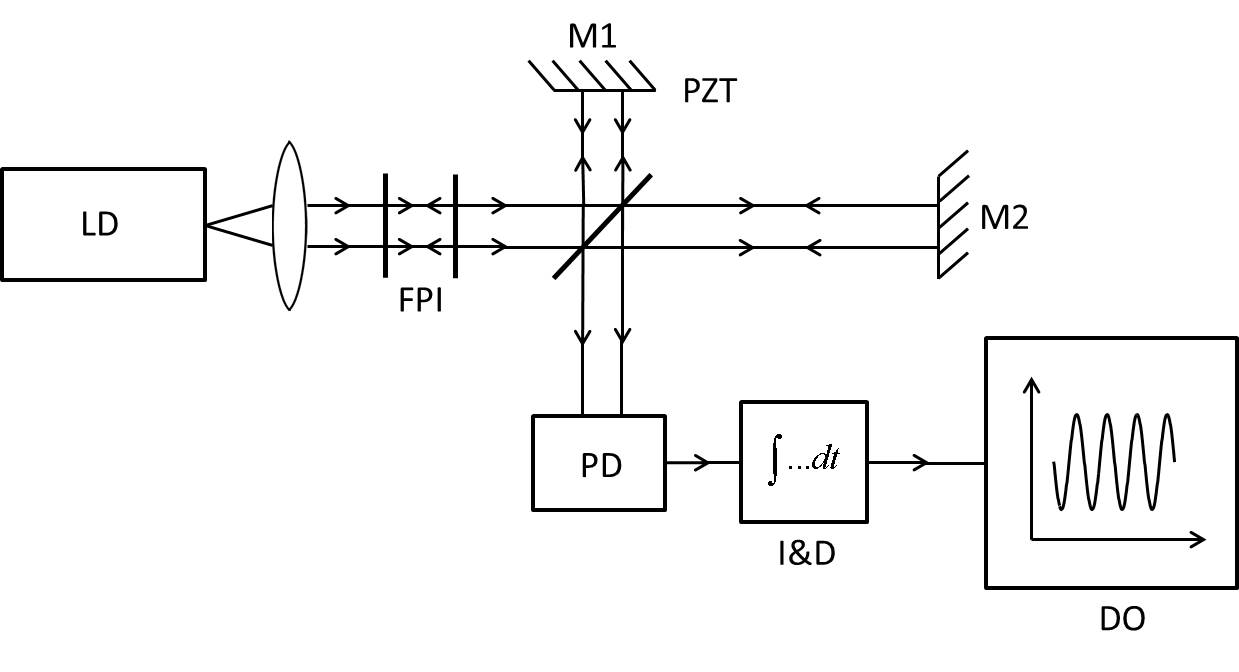}
\par\end{centering}

\caption{\selectlanguage{american}%
Experimental set-up for measuring the variance $\left\langle \left(\Delta\varphi\right)^{2}\right\rangle $
as a function of the delay between the laser beams in the interferometer.
LD: laser diode. FPI: Fabry-Perot interferometer. M1: PZT driven mirror.
M2: mirror that can be shifted on a larger scale. PD: photodiode.
I\&D: integrate-and-dump filter. DO: digital oscilloscope.\selectlanguage{english}%
}

\selectlanguage{american}%
\label{Expvariance}
\end{figure}

The variance can be measured by the experimental setup shown in Figure
\ref{Expvariance}. It is based on a Michelson interferometer where
the output field from the laser is added to a delayed version of the
same field and detected by a photodetector \cite{Daino-1983,Eichen-1984}.
If we ignore intensity fluctuations the complex field is proportional
to $E_{0}e^{j(\omega_{0}t+\psi(t))}$ and the delayed complex field
to $E_{0}e^{j(\omega_{0}(t-\tau)+\psi(t-\tau))}$. The time delay
$\tau$ is varied by changing the relative lengths of the two arms
of the interferometer. To obtain a delay of up to $\tau=10\ \textrm{ns}$
as in Figure \ref{Variance versus phase delay} the path-length difference
has to be $\tau c=3\ \textrm{m}$, i.e. one arm of the interferometer
has to be $1.5\ \textrm{m}$ longer than the other. The current from
the photodetector is proportional to
\begin{equation}
\biggl|E_{0}e^{j(\omega_{0}t+\psi(t))}+E_{0}e^{j(\omega_{0}(t-\tau)+\psi(t-\tau))}\biggr|^{2}=2E_{0}^{2}(1+\cos(\omega_{0}\tau+\Delta\psi(t)))
\end{equation}
where $\Delta\psi(t)=\psi(t)-\psi(t-\tau)$. The optical receiver
is assumed to be an integrate-and-dump receiver such that the output
signal is proportional to
\begin{equation}
V=\int_{0}^{T}h(T-t)(1+\cos(\omega_{0}\tau+\Delta\psi(t))dt\label{Int&dump-signal}
\end{equation}
where $T$ is the integration time and $h(t)$ is the impulse response
for the receiver. The signal $V$ is sampled by a digital oscilloscope.
The ensemble average of $V$ is
\begin{equation}
\langle V\rangle=\int_{0}^{T}h(T-t)(1+\langle\cos(\omega_{0}\tau+\Delta\psi(t))\rangle)dt
\end{equation}
and since by (\ref{eq:avr. ejdeltafi})
\begin{flalign}
\langle\cos(\omega_{0}\tau+\Delta\psi(t))\rangle & =\frac{1}{2}e^{j\omega_{0}\tau}\langle e^{j\Delta\psi(t)}\rangle+\frac{1}{2}e^{-j\omega_{0}\tau}\langle e^{-j\Delta\psi(t)}\rangle\nonumber \\
 & =e^{-\tfrac{1}{2}\langle(\Delta\psi)^{2}\rangle}\cos(\omega_{0}\tau)\label{Cos-ensemble-av}
\end{flalign}
the average is simply
\begin{equation}
\langle V\rangle=H\left(1+e^{-\tfrac{1}{2}\langle(\Delta\psi)^{2}\rangle}\cos(\omega_{0}\tau)\right)\label{<V>}
\end{equation}
where $H=\int_{0}^{T}h(t)dt$. By moving one of the mirrors of the
interferometer within a few optical wavelengths ($\lambda_{0}=2\pi c/\omega_{0}$),
e.g. by means of a piezoelectric transducer, the factor $\cos(\omega_{0}\tau)$
will oscillate over a few periods. The signal $\langle V\rangle$
oscillates sinusoidally between maximum $\langle V\rangle_{max}$
for $\cos(\omega_{0}\tau)=1$ and minimum $\langle V\rangle_{min}$
for $\cos(\omega_{0}\tau)=-1$ so $e^{-\tfrac{1}{2}\langle(\Delta\psi)^{2}\rangle}$
is derived as the fringe visibility
\begin{equation}
e^{-\tfrac{1}{2}\langle(\Delta\psi)^{2}\rangle}=\frac{\langle V\rangle_{max}-\langle V\rangle_{min}}{\langle V\rangle_{max}+\langle V\rangle_{min}}\label{Fringe-visibility}
\end{equation}
for given delay $\tau$. The variance $\langle(\Delta\psi)^{2}\rangle$
as a function of $\tau$ is obtained by repeating the fringe visibility
measurement for different positions of the other mirror. We refer
to \cite{Eichen-1984} for a measurement of the variance versus time
delay which agrees very well with a calculation based on the theoretical
expression (\ref{variance phi-7}) i.e. without including $1/f$-noise
and shot noise. The latter noise contributions may not be important
for the given integration time and measurement bandwidth.

\section{Postscript}
Professor Bjarne Tromborg, died on June 11, 2025, at the age of 84, after a short illness.
Throughout his long life, Bjarne had carried out highly respected work within first particle physics and later photonics. As a researcher, research leader and teacher, he left a clear mark on both colleagues and students.
I myself had the great pleasure of working with Bjarne as co-author of the book "Optical Communications from a Fourier Perspective: Fourier Theory and Optical Devices and Systems", Palle Jeppesen and Bjarne Tromborg, Elsevier 2024. After his passing, it was my colleague, Professor Jesper M{\o}rk, who adapted and finalized the manuscript for the present article "Spectra of laser diodes". I would like to express my warm thanks to Jesper for his work.
Working on the text reminded Jesper and me of Bjarne's great professional insight, thoroughness, precision and commitment to making the mathematical and physical foundations clear.
With Bjarne's passing, we lost a wise and committed physicist and a very kind person.
May his memory be honored,
Palle Jeppesen


\bibliographystyle{ieeetr}
\bibliography{references}

\end{document}